\documentclass[10pt]{article} 
\usepackage[preprint]{tmlr}

\usepackage{hyperref}
\usepackage{url}
\usepackage[T1]{fontenc}
\usepackage{amsmath,amssymb,bm,booktabs,graphicx,xcolor,microtype,tikz,placeins,adjustbox,float}
\usetikzlibrary{arrows.meta,positioning,calc}

\newcommand{\E}{\mathbb E}
\newcommand{\C}{\mathbb C}
\newcommand{\norm}[1]{\left\lVert #1\right\rVert}
\newcommand{\Oagg}{MaRFlow-O (aggregate)}

\newcommand{\fitTable}[1]{\begin{adjustbox}{max width=\linewidth}#1\end{adjustbox}}

\newcommand{\MainResults}{%
\begin{tabular}{lrrrrrr}
\toprule
Model & PESQ & STOI & ESTOI & SI-SDR & P.808 & OVRL \\
\midrule
\multicolumn{7}{l}{\emph{VoiceBank+DEMAND ($n=824$)}}\\
Noisy & 1.97 & 0.921 & 0.787 & 8.45 & 3.05 & 2.70 \\
FlowSE & 2.39 & 0.922 & 0.820 & 17.41 & 3.48 & 3.10 \\
MaRFlow-O (agg.) & 2.41 & 0.923 & 0.823 & 17.61 & 3.50 & 3.10 \\
MaRFlow-O & \textbf{2.62} & 0.931 & 0.840 & \textbf{17.85} & 3.52 & 3.14 \\
MaRFlow-R (agg.) & 2.34 & 0.920 & 0.817 & 17.36 & 3.47 & 3.07 \\
MaRFlow-R (local) & 2.59 & \textbf{0.932} & \textbf{0.840} & 17.84 & \textbf{3.55} & \textbf{3.17} \\
\midrule
\multicolumn{7}{l}{\emph{DNS 2020 ($n=150$)}}\\
Noisy & 1.58 & 0.915 & 0.810 & 9.23 & 3.16 & 2.48 \\
FlowSE & 1.70 & 0.905 & 0.810 & 12.64 & 3.78 & 3.09 \\
MaRFlow-O (agg.) & 1.74 & 0.912 & 0.821 & 12.94 & 3.80 & 3.14 \\
MaRFlow-O & 1.97 & 0.932 & 0.859 & 14.00 & 3.88 & 3.20 \\
MaRFlow-R (agg.) & 1.63 & 0.899 & 0.797 & 12.11 & 3.73 & 3.05 \\
MaRFlow-R (local) & \textbf{2.00} & \textbf{0.934} & \textbf{0.864} & \textbf{14.11} & \textbf{3.92} & \textbf{3.24} \\
\bottomrule
\end{tabular}
}
\newcommand{\SNRVoiceBank}{%
\begin{tabular}{llrrrrrr}
\toprule
Input SNR (dB) & Model & PESQ & STOI & ESTOI & SI-SDR & P.808 & OVRL \\
\midrule
$<0$ ($n=16$) & Noisy & 1.27 & 0.778 & 0.550 & -0.65 & 2.70 & 2.30 \\
 & FlowSE & 1.71 & 0.784 & 0.606 & 15.27 & 3.16 & 2.85 \\
 & MaRFlow-O (agg.) & 1.72 & 0.785 & 0.617 & 15.21 & 3.16 & 2.89 \\
 & MaRFlow-O & \textbf{2.01} & 0.803 & 0.639 & \textbf{15.81} & 3.24 & 2.94 \\
 & MaRFlow-R (agg.) & 1.66 & 0.781 & 0.599 & 14.93 & 3.13 & 2.81 \\
 & MaRFlow-R (local) & 1.92 & \textbf{0.806} & \textbf{0.646} & 15.63 & \textbf{3.28} & \textbf{2.95} \\
\midrule
$[0,5)$ ($n=202$) & Noisy & 1.47 & 0.876 & 0.681 & 1.29 & 2.79 & 2.27 \\
 & FlowSE & 1.89 & 0.883 & 0.743 & 13.93 & 3.37 & 2.99 \\
 & MaRFlow-O (agg.) & 1.89 & 0.884 & 0.748 & 14.06 & 3.39 & 2.99 \\
 & MaRFlow-O & \textbf{2.11} & 0.897 & 0.773 & \textbf{14.70} & 3.45 & 3.07 \\
 & MaRFlow-R (agg.) & 1.84 & 0.879 & 0.737 & 13.79 & 3.35 & 2.94 \\
 & MaRFlow-R (local) & 2.08 & \textbf{0.898} & \textbf{0.775} & 14.65 & \textbf{3.48} & \textbf{3.11} \\
\midrule
$[5,10)$ ($n=209$) & Noisy & 1.74 & 0.917 & 0.767 & 6.35 & 2.99 & 2.66 \\
 & FlowSE & 2.23 & 0.920 & 0.812 & 16.90 & 3.49 & 3.10 \\
 & MaRFlow-O (agg.) & 2.26 & 0.921 & 0.815 & 17.07 & 3.51 & 3.11 \\
 & MaRFlow-O & \textbf{2.47} & 0.930 & 0.833 & 17.32 & 3.54 & 3.14 \\
 & MaRFlow-R (agg.) & 2.18 & 0.918 & 0.808 & 16.85 & 3.48 & 3.07 \\
 & MaRFlow-R (local) & 2.45 & \textbf{0.931} & \textbf{0.834} & \textbf{17.33} & \textbf{3.56} & \textbf{3.18} \\
\midrule
$\geq10$ ($n=397$) & Noisy & 2.37 & 0.952 & 0.860 & 13.56 & 3.24 & 2.95 \\
 & FlowSE & 2.76 & 0.948 & 0.872 & 19.53 & 3.55 & 3.16 \\
 & MaRFlow-O (agg.) & 2.78 & 0.948 & 0.873 & 19.80 & 3.56 & 3.17 \\
 & MaRFlow-O & \textbf{2.98} & \textbf{0.955} & \textbf{0.886} & \textbf{19.82} & 3.57 & 3.18 \\
 & MaRFlow-R (agg.) & 2.72 & 0.948 & 0.872 & 19.54 & 3.54 & 3.14 \\
 & MaRFlow-R (local) & 2.94 & 0.954 & 0.885 & 19.81 & \textbf{3.58} & \textbf{3.21} \\
\bottomrule
\end{tabular}
}
\newcommand{\SNRDNS}{%
\begin{tabular}{llrrrrrr}
\toprule
Input SNR (dB) & Model & PESQ & STOI & ESTOI & SI-SDR & P.808 & OVRL \\
\midrule
$<0$ ($n=5$) & Noisy & 1.15 & 0.819 & 0.642 & 0.04 & 2.85 & 2.15 \\
 & FlowSE & 1.16 & 0.782 & 0.622 & 3.53 & 3.50 & 2.86 \\
 & MaRFlow-O (agg.) & 1.19 & 0.795 & 0.647 & 3.69 & 3.50 & 3.00 \\
 & MaRFlow-O & \textbf{1.29} & 0.834 & 0.705 & 4.85 & 3.66 & 3.09 \\
 & MaRFlow-R (agg.) & 1.16 & 0.774 & 0.615 & 3.34 & 3.43 & 2.90 \\
 & MaRFlow-R (local) & 1.29 & \textbf{0.846} & \textbf{0.722} & \textbf{5.23} & \textbf{3.78} & \textbf{3.19} \\
\midrule
$[0,5)$ ($n=36$) & Noisy & 1.33 & 0.868 & 0.723 & 3.23 & 2.93 & 2.29 \\
 & FlowSE & 1.38 & 0.845 & 0.713 & 8.50 & 3.58 & 2.99 \\
 & MaRFlow-O (agg.) & 1.43 & 0.855 & 0.729 & 8.97 & 3.60 & 3.04 \\
 & MaRFlow-O & 1.55 & \textbf{0.884} & 0.778 & 10.01 & 3.72 & 3.14 \\
 & MaRFlow-R (agg.) & 1.35 & 0.840 & 0.701 & 8.12 & 3.53 & 2.94 \\
 & MaRFlow-R (local) & \textbf{1.57} & 0.884 & \textbf{0.778} & \textbf{10.02} & \textbf{3.78} & \textbf{3.18} \\
\midrule
$[5,10)$ ($n=43$) & Noisy & 1.38 & 0.905 & 0.784 & 7.31 & 3.06 & 2.31 \\
 & FlowSE & 1.48 & 0.897 & 0.792 & 10.73 & 3.78 & 3.04 \\
 & MaRFlow-O (agg.) & 1.53 & 0.905 & 0.804 & 11.01 & 3.79 & 3.08 \\
 & MaRFlow-O & 1.73 & 0.927 & 0.845 & 11.99 & 3.86 & 3.16 \\
 & MaRFlow-R (agg.) & 1.43 & 0.891 & 0.779 & 10.36 & 3.70 & 2.99 \\
 & MaRFlow-R (local) & \textbf{1.77} & \textbf{0.931} & \textbf{0.854} & \textbf{12.24} & \textbf{3.89} & \textbf{3.20} \\
\midrule
$\geq10$ ($n=66$) & Noisy & 1.89 & 0.955 & 0.887 & 14.45 & 3.37 & 2.73 \\
 & FlowSE & 2.06 & 0.953 & 0.889 & 16.83 & 3.92 & 3.20 \\
 & MaRFlow-O (agg.) & 2.10 & 0.956 & 0.896 & 17.05 & 3.93 & 3.24 \\
 & MaRFlow-O & 2.42 & 0.968 & 0.923 & 18.17 & 4.00 & 3.28 \\
 & MaRFlow-R (agg.) & 1.94 & 0.946 & 0.876 & 16.09 & 3.88 & 3.16 \\
 & MaRFlow-R (local) & \textbf{2.45} & \textbf{0.970} & \textbf{0.928} & \textbf{18.24} & \textbf{4.02} & \textbf{3.30} \\
\bottomrule
\end{tabular}
}
\newcommand{\PairedResults}{%
\begin{tabular}{llrrrrr}
\toprule
Comparison & Metric & Mean gain & 95\% CI & Win (\%) & $p$ & $p_{\rm Holm}$ \\
\midrule
\multicolumn{7}{l}{\emph{VoiceBank+DEMAND}}\\
O vs. FlowSE & PESQ & +0.23 & [0.21, 0.24] & 92.8 & $5.00\times10^{-5}$ & $0.0014$ \\
 & SI-SDR & +0.44 & [0.40, 0.49] & 87.4 & $5.00\times10^{-5}$ & $0.0014$ \\
O-agg vs. FlowSE & PESQ & +0.02 & [0.01, 0.03] & 62.7 & $5.00\times10^{-5}$ & $0.0014$ \\
 & SI-SDR & +0.20 & [0.17, 0.24] & 79.4 & $5.00\times10^{-5}$ & $0.0014$ \\
R-agg vs. O-agg & PESQ & -0.07 & [-0.07, -0.06] & 26.6 & $5.00\times10^{-5}$ & $0.0014$ \\
 & SI-SDR & -0.25 & [-0.28, -0.22] & 18.7 & $5.00\times10^{-5}$ & $0.0014$ \\
O vs. O-agg & PESQ & +0.21 & [0.19, 0.22] & 92.0 & $5.00\times10^{-5}$ & $0.0014$ \\
 & SI-SDR & +0.24 & [0.19, 0.29] & 62.1 & $5.00\times10^{-5}$ & $0.0014$ \\
R-local vs. R-agg & PESQ & +0.24 & [0.23, 0.26] & 89.7 & $5.00\times10^{-5}$ & $0.0014$ \\
 & SI-SDR & +0.48 & [0.42, 0.53] & 77.2 & $5.00\times10^{-5}$ & $0.0014$ \\
R-local vs. FlowSE & PESQ & +0.20 & [0.18, 0.21] & 88.2 & $5.00\times10^{-5}$ & $0.0014$ \\
 & SI-SDR & +0.43 & [0.39, 0.47] & 87.1 & $5.00\times10^{-5}$ & $0.0014$ \\
R-local vs. O & PESQ & -0.03 & [-0.04, -0.02] & 40.5 & $5.00\times10^{-5}$ & $0.0014$ \\
 & SI-SDR & -0.02 & [-0.04, 0.01] & 50.0 & $0.2942$ & $0.3627$ \\
\midrule
\multicolumn{7}{l}{\emph{DNS 2020}}\\
O vs. FlowSE & PESQ & +0.28 & [0.24, 0.31] & 98.7 & $5.00\times10^{-5}$ & $0.0014$ \\
 & SI-SDR & +1.36 & [1.16, 1.57] & 91.3 & $5.00\times10^{-5}$ & $0.0014$ \\
O-agg vs. FlowSE & PESQ & +0.05 & [0.03, 0.07] & 70.7 & $5.00\times10^{-5}$ & $0.0014$ \\
 & SI-SDR & +0.30 & [0.16, 0.46] & 70.0 & $5.00\times10^{-5}$ & $0.0014$ \\
R-agg vs. O-agg & PESQ & -0.12 & [-0.14, -0.10] & 9.3 & $5.00\times10^{-5}$ & $0.0014$ \\
 & SI-SDR & -0.82 & [-1.01, -0.66] & 13.3 & $5.00\times10^{-5}$ & $0.0014$ \\
O vs. O-agg & PESQ & +0.23 & [0.20, 0.26] & 97.3 & $5.00\times10^{-5}$ & $0.0014$ \\
 & SI-SDR & +1.06 & [0.91, 1.22] & 90.0 & $5.00\times10^{-5}$ & $0.0014$ \\
R-local vs. R-agg & PESQ & +0.38 & [0.33, 0.42] & 97.3 & $5.00\times10^{-5}$ & $0.0014$ \\
 & SI-SDR & +2.00 & [1.74, 2.29] & 97.3 & $5.00\times10^{-5}$ & $0.0014$ \\
R-local vs. FlowSE & PESQ & +0.31 & [0.27, 0.35] & 96.0 & $5.00\times10^{-5}$ & $0.0014$ \\
 & SI-SDR & +1.48 & [1.25, 1.72] & 96.7 & $5.00\times10^{-5}$ & $0.0014$ \\
R-local vs. O & PESQ & +0.03 & [0.02, 0.05] & 59.3 & $1.50\times10^{-4}$ & $0.0014$ \\
 & SI-SDR & +0.12 & [-0.05, 0.29] & 56.7 & $0.1813$ & $0.3627$ \\
\bottomrule
\end{tabular}
}
\newcommand{\ComplexityResults}{%
\begin{tabular}{lrrrrrr}
\toprule
Model & VB ms & RTF & MiB & DNS ms & RTF & MiB \\
\midrule
FlowSE & 410.99 & 0.177 & 2404.0 & 1743.51 & 0.174 & 2404.0 \\
MaRFlow-O (agg.) & 412.33 & 0.176 & 2402.0 & 1748.92 & 0.175 & 2402.0 \\
MaRFlow-O & 412.65 & 0.176 & 2402.0 & 1750.07 & 0.175 & 2402.0 \\
MaRFlow-R (agg.) & 412.36 & 0.176 & 2402.0 & 1749.03 & 0.175 & 2402.0 \\
MaRFlow-R (local) & 411.65 & 0.176 & 2404.0 & 1751.34 & 0.175 & 2404.0 \\
\bottomrule
\end{tabular}
}
\newcommand{\NFEPESQ}{%
\begin{tabular}{lrrrrr}
\toprule
Model & 1 & 2 & 3 & 5 & 10 \\
\midrule
\multicolumn{6}{l}{\emph{VoiceBank+DEMAND}}\\
FlowSE & 2.37 & 2.44 & 2.39 & 2.39 & 2.35 \\
MaRFlow-O (agg.) & 2.45 & 2.48 & 2.44 & 2.41 & 2.39 \\
MaRFlow-O & 2.50 & 2.62 & 2.64 & 2.62 & 2.58 \\
MaRFlow-R (agg.) & 2.37 & 2.40 & 2.37 & 2.34 & 2.33 \\
MaRFlow-R (local) & 2.52 & 2.57 & 2.58 & 2.59 & 2.57 \\
\midrule
\multicolumn{6}{l}{\emph{DNS 2020}}\\
FlowSE & 1.85 & 1.81 & 1.73 & 1.70 & 1.70 \\
MaRFlow-O (agg.) & 1.95 & 1.88 & 1.79 & 1.74 & 1.74 \\
MaRFlow-O & 2.10 & 2.06 & 1.99 & 1.97 & 1.96 \\
MaRFlow-R (agg.) & 1.78 & 1.76 & 1.67 & 1.63 & 1.62 \\
MaRFlow-R (local) & 2.12 & 2.07 & 2.03 & 2.00 & 1.99 \\
\bottomrule
\end{tabular}
}
\newcommand{\NFESDR}{%
\begin{tabular}{lrrrrr}
\toprule
Model & 1 & 2 & 3 & 5 & 10 \\
\midrule
\multicolumn{6}{l}{\emph{VoiceBank+DEMAND}}\\
FlowSE & 17.97 & 18.20 & 18.09 & 17.41 & 16.68 \\
MaRFlow-O (agg.) & 18.07 & 18.40 & 18.25 & 17.61 & 16.89 \\
MaRFlow-O & 18.56 & 18.84 & 18.61 & 17.85 & 17.08 \\
MaRFlow-R (agg.) & 17.62 & 17.96 & 17.99 & 17.36 & 16.70 \\
MaRFlow-R (local) & 17.42 & 18.35 & 18.48 & 17.84 & 17.09 \\
\midrule
\multicolumn{6}{l}{\emph{DNS 2020}}\\
FlowSE & 12.89 & 13.02 & 13.15 & 12.64 & 12.20 \\
MaRFlow-O (agg.) & 13.09 & 13.33 & 13.49 & 12.94 & 12.47 \\
MaRFlow-O & 14.46 & 14.67 & 14.55 & 14.00 & 13.48 \\
MaRFlow-R (agg.) & 12.32 & 12.50 & 12.63 & 12.11 & 11.67 \\
MaRFlow-R (local) & 13.97 & 14.62 & 14.65 & 14.11 & 13.58 \\
\bottomrule
\end{tabular}
}
\newcommand{\NFESTOI}{%
\begin{tabular}{lrrrrr}
\toprule
Model & 1 & 2 & 3 & 5 & 10 \\
\midrule
\multicolumn{6}{l}{\emph{VoiceBank+DEMAND}}\\
FlowSE & 0.925 & 0.924 & 0.924 & 0.922 & 0.919 \\
MaRFlow-O (agg.) & 0.927 & 0.926 & 0.925 & 0.923 & 0.921 \\
MaRFlow-O & 0.934 & 0.934 & 0.934 & 0.931 & 0.929 \\
MaRFlow-R (agg.) & 0.924 & 0.924 & 0.923 & 0.920 & 0.918 \\
MaRFlow-R (local) & 0.931 & 0.933 & 0.933 & 0.932 & 0.930 \\
\midrule
\multicolumn{6}{l}{\emph{DNS 2020}}\\
FlowSE & 0.917 & 0.917 & 0.911 & 0.905 & 0.902 \\
MaRFlow-O (agg.) & 0.920 & 0.920 & 0.917 & 0.912 & 0.909 \\
MaRFlow-O & 0.935 & 0.936 & 0.935 & 0.932 & 0.930 \\
MaRFlow-R (agg.) & 0.913 & 0.913 & 0.905 & 0.899 & 0.896 \\
MaRFlow-R (local) & 0.936 & 0.938 & 0.937 & 0.934 & 0.932 \\
\bottomrule
\end{tabular}
}
\newcommand{\NFEMOS}{%
\begin{tabular}{lrrrrr}
\toprule
Model & 1 & 2 & 3 & 5 & 10 \\
\midrule
\multicolumn{6}{l}{\emph{VoiceBank+DEMAND}}\\
FlowSE & 3.00 & 3.06 & 3.09 & 3.10 & 3.09 \\
MaRFlow-O (agg.) & 3.02 & 3.07 & 3.09 & 3.10 & 3.10 \\
MaRFlow-O & 3.06 & 3.12 & 3.14 & 3.14 & 3.14 \\
MaRFlow-R (agg.) & 2.97 & 3.02 & 3.07 & 3.07 & 3.07 \\
MaRFlow-R (local) & 3.08 & 3.15 & 3.17 & 3.17 & 3.17 \\
\midrule
\multicolumn{6}{l}{\emph{DNS 2020}}\\
FlowSE & 2.99 & 3.06 & 3.09 & 3.09 & 3.08 \\
MaRFlow-O (agg.) & 3.02 & 3.11 & 3.14 & 3.14 & 3.12 \\
MaRFlow-O & 3.12 & 3.18 & 3.21 & 3.20 & 3.20 \\
MaRFlow-R (agg.) & 2.98 & 3.04 & 3.06 & 3.05 & 3.03 \\
MaRFlow-R (local) & 3.12 & 3.22 & 3.23 & 3.24 & 3.24 \\
\bottomrule
\end{tabular}
}
\newcommand{\InterventionResults}{%
\begin{tabular}{llrrrr}
\toprule
Model & Disabled & VB PESQ & SI-SDR & DNS PESQ & SI-SDR \\
\midrule
MaRFlow-O & Multiplicative & 0.19 & 7.39 & 0.13 & 2.79 \\
 & Additive & 1.58 & 27.21 & 0.95 & 24.63 \\
MaRFlow-O (agg.) & Multiplicative & 0.13 & 4.18 & 0.28 & 3.25 \\
 & Additive & 1.37 & 23.91 & 0.72 & 20.50 \\
MaRFlow-R (agg.) & Multiplicative & 1.13 & 10.81 & 0.48 & 5.38 \\
 & Additive & 1.31 & 22.81 & 0.60 & 19.62 \\
MaRFlow-R (local) & Multiplicative & 1.55 & 26.06 & 0.98 & 23.41 \\
 & Additive & 1.02 & 2.77 & 0.62 & 2.95 \\
\bottomrule
\end{tabular}
}

\newcommand{\ReverbResults}{%
\begin{tabular}{lrrrrrr}
\toprule
Model & PESQ & STOI & ESTOI & SI-SDR & P.808 & OVRL \\
\midrule
Noisy & \textbf{1.82} & \textbf{0.866} & \textbf{0.776} & \textbf{9.19} & 2.74 & 1.39 \\
FlowSE & 1.07 & 0.606 & 0.402 & 3.15 & 3.22 & 2.32 \\
O-agg & 1.08 & 0.597 & 0.388 & 2.99 & 3.19 & 2.37 \\
O & 1.09 & 0.583 & 0.372 & 2.46 & 3.18 & 2.42 \\
R-agg & 1.08 & 0.616 & 0.412 & 3.73 & 3.19 & 2.26 \\
R-local & 1.09 & 0.587 & 0.381 & 2.01 & \textbf{3.31} & \textbf{2.56} \\
\bottomrule
\end{tabular}}

\newcommand{\WERResults}{%
\begin{tabular}{lrrr}
\toprule
Model & VoiceBank & DNS non-reverb & DNS reverb \\
\midrule
Clean reference & 1.39 & 2.36 & 4.35 \\
Noisy & 2.42 & 4.33 & 10.00 \\
FlowSE & 3.67 & 8.78 & 95.45 \\
O-agg & 3.66 & 9.22 & 90.38 \\
O & \textbf{3.18} & \textbf{6.69} & 94.18 \\
R-agg & 3.45 & 10.00 & \textbf{83.54} \\
R-local & 3.26 & 7.16 & 105.00 \\
\bottomrule
\end{tabular}}

\newcommand{\WERBreakdown}{%
\begin{tabular}{lrrrr}
\toprule
Model & Sub. & Del. & Ins. & WER \\
\midrule
\multicolumn{5}{l}{\emph{VoiceBank+DEMAND}} \\
Clean reference & 0.90 & 0.24 & 0.25 & 1.39 \\
Noisy & 1.66 & 0.40 & 0.36 & 2.42 \\
FlowSE & 2.44 & 0.44 & 0.79 & 3.67 \\
O-agg & 2.60 & 0.49 & 0.57 & 3.66 \\
O & 2.28 & 0.43 & 0.47 & 3.18 \\
R-agg & 2.63 & 0.43 & 0.40 & 3.45 \\
R-local & 2.41 & 0.38 & 0.47 & 3.26 \\
\multicolumn{5}{l}{\emph{DNS non-reverb}} \\
Clean reference & 0.90 & 0.67 & 0.80 & 2.36 \\
Noisy & 1.72 & 1.44 & 1.17 & 4.33 \\
FlowSE & 5.20 & 1.99 & 1.59 & 8.78 \\
O-agg & 4.67 & 2.93 & 1.62 & 9.22 \\
O & 3.80 & 1.49 & 1.39 & 6.69 \\
R-agg & 5.10 & 3.26 & 1.64 & 10.00 \\
R-local & 4.20 & 1.49 & 1.47 & 7.16 \\
\multicolumn{5}{l}{\emph{DNS reverb}} \\
Clean reference & 2.11 & 0.82 & 1.42 & 4.35 \\
Noisy & 5.57 & 2.11 & 2.31 & 10.00 \\
FlowSE & 55.40 & 20.66 & 19.39 & 95.45 \\
O-agg & 52.41 & 24.37 & 13.60 & 90.38 \\
O & 62.38 & 18.25 & 13.55 & 94.18 \\
R-agg & 51.27 & 23.42 & 8.85 & 83.54 \\
R-local & 65.49 & 13.10 & 26.40 & 105.00 \\
\bottomrule
\end{tabular}}

\newcommand{\WERPaired}{%
\begin{tabular}{lrrr}
\toprule
Comparison & VoiceBank & DNS non-reverb & DNS reverb \\
\midrule
O vs. FlowSE & -0.49 [-1.64, +0.30] & -2.09 [-3.14, -1.08] & -1.27 [-13.32, +9.08] \\
\quad $p$ [$p_{\rm Holm}$] & $0.4244$ [$1$] & $1.00\times10^{-4}$ [$0.0014$] & $0.8476$ [$1$] \\
\addlinespace[2pt]
R-local vs. FlowSE & -0.41 [-1.02, +0.12] & -1.62 [-2.68, -0.58] & +9.55 [-4.96, +24.68] \\
\quad $p$ [$p_{\rm Holm}$] & $0.1705$ [$1$] & $0.0029$ [$0.029$] & $0.2283$ [$1$] \\
\addlinespace[2pt]
R-local vs. R-agg & -0.19 [-0.78, +0.37] & -2.83 [-4.09, -1.64] & +21.46 [+9.70, +35.43] \\
\quad $p$ [$p_{\rm Holm}$] & $0.5696$ [$1$] & $5.00\times10^{-5}$ [$9.00\times10^{-4}$] & $3.00\times10^{-4}$ [$0.0033$] \\
\addlinespace[2pt]
R-local vs. O & +0.08 [-0.53, +0.82] & +0.47 [-0.37, +1.36] & +10.82 [+1.53, +22.67] \\
\quad $p$ [$p_{\rm Holm}$] & $0.8721$ [$1$] & $0.322$ [$1$] & $0.0399$ [$0.3192$] \\
\addlinespace[2pt]
O vs. Noisy & +0.76 [+0.41, +1.17] & +2.36 [+1.15, +3.60] & +84.19 [+78.35, +90.74] \\
\quad $p$ [$p_{\rm Holm}$] & $5.00\times10^{-5}$ [$9.00\times10^{-4}$] & $1.00\times10^{-4}$ [$0.0014$] & $5.00\times10^{-5}$ [$9.00\times10^{-4}$] \\
\addlinespace[2pt]
R-local vs. Noisy & +0.84 [+0.23, +1.64] & +2.83 [+1.43, +4.47] & +95.00 [+83.76, +108.49] \\
\quad $p$ [$p_{\rm Holm}$] & $0.0045$ [$0.0409$] & $1.00\times10^{-4}$ [$0.0014$] & $5.00\times10^{-5}$ [$9.00\times10^{-4}$] \\
\bottomrule
\end{tabular}}

\newcommand{\ReverbPaired}{%
\begin{tabular}{lrrr}
\toprule
Comparison & PESQ & SI-SDR & OVRL \\
\midrule
O vs. FlowSE & +0.02 [+0.01, +0.02] & -0.69 [-0.92, -0.47] & +0.10 [+0.06, +0.13] \\
\quad $p$ [$p_{\rm Holm}$] & $5.00\times10^{-5}$ [$9.00\times10^{-4}$] & $5.00\times10^{-5}$ [$9.00\times10^{-4}$] & $5.00\times10^{-5}$ [$9.00\times10^{-4}$] \\
\addlinespace[2pt]
R-local vs. FlowSE & +0.01 [+0.01, +0.02] & -1.15 [-1.42, -0.88] & +0.24 [+0.21, +0.28] \\
\quad $p$ [$p_{\rm Holm}$] & $5.00\times10^{-5}$ [$9.00\times10^{-4}$] & $5.00\times10^{-5}$ [$9.00\times10^{-4}$] & $5.00\times10^{-5}$ [$9.00\times10^{-4}$] \\
\addlinespace[2pt]
R-local vs. R-agg & +0.01 [+0.00, +0.01] & -1.72 [-2.02, -1.43] & +0.31 [+0.27, +0.34] \\
\quad $p$ [$p_{\rm Holm}$] & $0.0023$ [$0.0046$] & $5.00\times10^{-5}$ [$9.00\times10^{-4}$] & $5.00\times10^{-5}$ [$9.00\times10^{-4}$] \\
\addlinespace[2pt]
R-local vs. O & -0.00 [-0.01, +0.00] & -0.45 [-0.64, -0.26] & +0.14 [+0.11, +0.17] \\
\quad $p$ [$p_{\rm Holm}$] & $0.3719$ [$0.3719$] & $5.00\times10^{-5}$ [$9.00\times10^{-4}$] & $5.00\times10^{-5}$ [$9.00\times10^{-4}$] \\
\addlinespace[2pt]
O vs. Noisy & -0.73 [-0.82, -0.64] & -6.74 [-7.54, -5.91] & +1.03 [+0.96, +1.09] \\
\quad $p$ [$p_{\rm Holm}$] & $5.00\times10^{-5}$ [$9.00\times10^{-4}$] & $5.00\times10^{-5}$ [$9.00\times10^{-4}$] & $5.00\times10^{-5}$ [$9.00\times10^{-4}$] \\
\addlinespace[2pt]
R-local vs. Noisy & -0.73 [-0.82, -0.64] & -7.19 [-8.05, -6.31] & +1.17 [+1.10, +1.24] \\
\quad $p$ [$p_{\rm Holm}$] & $5.00\times10^{-5}$ [$9.00\times10^{-4}$] & $5.00\times10^{-5}$ [$9.00\times10^{-4}$] & $5.00\times10^{-5}$ [$9.00\times10^{-4}$] \\
\bottomrule
\end{tabular}}

\newcommand{\ReverbTrainedResults}{%
\begin{tabular}{lrrrrrr}
\toprule
Model & PESQ & STOI & ESTOI & SI-SDR & P.808 & OVRL \\
\midrule
\multicolumn{7}{l}{\emph{Synthetic reverberant VoiceBank ($n=824$)}}\\
Noisy & 1.88 & 0.886 & 0.792 & 7.55 & 2.66 & 1.56 \\
FlowSE & 1.84 & 0.874 & 0.771 & 13.63 & 2.82 & 2.01 \\
O & 2.16 & 0.897 & 0.806 & 15.41 & 3.05 & 2.22 \\
R-local & \textbf{2.39} & \textbf{0.911} & \textbf{0.826} & \textbf{16.29} & \textbf{3.11} & \textbf{2.32} \\
\midrule
\multicolumn{7}{l}{\emph{Reverberant DNS 2020 ($n=150$)}}\\
Noisy & 1.82 & 0.866 & 0.776 & 9.19 & 2.74 & 1.39 \\
FlowSE & 1.66 & 0.829 & 0.722 & 9.67 & 2.97 & 1.73 \\
O & 2.09 & 0.884 & 0.792 & \textbf{12.73} & 3.11 & 1.90 \\
R-local & \textbf{2.16} & \textbf{0.890} & \textbf{0.802} & 12.44 & \textbf{3.17} & \textbf{1.97} \\
\bottomrule
\end{tabular}
}

\newcommand{\ReverbTrainedPairedVB}{%
\begin{tabular}{llrrrrr}
\toprule
Comparison & Metric & Mean gain & 95\% CI & Win (\%) & $p$ & $p_{\rm Holm}$ \\
\midrule
O vs. FlowSE & PESQ & +0.32 & [+0.30, +0.34] & 91.1 & $5.00\times10^{-5}$ & 0.0018 \\
 & STOI & +0.024 & [+0.022, +0.025] & 92.4 & $5.00\times10^{-5}$ & 0.0018 \\
 & ESTOI & +0.034 & [+0.032, +0.037] & 91.9 & $5.00\times10^{-5}$ & 0.0018 \\
 & SI-SDR & +1.79 & [+1.64, +1.94] & 91.0 & $5.00\times10^{-5}$ & 0.0018 \\
 & P.808 & +0.23 & [+0.21, +0.25] & 84.6 & $5.00\times10^{-5}$ & 0.0018 \\
 & OVRL & +0.21 & [+0.19, +0.23] & 84.2 & $5.00\times10^{-5}$ & 0.0018 \\
\addlinespace[2pt]
R-local vs. FlowSE & PESQ & +0.55 & [+0.52, +0.58] & 96.5 & $5.00\times10^{-5}$ & 0.0018 \\
 & STOI & +0.037 & [+0.035, +0.040] & 97.9 & $5.00\times10^{-5}$ & 0.0018 \\
 & ESTOI & +0.054 & [+0.052, +0.057] & 97.5 & $5.00\times10^{-5}$ & 0.0018 \\
 & SI-SDR & +2.66 & [+2.49, +2.84] & 96.2 & $5.00\times10^{-5}$ & 0.0018 \\
 & P.808 & +0.29 & [+0.27, +0.31] & 82.6 & $5.00\times10^{-5}$ & 0.0018 \\
 & OVRL & +0.31 & [+0.29, +0.33] & 86.4 & $5.00\times10^{-5}$ & 0.0018 \\
\addlinespace[2pt]
R-local vs. O & PESQ & +0.23 & [+0.21, +0.25] & 88.6 & $5.00\times10^{-5}$ & 0.0018 \\
 & STOI & +0.014 & [+0.012, +0.015] & 90.3 & $5.00\times10^{-5}$ & 0.0018 \\
 & ESTOI & +0.020 & [+0.018, +0.021] & 90.3 & $5.00\times10^{-5}$ & 0.0018 \\
 & SI-SDR & +0.88 & [+0.78, +0.98] & 86.5 & $5.00\times10^{-5}$ & 0.0018 \\
 & P.808 & +0.06 & [+0.05, +0.07] & 58.6 & $5.00\times10^{-5}$ & 0.0018 \\
 & OVRL & +0.10 & [+0.09, +0.11] & 73.7 & $5.00\times10^{-5}$ & 0.0018 \\
\bottomrule
\end{tabular}
}

\newcommand{\ReverbTrainedPairedDNS}{%
\begin{tabular}{llrrrrr}
\toprule
Comparison & Metric & Mean gain & 95\% CI & Win (\%) & $p$ & $p_{\rm Holm}$ \\
\midrule
O vs. FlowSE & PESQ & +0.43 & [+0.37, +0.48] & 98.0 & $5.00\times10^{-5}$ & 0.0018 \\
 & STOI & +0.055 & [+0.049, +0.061] & 99.3 & $5.00\times10^{-5}$ & 0.0018 \\
 & ESTOI & +0.069 & [+0.063, +0.075] & 99.3 & $5.00\times10^{-5}$ & 0.0018 \\
 & SI-SDR & +3.06 & [+2.69, +3.46] & 96.7 & $5.00\times10^{-5}$ & 0.0018 \\
 & P.808 & +0.14 & [+0.10, +0.17] & 73.3 & $5.00\times10^{-5}$ & 0.0018 \\
 & OVRL & +0.17 & [+0.11, +0.23] & 68.0 & $5.00\times10^{-5}$ & 0.0018 \\
\addlinespace[2pt]
R-local vs. FlowSE & PESQ & +0.50 & [+0.44, +0.56] & 97.3 & $5.00\times10^{-5}$ & 0.0018 \\
 & STOI & +0.061 & [+0.055, +0.068] & 97.3 & $5.00\times10^{-5}$ & 0.0018 \\
 & ESTOI & +0.080 & [+0.073, +0.087] & 98.0 & $5.00\times10^{-5}$ & 0.0018 \\
 & SI-SDR & +2.77 & [+2.34, +3.20] & 91.3 & $5.00\times10^{-5}$ & 0.0018 \\
 & P.808 & +0.19 & [+0.16, +0.23] & 80.0 & $5.00\times10^{-5}$ & 0.0018 \\
 & OVRL & +0.25 & [+0.19, +0.31] & 74.7 & $5.00\times10^{-5}$ & 0.0018 \\
\addlinespace[2pt]
R-local vs. O & PESQ & +0.08 & [+0.04, +0.11] & 74.7 & $1.00\times10^{-4}$ & 0.0018 \\
 & STOI & +0.006 & [+0.003, +0.009] & 71.3 & $1.50\times10^{-4}$ & 0.0018 \\
 & ESTOI & +0.011 & [+0.007, +0.014] & 76.0 & $5.00\times10^{-5}$ & 0.0018 \\
 & SI-SDR & -0.29 & [-0.63, +0.02] & 56.0 & 0.0744 & 0.0744 \\
 & P.808 & +0.06 & [+0.04, +0.08] & 71.3 & $5.00\times10^{-5}$ & 0.0018 \\
 & OVRL & +0.08 & [+0.04, +0.12] & 63.3 & $3.00\times10^{-4}$ & 0.0018 \\
\bottomrule
\end{tabular}
}

\newcommand{\ReverbTrainedWER}{%
\begin{tabular}{lrr}
\toprule
Model & Synthetic VoiceBank & DNS reverb \\
\midrule
Clean reference & 3.31 & 4.35 \\
Noisy & 12.97 & 10.00 \\
FlowSE & 21.56 & 26.28 \\
O & 22.74 & \textbf{18.27} \\
R-local & \textbf{20.22} & 18.80 \\
\bottomrule
\end{tabular}
}

\newcommand{\ReverbTrainedWERPaired}{%
\begin{tabular}{llrrrr}
\toprule
Dataset & Comparison & Mean $\Delta$WER & 95\% CI & $p$ & $p_{\rm Holm}$ \\
\midrule
Synthetic VB & O vs. FlowSE & +1.17 & [-1.93, +5.54] & 0.6061 & 1.0000 \\
 & R-local vs. FlowSE & -1.35 & [-3.43, +0.98] & 0.2424 & 0.9446 \\
 & R-local vs. O & -2.52 & [-6.94, +0.96] & 0.2361 & 0.9446 \\
\midrule
DNS reverb & O vs. FlowSE & -8.01 & [-10.59, -5.50] & $5.00\times10^{-5}$ & $3.00\times10^{-4}$ \\
 & R-local vs. FlowSE & -7.48 & [-9.93, -5.15] & $5.00\times10^{-5}$ & $3.00\times10^{-4}$ \\
 & R-local vs. O & +0.52 & [-1.22, +2.25] & 0.5714 & 1.0000 \\
\bottomrule
\end{tabular}
}

\newcommand{\ReverbTrainedIntervention}{%
\begin{tabular}{llrrrr}
\toprule
Model & Disabled & VB PESQ & SI-SDR & DNS PESQ & SI-SDR \\
\midrule
O & Multiplicative & 0.13 & 5.78 & 0.11 & 2.31 \\
 & Additive & 1.12 & 25.10 & 1.05 & 22.70 \\
R-local & Multiplicative & 1.35 & 24.63 & 1.13 & 20.77 \\
 & Additive & 0.80 & 3.12 & 0.53 & 2.18 \\
\bottomrule
\end{tabular}
}

\newcommand{\ReverbTrainedSecondary}{%
\begin{tabular}{lrrrrr}
\toprule
Restored branch & PESQ-WB & ESTOI & SIG & BAK & OVRL \\
\midrule
O: multiplicative & 0.13/0.11 & 0.006/0.004 & 0.57/0.53 & 1.40/0.79 & 0.57/0.41 \\
R-local: additive & 0.80/0.53 & 0.062/0.074 & 0.63/0.74 & 1.71/1.08 & 0.77/0.62 \\
\bottomrule
\end{tabular}
}

\newcommand{\ReverbTrainedSuppressionVB}{%
\begin{tabular}{lrrrr}
\toprule
Metric & O: no mult. & O: no add. & R-local: no mult. & R-local: no add. \\
\midrule
PESQ & +0.13 [+0.11, +0.16] & +1.12 [+1.07, +1.18] & +1.35 [+1.29, +1.41] & +0.80 [+0.75, +0.84] \\
STOI & +0.006 [+0.005, +0.008] & +0.462 [+0.455, +0.469] & +0.446 [+0.440, +0.453] & +0.029 [+0.028, +0.030] \\
ESTOI & +0.006 [+0.004, +0.008] & +0.554 [+0.545, +0.563] & +0.540 [+0.531, +0.548] & +0.062 [+0.060, +0.064] \\
SI-SDR & +5.78 [+5.46, +6.10] & +25.10 [+24.67, +25.52] & +24.63 [+24.20, +25.04] & +3.12 [+2.93, +3.30] \\
P.808 & +0.20 [+0.19, +0.22] & +0.98 [+0.96, +1.00] & +1.01 [+0.99, +1.04] & +0.55 [+0.53, +0.57] \\
OVRL & +0.57 [+0.54, +0.60] & +1.13 [+1.10, +1.16] & +1.23 [+1.19, +1.26] & +0.77 [+0.75, +0.79] \\
SIG & +0.57 [+0.52, +0.61] & +1.30 [+1.26, +1.33] & +1.42 [+1.39, +1.46] & +0.63 [+0.59, +0.66] \\
BAK & +1.40 [+1.35, +1.46] & +2.21 [+2.17, +2.25] & +2.26 [+2.23, +2.30] & +1.71 [+1.67, +1.74] \\
\bottomrule
\end{tabular}
}

\newcommand{\ReverbTrainedSuppressionDNS}{%
\begin{tabular}{lrrrr}
\toprule
Metric & O: no mult. & O: no add. & R-local: no mult. & R-local: no add. \\
\midrule
PESQ & +0.11 [+0.06, +0.17] & +1.05 [+0.95, +1.16] & +1.13 [+1.02, +1.23] & +0.53 [+0.46, +0.60] \\
STOI & +0.008 [+0.004, +0.012] & +0.507 [+0.497, +0.518] & +0.476 [+0.467, +0.486] & +0.039 [+0.037, +0.040] \\
ESTOI & +0.004 [-0.001, +0.008] & +0.583 [+0.566, +0.600] & +0.553 [+0.537, +0.568] & +0.074 [+0.071, +0.077] \\
SI-SDR & +2.31 [+1.69, +2.93] & +22.70 [+21.95, +23.44] & +20.77 [+20.00, +21.51] & +2.18 [+1.90, +2.47] \\
P.808 & +0.24 [+0.21, +0.27] & +1.00 [+0.96, +1.05] & +1.07 [+1.02, +1.12] & +0.55 [+0.52, +0.58] \\
OVRL & +0.41 [+0.34, +0.48] & +0.82 [+0.73, +0.90] & +0.89 [+0.80, +0.98] & +0.62 [+0.55, +0.69] \\
SIG & +0.53 [+0.42, +0.64] & +1.18 [+1.06, +1.29] & +1.26 [+1.14, +1.38] & +0.74 [+0.64, +0.84] \\
BAK & +0.79 [+0.68, +0.91] & +1.29 [+1.16, +1.42] & +1.37 [+1.23, +1.50] & +1.08 [+0.96, +1.19] \\
\bottomrule
\end{tabular}
}

\newcommand{\ReverbTrainedSuppAbsVB}{%
\begin{tabular}{llrrrrrrrr}
\toprule
Model & Condition & PESQ & STOI & ESTOI & SI-SDR & P.808 & OVRL & SIG & BAK \\
\midrule
Noisy & -- & 1.88 & 0.886 & 0.792 & 7.55 & 2.66 & 1.56 & 1.93 & 1.79 \\
O & Full & 2.16 & 0.897 & 0.806 & 15.41 & 3.05 & 2.22 & 2.57 & 3.40 \\
O & Add. only & 2.03 & 0.891 & 0.800 & 9.64 & 2.85 & 1.65 & 2.00 & 1.99 \\
O & Mult. only & 1.04 & 0.436 & 0.251 & -9.69 & 2.07 & 1.09 & 1.27 & 1.19 \\
R-local & Full & 2.39 & 0.911 & 0.826 & 16.29 & 3.11 & 2.32 & 2.69 & 3.45 \\
R-local & Add. only & 1.04 & 0.465 & 0.286 & -8.34 & 2.10 & 1.09 & 1.26 & 1.18 \\
R-local & Mult. only & 1.60 & 0.882 & 0.764 & 13.17 & 2.56 & 1.54 & 2.06 & 1.74 \\
\bottomrule
\end{tabular}
}

\newcommand{\ReverbTrainedSuppAbsDNS}{%
\begin{tabular}{llrrrrrrrr}
\toprule
Model & Condition & PESQ & STOI & ESTOI & SI-SDR & P.808 & OVRL & SIG & BAK \\
\midrule
Noisy & -- & 1.82 & 0.866 & 0.776 & 9.19 & 2.74 & 1.39 & 1.76 & 1.50 \\
O & Full & 2.09 & 0.884 & 0.792 & 12.73 & 3.11 & 1.90 & 2.40 & 2.44 \\
O & Add. only & 1.98 & 0.876 & 0.788 & 10.42 & 2.87 & 1.48 & 1.87 & 1.65 \\
O & Mult. only & 1.04 & 0.377 & 0.209 & -9.97 & 2.11 & 1.08 & 1.22 & 1.15 \\
R-local & Full & 2.16 & 0.890 & 0.802 & 12.44 & 3.17 & 1.97 & 2.49 & 2.53 \\
R-local & Add. only & 1.04 & 0.414 & 0.250 & -8.33 & 2.10 & 1.08 & 1.23 & 1.15 \\
R-local & Mult. only & 1.63 & 0.852 & 0.728 & 10.26 & 2.62 & 1.35 & 1.75 & 1.45 \\
\bottomrule
\end{tabular}
}

\newcommand{\ReverbTrainedWERBreakdown}{%
\begin{tabular}{lrrrr}
\toprule
Model & Sub. & Del. & Ins. & WER \\
\midrule
\multicolumn{5}{l}{\emph{Synthetic reverberant VoiceBank ($n=824$)}}\\
Clean reference & 2.55 & 0.41 & 0.35 & 3.31 \\
Noisy & 8.42 & 2.56 & 1.98 & 12.97 \\
FlowSE & 14.23 & 3.70 & 3.63 & 21.56 \\
O & 13.98 & 3.02 & 5.73 & 22.74 \\
R-local & 12.86 & 3.25 & 4.12 & 20.22 \\
\midrule
\multicolumn{5}{l}{\emph{Reverberant DNS 2020 ($n=150$)}}\\
Clean reference & 2.11 & 0.82 & 1.42 & 4.35 \\
Noisy & 5.57 & 2.11 & 2.31 & 10.00 \\
FlowSE & 15.04 & 7.09 & 4.15 & 26.28 \\
O & 10.84 & 4.57 & 2.86 & 18.27 \\
R-local & 11.64 & 4.53 & 2.64 & 18.80 \\
\bottomrule
\end{tabular}
}

\newcommand{\ReverbTrainedStrataVB}{%
\begin{tabular}{lrrrr}
\toprule
Condition & $n$ & PESQ (N/F/O/R) & SI-SDR (N/F/O/R) & OVRL (N/F/O/R) \\
\midrule
\multicolumn{5}{l}{\emph{Input SNR (dB)}}\\
0 & 203 & 1.38/1.46/1.70/1.88 & 0.00/8.10/9.80/10.96 & 1.24/1.78/2.07/2.24 \\
5 & 207 & 1.65/1.68/2.00/2.22 & 5.01/11.87/13.75/14.66 & 1.38/1.94/2.17/2.29 \\
10 & 205 & 2.00/1.94/2.29/2.54 & 10.00/15.70/17.38/18.18 & 1.64/2.10/2.29/2.34 \\
15 & 209 & 2.47/2.28/2.64/2.92 & 15.00/18.69/20.58/21.23 & 1.96/2.23/2.34/2.39 \\
\addlinespace[2pt]
\multicolumn{5}{l}{\emph{RIR identifier}}\\
0.1 & 138 & 1.70/1.60/1.94/2.18 & 7.51/12.11/14.80/15.88 & 2.08/2.47/2.70/2.85 \\
0.2 & 137 & 1.84/1.79/2.09/2.35 & 7.56/14.42/16.25/17.33 & 1.85/2.28/2.56/2.70 \\
0.4 & 136 & 1.89/1.89/2.22/2.42 & 7.58/14.40/15.82/16.67 & 1.61/2.03/2.25/2.36 \\
0.5 & 139 & 1.86/1.90/2.22/2.41 & 7.52/13.53/14.77/15.38 & 1.28/1.74/1.90/1.99 \\
0.7 & 136 & 1.96/1.92/2.25/2.49 & 7.61/13.68/15.37/16.23 & 1.33/1.90/2.11/2.18 \\
0.8 & 138 & 2.00/1.95/2.26/2.51 & 7.53/13.63/15.48/16.26 & 1.20/1.65/1.80/1.83 \\
\addlinespace[2pt]
\multicolumn{5}{l}{\emph{Noise recording}}\\
Babble & 118 & 1.56/1.52/1.63/1.71 & 7.55/8.91/9.80/10.37 & 1.26/1.66/1.87/1.98 \\
Cafeteria & 118 & 1.49/1.46/1.57/1.69 & 7.54/10.76/12.08/12.92 & 1.42/1.85/2.11/2.26 \\
Car & 118 & 3.12/2.99/3.45/3.77 & 7.58/21.13/24.03/24.90 & 2.05/2.38/2.42/2.43 \\
Livingroom & 118 & 1.61/1.58/1.88/2.09 & 7.47/11.50/13.82/14.62 & 1.54/1.92/2.22/2.34 \\
Shopping & 118 & 1.57/1.48/1.66/1.92 & 7.58/10.53/12.04/13.88 & 1.43/1.90/2.19/2.33 \\
Traffic & 115 & 1.67/1.77/2.33/2.64 & 7.57/15.48/17.01/17.60 & 1.58/2.16/2.37/2.45 \\
Train\_Station & 119 & 2.11/2.07/2.62/2.92 & 7.56/17.09/19.11/19.72 & 1.62/2.20/2.36/2.43 \\
\addlinespace[2pt]
\multicolumn{5}{l}{\emph{Speaker}}\\
p232 & 393 & 2.10/1.93/2.27/2.44 & 7.76/12.42/14.54/15.19 & 1.60/2.03/2.25/2.36 \\
p257 & 431 & 1.68/1.76/2.06/2.35 & 7.36/14.73/16.20/17.29 & 1.52/1.99/2.20/2.27 \\
\addlinespace[2pt]
\bottomrule
\end{tabular}
}

\newcommand{\ReverbTrainedStrataDNS}{%
\begin{tabular}{lrrrr}
\toprule
Condition & $n$ & PESQ (N/F/O/R) & SI-SDR (N/F/O/R) & OVRL (N/F/O/R) \\
\midrule
\multicolumn{5}{l}{\emph{Input SNR (dB)}}\\
$<0$ & 7 & 1.18/1.15/1.50/1.59 & -0.08/1.54/7.16/7.95 & 1.12/1.59/1.95/2.07 \\
$[0,5)$ & 36 & 1.49/1.48/1.74/1.78 & 3.48/7.26/10.00/9.76 & 1.21/1.69/1.78/1.93 \\
$[5,10)$ & 42 & 1.62/1.49/1.80/1.92 & 7.47/8.43/10.76/10.73 & 1.26/1.65/1.90/1.91 \\
$\geq10$ & 65 & 2.21/1.93/2.54/2.59 & 14.47/12.68/16.12/15.51 & 1.60/1.81/1.95/2.03 \\
\addlinespace[2pt]
\bottomrule
\end{tabular}
}

\newcommand{\ReverbTrainedValidation}{%
\begin{tabular}{lrrrr}
\toprule
Model & Epochs & Best epoch & Best SI-SDR & Final SI-SDR \\
\midrule
FlowSE & 30 & 29 & 11.61 & 11.13 \\
O & 30 & 30 & 14.99 & 14.99 \\
R-local & 30 & 29 & 15.82 & 15.50 \\
\bottomrule
\end{tabular}
}

\title{MaRFlow: Observation-Mediated and Recursive Multiplicative--Additive Velocity Factorization for Flow-Matching Speech Enhancement}

\author{\name Adhiraj Banerjee \email adhirajbanerjee35@gmail.com\\
\addr Department of Electrical Engineering\\
Indian Institute of Technology, Kanpur}

\def\month{MM}  
\def\year{YYYY} 
\def\openreview{\url{https://openreview.net/forum?id=XXXX}} 

\begin{document}

\maketitle

\begin{abstract}
Conditional flow matching (CFM) models speech enhancement through a learned spectral velocity field, typically predicted directly by a neural network. We introduce \textbf{MaRFlow}, a multiplicative--additive velocity factorization with two reference-specific formulations: observation-mediated restoration (O), which modulates the fixed noisy observation, and recursive restoration (R), which modulates the evolving state, including earlier additive updates. A regularized terminal-displacement oracle for O and a local instantaneous-velocity oracle for R specify complementary branch targets while preserving the conditional path and its unrestricted population-optimal aggregate velocity. Both add only 258 parameters to FlowSE.

Under matched 30-epoch training and five network evaluations, O and locally supervised R attain 2.62/17.85 and 2.59/17.84 PESQ-WB/SI-SDR (dB) on VoiceBank+DEMAND versus 2.39/17.41 for FlowSE, and 1.97/14.00 and 2.00/14.11 on non-reverberant DNS versus 1.70/12.64. Local supervision improves recursive SI-SDR by 2.00 dB on DNS over aggregate-only training. Diagnostics distinguish reference-dependent energy allocation from supervision-dependent functional reliance: O is additive-dominant, whereas R-local sustains early state modulation, exhibits heterogeneous cumulative gains, and relies on complementary additive correction for perceptual quality. All five additive-noise-trained models transfer poorly to unseen reverberation. Under matched reverberation-aware training, MaRFlow-O and
MaRFlow-R-local outperform FlowSE across all six acoustic
metrics on synthetic reverberant VoiceBank and reverberant DNS,
with paired bootstrap intervals and Holm-adjusted tests
supporting all 24 improvements.
Specifically, they improve PESQ-WB/SI-SDR from 1.84/13.63
to 2.16/15.41 and 2.39/16.29 on VoiceBank, and from
1.66/9.67 to 2.09/12.73 and 2.16/12.44 on DNS, respectively.
Both formulations retain their characteristic branch organization
across training regimes and improve linguistic recoverability
on DNS relative to FlowSE, though not relative to noisy speech.
These findings support reference-specific velocity factorization
and supervision as effective inductive biases for acoustic restoration.
\end{abstract}

\section{Introduction}\label{sec:intro}
We propose \textbf{MaRFlow}, a conditional flow-matching (CFM) speech enhancement framework that structures the instantaneous restoration field through \emph{multiplicative--additive velocity factorization}. Rather than predict one unconstrained spectral rate, MaRFlow learns a complex gain rate and an additive spectral rate, jointly evaluated within a single ordinary differential equation (ODE). We develop two forms of this interface: a fixed-observation formulation and a recursive state-mediated formulation. Crucially, we derive distinct supervision targets that are matched to the reference used in each formulation.

Observation-dependent modulation is a longstanding principle in speech enhancement, from statistical gains \citep{ephraim} to phase-sensitive and complex ratio masks \citep{erdogan,williamson}. Direct spectral prediction, by contrast, can express arbitrary corrections including at observation nulls \citep{wangchen}. Their complementary algebraic properties motivate structuring the \emph{velocity} of restoration, rather than only the final enhanced spectrum. In generative enhancement, FlowSE \citep{lee} learns a conditional velocity through CFM \citep{lipman} and integrates the resulting ODE from a perturbed observation to clean speech. Although the noisy signal conditions the network at each step, its role in the predicted velocity is implicit. Our question is whether exposing a reference-dependent multiplicative coordinate alongside an additive coordinate makes the restoration field easier to learn under a fixed training budget.


In \textbf{MaRFlow-O}, the gain acts on the original noisy
spectrum $X$ throughout the trajectory, providing a stable
reference for successive relative corrections. In
\textbf{MaRFlow-R}, it acts on the evolving state $Y_t$,
introducing an inductive bias toward progressive relative
correction. The resulting Euler rule is a recursive affine
update in which successive gains compose multiplicatively
and directly reweight the initial state, including its
artificial Gaussian perturbation, and previously introduced
additive increments (Fig.~\ref{fig:recursive_schematic}).
This provides an explicit mechanism for refining previously
restored spectral content, although subsequent modulation
may also attenuate useful corrections or amplify existing
errors. These restoration dynamics are distinct even though
the two models share the same NCSN++ backbone, two output
heads and a single network evaluation per Euler step.
All operations are defined in compressed complex spectral
coordinates; the factorization specifies how the restoration
field is represented rather than imposing a physical
partition into speech and noise.

A central issue is that aggregate velocity supervision alone leaves the two outputs underdetermined. For O, we derive an identity-regularized terminal complex-mask oracle that allocates the paired endpoint displacement and assigns the artificial Gaussian velocity to the additive coordinate. For R, terminal allocation is no longer naturally aligned with the changing reference. We instead derive a local regularized oracle that allocates the entire instantaneous conditional velocity relative to the current state, with an exact additive complement. Both constructions preserve the linear FlowSE path, resolve the ambiguity of the aggregate-only objective in unrestricted output space and preserve the population-optimal aggregate field. The local construction complements the recursive inductive
bias by allocating relative correction according to the
instantaneous state power, while assigning the remaining
velocity to the additive coordinate. This allows the
supervision to adapt to the changing reference without
requiring a terminal mask or an explicitly prescribed
temporal allocation schedule. The choice of reference
therefore determines how restoration operations compose,
whereas the reference-specific oracle determines how the
instantaneous restoration velocity is allocated.

We evaluate MaRFlow along eight complementary axes: (i) \emph{in-domain enhancement and cross-domain generalization} on VoiceBank+DEMAND and non-reverberant DNS 2020; (ii) \emph{architectural analysis}, contrasting observation-mediated and recursive velocity parameterizations under aggregate-only and reference-specific supervision; (iii) \emph{robustness across input SNR and few-step inference}, studying enhancement across acoustic difficulty and solver budgets; (iv) \emph{diagnostic analysis} of the learned velocity decomposition, including oracle agreement, branch-energy allocation and branch suppression; (v) \emph{recursive restoration dynamics}, examining cumulative gains and the propagation of earlier additive updates; (vi) \emph{reverberant enhancement}, comparing zero-shot transfer with matched reverberation-aware training on synthetic reverberant VoiceBank and reverberant DNS; (vii) \emph{linguistic recoverability}, using a frozen ASR model to assess whether acoustic restoration preserves recognizable speech content across both training regimes; and (viii) \emph{inference efficiency} in terms of latency, real-time factor and memory footprint.

We compare primarily against FlowSE~\citep{lee} under matched training data, backbone, conditional path and inference protocols, with a fixed training budget of \textbf{30 epochs for every configuration}. Across the additive-noise and reverberation-aware training settings, both reference-specifically supervised variants consistently improve over FlowSE in all six mean acoustic metrics on the corresponding in-domain and cross-domain benchmarks. The reverberant DNS stress test reveals a shared failure under zero-shot transfer, but under matched reverberant training the two MaRFlow formulations recover reference fidelity and outperform the retrained FlowSE, with R-local attaining the strongest mean PESQ and OVRL on both reverberant benchmarks. Importantly, the diagnostic analyses show that the characteristic branch organization of the two formulations persists across training regimes. The linguistic recoverability study complements these acoustic findings: both supervised models reduce WER relative to FlowSE on non-reverberant DNS and, after reverberation-aware training, on reverberant DNS, although neither improves upon the noisy-input WER. Our analyses thus focus on the effect of structuring the restoration field, rather than a ranking of unrelated enhancement systems.

Our main contributions are \textbf{fourfold}.

\textbf{First}, we introduce observation-mediated and
recursive multiplicative--additive velocity factorizations
of a single conditional restoration field. We mathematically
characterize their distinct inductive biases: stable-reference
accumulation in O and progressive relative correction in R.
In particular, we establish how recursive gain composition
explicitly modulates the initial perturbation and earlier
additive increments, while retaining a common backbone and
one network call per Euler step.

\textbf{Second}, we derive \emph{reference-specific structured
supervision}: regularized terminal-displacement allocation
for O and local instantaneous-velocity allocation for R.
Both follow a common regularized allocation principle
adapted to their respective multiplicative references.
We establish exact target consistency and preservation
of the unrestricted population-optimal aggregate field,
and characterize the population-optimal state-dependent
branch allocation for R.

\textbf{Third}, under matched training, both supervised formulations improve the six primary enhancement metrics over FlowSE on VoiceBank and non-reverberant DNS. The recursive local formulation recovers the performance lost under aggregate-only recursive supervision, and leads all six mean scores on non-reverberant DNS at five NFE. The formulations are effective at low NFE, with complementary quality/fidelity trade-offs and only 258 extra parameters. Under a second, matched reverberant training protocol, both outperform direct velocity prediction on all six acoustic metrics on synthetic reverberant VoiceBank and DNS, with all 24 model-versus-FlowSE paired intervals excluding zero and all corresponding tests significant after Holm correction. R-local leads all six means on synthetic reverberant VoiceBank and five of six on reverberant DNS, while O leads DNS SI-SDR. In a separate linguistic recoverability analysis, both supervised models reduce WER relative to FlowSE on DNS under the two matched acoustic training regimes, including reductions of 8.01/7.48 percentage points on reverberant DNS.

\textbf{Fourth}, our diagnostic analyses distinguish
reference-dependent branch-energy allocation from
supervision-dependent functional reliance. Both
observation-mediated variants are additive-dominant in
squared velocity magnitude, whereas both recursive variants
exhibit early multiplicative dominance followed by increasing
additive allocation. Local supervision changes recursive
modulation from relatively uniform attenuation to more
heterogeneous cumulative gains, shifts functional reliance
toward the multiplicative branch, and preserves a greater
fraction of substantially smaller early additive increments.
Branch interventions further reveal perceptually important
complementary corrections in both supervised formulations.
These findings characterize the learned restoration dynamics
in compressed coordinates without assigning physical speech
or noise components to individual branches. Evaluations with
a frozen recognizer and on reverberant DNS delineate the
distinction between perceptual quality, signal fidelity and
linguistic recoverability. The characteristic energy profiles,
functional branch dependence and complementary corrections
persist after matched reverberant training, extending the
mechanistic evidence to a second acoustic training regime.

\section{Related Work}\label{sec:related}

\paragraph{Speech enhancement.}
Statistical speech enhancement methods estimate observation-dependent
spectral gains \citep{ephraim}, whereas neural approaches learn
phase-sensitive or complex ratio masks \citep{erdogan,williamson},
or directly estimate clean spectral representations \citep{wangchen}.
Masking exploits information retained in the noisy observation,
while direct spectral mapping permits reconstruction beyond
multiplicative modulation. Hybrid architectures such as CMGAN
and D2Former combine these complementary restoration mechanisms
\citep{cmgan,d2former}. MaRFlow introduces this decomposition
at the level of the instantaneous restoration velocity rather
than as complementary endpoint predictions, allowing modulation
and additive reconstruction to operate jointly throughout
the restoration trajectory.

\paragraph{Audio source separation.}
Masking is also a central mechanism in source separation,
where the objective is to recover individual constituent sources
from an acoustic mixture. Conv-TasNet \citep{convtasnet}
performs masking in a learned representation space, while
AudioSep \citep{audiosep} extends this paradigm to
language-queried universal sound separation.
Beyond introducing an efficient codec-native source separation
framework, CodecSep \citep{codecsep} provides representational
and mechanistic insights into the source-specific structure
inherent in neural audio codec latents. Through architectural
comparisons and latent-space analyses, it demonstrates that
this structure can be exploited through source-conditioned
masking, predominantly via channel-wise modulation, without
requiring explicit latent generation.
In contrast, FlowSep \citep{flowsep} formulates
language-queried source separation as conditional generative
transport in a pretrained audio latent space.
FlowSep~2 \citep{flowsep2} extends this formulation through
self-supervised flow matching, encouraging semantically
structured internal representations during generative
source recovery.
These works highlight two complementary mechanisms for
audio recovery: exploiting source-related structure already
present in the input representation through modulation,
and reconstructing target content through conditional
generation. MaRFlow investigates their integration within
a single continuous restoration field.

\paragraph{Generative restoration and flow matching.}
Score-based generative models \citep{richter} and
Schr\"odinger bridges \citep{jukic} formulate speech
enhancement through continuous restoration dynamics.
Flow matching \citep{lipman} and rectified flow \citep{liu}
provide simulation-free training objectives for learning
transport velocity fields along prescribed probability
paths. FlowSE \citep{lee} applies conditional flow matching
(CFM) to speech enhancement by learning a spectral
velocity field conditioned on the noisy observation.
Subsequent approaches investigate alternative probability
paths \citep{cross} and cascaded restoration flows
\citep{cascade}.
While these developments primarily concern the construction
of restoration trajectories and their associated transport
processes, MaRFlow investigates the internal parameterization
and supervision of the velocity field itself.
The original conditional probability path is retained,
allowing the effect of structured velocity factorization
to be studied independently of changes to the transport
path or numerical solver.

\paragraph{Structured and factorized flow matching.}
Several approaches introduce structure into generative
dynamics by decomposing the underlying transport or its
parameterization. Residual-decoupled diffusion \citep{rddm}
distinguishes degradation-related residuals from artificial
noise, while FRSE \citep{frse2026} separates restoration
and perturbation velocity fields in flow-matching speech
enhancement. Decomposable Flow Matching \citep{hajiali2025}
factorizes generation across multiple representation scales,
learning coarse-to-fine transport within a shared model.
Blockwise Flow Matching \citep{blockwise2025} partitions
the transport trajectory into temporal segments governed
by specialized velocity predictors.
Factored velocity fields have also been investigated for
compositional conditioning \citep{ryoo2026} and disentangled
representation learning \citep{chi2026}, where individual
velocity contributions are associated with semantic
attributes or factors of variation.
From a numerical integration perspective, AREX
\citep{lin2026arex} decomposes pretrained flow dynamics
into an analytically tractable affine component and a
complementary neural residual.

MaRFlow differs from these approaches in both the
construction and purpose of its factorization.
Rather than decomposing the transport across semantic
attributes, representation scales, temporal segments,
or separately defined restoration processes, MaRFlow
expresses a single instantaneous spectral velocity field
through two algebraically complementary coordinates:
a reference-dependent multiplicative modulation and an
additive residual.
Both coordinates are learned jointly under a common
conditional probability path, with explicit supervision
derived from regularized oracle decompositions.
The choice of reference further distinguishes
observation-mediated restoration, which modulates the
fixed noisy observation, from recursive restoration,
which modulates the evolving state and consequently
reweights previously introduced additive content.
The resulting formulation introduces a structured
inductive bias into velocity learning without changing
the unrestricted population-optimal aggregate velocity
associated with the original conditional path.

\section{Method}\label{sec:method}
\subsection{Task formulation and FlowSE background}
\label{subsec:background}

Let $x,s\in\mathbb R^N$ denote aligned noisy and clean
waveforms of $N$ samples, respectively. Following FlowSE,
both are normalized by the noisy peak
$g=\max(\norm{x}_\infty,10^{-8})$, where
$\norm{x}_\infty=\max_n|x_n|$.
For a waveform $u\in\mathbb R^N$, define the compressed
complex spectral representation
\begin{align}
\mathcal T_g(u)
&=c|\mathrm{STFT}(u/g)|^p
e^{j\angle\mathrm{STFT}(u/g)},\\
X&=\mathcal T_g(x),\qquad S=\mathcal T_g(s),
\qquad X,S\in\C^{F\times L}.
\end{align}
Here, $\mathrm{STFT}$ denotes the short-time Fourier
transform, $j=\sqrt{-1}$ is the imaginary unit,
$|\cdot|$ and $\angle(\cdot)$ denote complex magnitude
and phase, and $F,L$ denote the numbers of frequency
bins and temporal frames, respectively.
We use the unnormalized 510-point STFT ($F=256$),
amplitude compression exponent $p=0.5$, and spectral
scaling coefficient $c=0.15$.

Throughout this work, $\odot$ denotes element-wise
multiplication, $(\cdot)^*$ denotes complex conjugation
for scalar and spectral-array quantities, and
$\norm{\cdot}_F$ denotes the Frobenius norm.
Magnitudes, powers, divisions and products between
spectral arrays are interpreted element-wise unless
otherwise specified. Scalar constants are broadcast
across the spectral dimensions. In per-bin derivations,
the same symbols denote the corresponding scalar entries.

For a fixed normalization factor $g$, the transform
$\mathcal T_g$ is nonlinear when $p\neq1$. In particular,
\begin{equation}
\mathcal T_g(x-s)\neq X-S
\end{equation}
in general. Throughout this work, $S-X$ denotes the
algebraic displacement between paired representations
in compressed complex spectral coordinates, rather
than the transform of a physical waveform residual.

Both the multiplicative and additive operations introduced
below are defined exclusively in these modeled coordinates.
They do not assume linear mixing after spectral compression
or a physical decomposition into speech and noise components.
The clean representation $S$ depends on the noisy utterance
through the shared normalization factor $g$.

\paragraph{Conditional probability path.}
Let $t\in[0,1]$ denote the continuous transport time.
Under the reverse-time convention of FlowSE, $t=1$
corresponds to the perturbed noisy source and $t=0$
to the clean endpoint.
Let $Z\in\C^{F\times L}$ be a standard circularly
symmetric complex Gaussian perturbation,
$Z\sim\mathcal{CN}(0,I)$, where $I$ is the identity
covariance on the vectorized spectral coordinates.
The paired conditional probability path and its
instantaneous velocity are
\begin{align}
Y_t&=(1-t)S+tX+\sigma(t)Z,
\label{eq:flowse_path}\\
\sigma(t)&=\sigma_{\min}
+t(\sigma_{\max}-\sigma_{\min}),\\
U_t&=\frac{dY_t}{dt}
=X-S+\sigma'(t)Z.
\label{eq:flowse_velocity}
\end{align}
Here, $Y_t\in\C^{F\times L}$ is the sampled conditional
state, $U_t\in\C^{F\times L}$ is its target velocity,
$\sigma(t)$ is the perturbation standard deviation,
and $\sigma'(t)=d\sigma(t)/dt$.
The real and imaginary components of each entry of $Z$
are independently sampled with variance $1/2$.
We use $\sigma_{\min}=0$ and $\sigma_{\max}=0.487$.
Consequently, the conditional source distribution is
Gaussian centered on $X$, and the target is the
conditional clean-speech distribution.

\paragraph{Conditional flow matching.}
Let $W=(Y_t,X,t)$ denote the network conditioning
variables, and let $\theta$ collect the trainable
network parameters.
FlowSE learns a velocity predictor
$v_\theta(W)\in\C^{F\times L}$ through squared-error
regression on $U_t$.
The expectation $\E[\cdot]$ is taken over paired
training examples, sampled transport times and
Gaussian perturbations, unless otherwise specified.
The unrestricted population-optimal velocity is
\begin{equation}
v^\star(W)=\E[U_t\mid W].
\end{equation}

To distinguish the learned inference trajectory from
the prescribed conditional training path $Y_t$,
let $\widetilde Y_t$ denote the state generated by
the learned ordinary differential equation (ODE).
Inference starts from
$\widetilde Y_1=X+\sigma_{\max}Z$ and proceeds toward
$t=0$ according to
\begin{equation}
\frac{d\widetilde Y_t}{dt}
=v_\theta(\widetilde Y_t,X,t).
\end{equation}
The linear paired conditional path does not imply
that the learned velocity remains constant along
inference trajectories.
Training does not require ODE rollouts.

\subsection{Multiplicative--additive velocity factorization}
\label{subsec:factorization}

With $W=(Y_t,X,t)$ denoting the conditioning variables,
MaRFlow parameterizes the instantaneous restoration
velocity through two jointly predicted outputs:
\begin{equation}
\boxed{
v_\theta(W)=\Phi_t\odot v_M(W)+v_R(W),
}
\label{eq:unified}
\end{equation}
where $\Phi_t\in\C^{F\times L}$ denotes the
multiplicative reference,
$v_M(W)\in\C^{F\times L}$ is the predicted complex
gain rate, and $v_R(W)\in\C^{F\times L}$ is the
predicted additive spectral velocity.
Both outputs depend on the shared network parameters
$\theta$, whose dependence is suppressed in the
branch notation for brevity.

The product $\Phi_t\odot v_M(W)$ defines the
multiplicative velocity contribution, whereas
$v_R(W)$ directly defines the additive velocity
contribution. Their sum constitutes the aggregate
velocity $v_\theta(W)$ supplied to the ODE solver.
The gain rate $v_M$ parameterizes instantaneous
relative correction with respect to transport time
rather than a terminal complex mask. The additive
coordinate permits complementary corrections that
are not constrained to be proportional to the
multiplicative reference.

We investigate two choices of multiplicative reference:
the fixed noisy observation $\Phi_t=X$ in
MaRFlow-O (O), and the current transport state
$\Phi_t=Y_t$ in MaRFlow-R (R).
During inference, the latter is evaluated using the
learned state $\widetilde Y_t$.
Both formulations retain identical conditioning
variables, network architecture and output dimensions,
but differ in how the predicted gain rate enters
the assembled velocity.
Only the spectral transport state is integrated;
neither predicted coordinate defines an independently
evolving state.

All multiplicative and additive operations are defined in
compressed complex spectral coordinates. Accordingly, the
additive contribution represents a complementary
coordinate-space velocity rather than a physical waveform
residual, and the gain rate is not interpreted as a mask
in the uncompressed linear STFT domain.


Here, factorization denotes a structured output
parameterization, affine in the reference for fixed head
outputs. The two heads share features and conditioning
variables and jointly parameterize a single vector field;
the formulation assumes neither statistical independence
of the branches nor separate probability flows. It therefore
modifies the parameterization of the FlowSE vector field
while retaining the original conditional probability path
and aggregate velocity target.

\subsubsection{Observation-mediated restoration (MaRFlow-O)}
\label{subsec:observation}

MaRFlow-O uses the fixed noisy observation as its
multiplicative reference:
\begin{equation}
v_\theta(W)=X\odot v_M(W)+v_R(W).
\label{eq:factor}
\end{equation}
Along a learned inference trajectory, integration
from source time $1$ to time $t$ gives
\begin{equation}
\begin{aligned}
\widetilde Y_t
&=\widetilde Y_1
+X\odot\int_1^t
v_M(\widetilde Y_\tau,X,\tau)\,d\tau\\
&\quad+\int_1^t
v_R(\widetilde Y_\tau,X,\tau)\,d\tau,
\end{aligned}
\label{eq:integrated}
\end{equation}
where $\tau$ is the integration-time variable.

Define the effective accumulated complex gain
$M_t\in\C^{F\times L}$ and the accumulated additive
correction $C_t\in\C^{F\times L}$ as
\begin{align}
M_t&=1+\int_1^t
v_M(\widetilde Y_\tau,X,\tau)\,d\tau,\\
C_t&=\int_1^t
v_R(\widetilde Y_\tau,X,\tau)\,d\tau.
\end{align}
Using the initial state
$\widetilde Y_1=X+\sigma_{\max}Z$, we obtain
\begin{equation}
\widetilde Y_t
=M_t\odot X+C_t+\sigma_{\max}Z.
\end{equation}

The effective gain therefore accumulates relative
to a stable observation. Both $M_t$ and $C_t$ are
derived quantities describing the integrated velocity
contributions along a realized trajectory; they are
not separately integrated by the implementation.

The branches remain coupled through their conditioning
on the current state. In particular, the multiplicative
term does not directly multiply the artificial Gaussian
perturbation, although its prediction may depend on it.

\subsubsection{Recursive state-mediated restoration (MaRFlow-R)}\label{subsec:recursive}
Alternatively, the gain acts on the evolving state:
\begin{equation}
 v_\theta(W)=Y_t\odot v_M(W)+v_R(W).\label{eq:recursive}
\end{equation}
Let $K$ denote the total number of Euler updates
and let
\begin{equation}
1=t_0>t_1>\cdots>t_K=0
\end{equation}
be the reverse-time integration grid.
For $k=0,\ldots,K-1$, define
$h_k=t_k-t_{k+1}>0$ as the positive step size and
$\mathsf Y_k\in\C^{F\times L}$ as the numerical
approximation to the learned transport state at
time $t_k$, initialized by
$\mathsf Y_0=X+\sigma_{\max}Z$.

For notational convenience, define the predicted
gain rate and additive velocity at the $k$-th
solver state as
\begin{align}
v_{M,k}&=v_M(\mathsf Y_k,X,t_k),\\
v_{R,k}&=v_R(\mathsf Y_k,X,t_k).
\end{align}
Both quantities belong to $\C^{F\times L}$.
The reverse Euler update becomes
\begin{align}
\mathsf Y_{k+1}&=\mathsf Y_k-h_k(\mathsf Y_k\odot v_{M,k}+v_{R,k})\nonumber\\
&=G_k\odot\mathsf Y_k+D_k,\label{eq:recursive_update}\\
G_k&=1-h_kv_{M,k},\qquad D_k=-h_kv_{R,k}.
\end{align}
Here, $G_k\in\C^{F\times L}$ denotes the effective
complex multiplicative gain applied during the
$k$-th Euler update, and $D_k\in\C^{F\times L}$
denotes the additive spectral increment introduced
during that update. Unlike the predicted gain rate
$v_{M,k}$ and additive velocity $v_{R,k}$,
these quantities incorporate the Euler step size.

Expanding $K$ updates along a realized trajectory yields
\begin{equation}
\mathsf Y_K=\left(\prod_{k=0}^{K-1}G_k\right)\odot\mathsf Y_0+
\sum_{i=0}^{K-1}\left(\prod_{j=i+1}^{K-1}G_j\right)\odot D_i.\label{eq:recursive_expansion}
\end{equation}
The indices $i$ and $j$ enumerate Euler updates.
All gain products are element-wise, and an empty
product equals the multiplicative identity.
The initial numerical state is
$\mathsf Y_0=\widetilde Y_1=X+\sigma_{\max}Z$.
 The noisy spectrum remains a conditioning input, but the multiplicative reference contains the artificial perturbation at initialization and evolves with restoration.

\paragraph{What makes the state-mediated factorization recursive?}
To compare the two formulations, let superscripts
$\mathrm O$ and $\mathrm R$ identify quantities
evaluated along the observation-mediated and recursive
inference trajectories, respectively.
For notational brevity, the superscript $\mathrm R$
is omitted when discussing the recursive formulation
alone.

For observation-mediated restoration, define
\begin{align}
A_k&=-h_kX\odot v_{M,k}^{\mathrm O},\\
D_k^{\mathrm O}&=-h_kv_{R,k}^{\mathrm O},
\end{align}
where $A_k,D_k^{\mathrm O}\in\C^{F\times L}$
denote the multiplicative-reference-mediated and
additive spectral increments, respectively,
introduced during the $k$-th Euler update.
Then
\begin{align}
\mathsf Y_{k+1}^{\mathrm O}&=\mathsf Y_k^{\mathrm O}+A_k+D_k^{\mathrm O},\label{eq:observation_update}\\
\mathsf Y_K^{\mathrm O}&=\mathsf Y_0^{\mathrm O}+\sum_{k=0}^{K-1}A_k+\sum_{k=0}^{K-1}D_k^{\mathrm O}.\label{eq:observation_accumulation}
\end{align}
In contrast, expanding the first three updates of R (omitting the superscript $\mathrm R$ for brevity) gives
\begin{align}
\mathsf Y_1&=G_0\odot\mathsf Y_0+D_0,\nonumber\\
\mathsf Y_2&=G_1\odot G_0\odot\mathsf Y_0+G_1\odot D_0+D_1,\nonumber\\
\mathsf Y_3&=G_2\odot G_1\odot G_0\odot\mathsf Y_0\nonumber\\
&\quad+G_2\odot G_1\odot D_0+G_2\odot D_1+D_2.\label{eq:recursive_three_step}
\end{align}
Thus, in O, the first additive increment enters the expanded
endpoint directly as $D_0^{\mathrm O}$, whereas in R it enters
as $G_2\odot G_1\odot D_0$ after three steps
(Fig.~\ref{fig:recursive_schematic}).

\paragraph{Inductive bias toward progressive relative correction.}
To isolate the structural effect of the multiplicative reference,
consider the accumulation of multiplicative corrections in the
absence of additive updates. Equations~\eqref{eq:observation_accumulation}
and~\eqref{eq:recursive_expansion} reduce to
\begin{align}
\mathsf Y_K^{\mathrm O}
&=\mathsf Y_0^{\mathrm O}
-X\odot\sum_{k=0}^{K-1}h_kv_{M,k}^{\mathrm O},
\label{eq:observation_gain_accumulation}\\
\mathsf Y_K^{\mathrm R}
&=\left[
\prod_{k=0}^{K-1}(1-h_kv_{M,k}^{\mathrm R})
\right]\odot\mathsf Y_0^{\mathrm R}.
\label{eq:recursive_gain_accumulation}
\end{align}
The effective gain corrections in O accumulate additively
relative to a fixed observation, whereas those in R compose
multiplicatively relative to the evolving state.
Consequently, R introduces an \emph{inductive bias toward
progressive relative correction}: for a fixed predicted gain rate,
the magnitude of the multiplicative contribution scales directly
with the current spectral magnitude rather than that of the
original observation.

This distinction is particularly relevant to conditional flow
matching, where the initial state contains an artificial Gaussian
perturbation in addition to the observed mixture.
Substituting $\mathsf Y_0=X+\sigma_{\max}Z$ into
Eq.~\eqref{eq:recursive_expansion} and defining the cumulative complex gain
\begin{equation}
\mathcal G_{a:b}
=\prod_{j=a}^{b}G_j
\in\C^{F\times L},
\end{equation}
where $a$ and $b$ denote the first and last update
indices included in the element-wise product, we obtain
\begin{align}
\mathsf Y_K^{\mathrm R}
&=\mathcal G_{0:K-1}\odot X
+\sigma_{\max}\mathcal G_{0:K-1}\odot Z
\nonumber\\
&\quad+
\sum_{i=0}^{K-1}\mathcal G_{i+1:K-1}\odot D_i,
\label{eq:recursive_component_expansion}
\end{align}
where an empty gain product equals one.
In contrast, the observation-mediated expansion is
\begin{align}
\mathsf Y_K^{\mathrm O}
&=\left(1-\sum_{k=0}^{K-1}h_kv_{M,k}^{\mathrm O}\right)
\odot X
+\sigma_{\max}Z
\nonumber\\
&\quad+\sum_{k=0}^{K-1}D_k^{\mathrm O}.
\label{eq:observation_component_expansion}
\end{align}
Thus, R provides a direct multiplicative mechanism for modulating
not only the observed spectrum but also the artificial perturbation
and previously introduced additive content.
In O, these components can be compensated through subsequently
predicted velocity contributions, but are not themselves directly
subject to multiplicative modulation.

\paragraph{Recursive refinement of earlier additive predictions.}
For either formulation, consider an additive increment
$D_i=D_i^\star+E_i$, where
$D_i^\star\in\C^{F\times L}$ denotes a hypothetical
reference increment and $E_i\in\C^{F\times L}$
denotes its deviation from that reference.
The quantity $D_i^\star$ is introduced solely for
this analytical comparison and is not an additional
oracle or training target.

Let $\Delta\mathsf Y_K$ denote the difference
between the expanded endpoints with increments
$D_i^\star+E_i$ and $D_i^\star$, respectively,
while holding all other increments and subsequent
predicted gains fixed.
The explicit endpoint contribution of $E_i$
in each formulation is then
\begin{align}
\Delta\mathsf Y_K^{\mathrm O}
&=E_i^{\mathrm O},\\
\Delta\mathsf Y_K^{\mathrm R}
&=\mathcal G_{i+1:K-1}\odot E_i^{\mathrm R}.
\label{eq:recursive_error_propagation}
\end{align}
For an individual time--frequency bin, recursive modulation
attenuates this deviation whenever
\begin{equation}
\left|\mathcal G_{i+1:K-1}\right|<1.
\label{eq:recursive_attenuation_condition}
\end{equation}
In particular, the magnitude of an individual complex Euler gain
satisfies
\begin{align}
|G_k|^2
&=|1-h_kv_{M,k}|^2\\
&=1-2h_k\operatorname{Re}(v_{M,k})
+h_k^2|v_{M,k}|^2.
\end{align}
For $h_k>0$, the corresponding condition for strict attenuation
by the multiplicative factor, i.e., $|G_k|<1$, is
\begin{equation}
2\operatorname{Re}(v_{M,k})
>h_k|v_{M,k}|^2.
\label{eq:recursive_contraction_condition}
\end{equation}
Successive recursive gains can therefore directly attenuate
the algebraic contributions of earlier reconstruction errors
when the corresponding cumulative gain magnitude is less
than one. More generally, complex gain products permit
attenuation, amplification and phase rotation of previously
restored spectral content.

This mechanism may be particularly useful under challenging
acoustic conditions, where intermediate restoration estimates
remain imperfect and subsequently benefit from refinement.
Conversely, recursive modulation can also attenuate useful
additive content or amplify existing errors. Observation-mediated
factorization avoids such direct remodulation and instead
provides a stable reference for successive corrections.

Importantly, these structural properties arise from the choice
of multiplicative reference, independently of explicit branch
supervision. Both formulations repeatedly evaluate a
state-conditioned field; the distinction is not recursive
numerical integration itself, but the \emph{explicit
multiplicative recurrence} induced by the state-mediated
factorization.

The above expansions describe algebraic contributions along
realized trajectories. Since subsequent head predictions depend
on the evolving state, Eq.~\eqref{eq:recursive_error_propagation}
does not characterize the complete perturbation response of
the nonlinear learned dynamics. Moreover, attenuation of the multiplicative factor in
Eq.~\eqref{eq:recursive_contraction_condition} does not imply
contraction of the complete state-dependent Euler map.
These distinctions motivate our cumulative-gain and
additive-update propagation analyses in
Section~\ref{subsec:recursive_results}.

\begin{figure}[t]
\centering
\begin{adjustbox}{max width=\linewidth}
\begin{tikzpicture}[
 >=Latex,font=\small,
 state/.style={draw=black!65,rounded corners=2pt,fill=black!3,minimum width=14mm,minimum height=8mm,inner sep=3pt},
 gain/.style={draw=blue!65!black,rounded corners=2pt,fill=blue!5,minimum height=7mm,inner sep=4pt},
 add/.style={draw=teal!65!black,rounded corners=2pt,fill=teal!5,minimum height=7mm,inner sep=4pt},
 op/.style={circle,draw=black!65,minimum size=6mm,inner sep=0pt},
 mul/.style={op,draw=blue!65!black,fill=blue!5},
 a/.style={->,semithick},m/.style={->,semithick,blue!65!black},r/.style={->,semithick,teal!65!black}]
\node[anchor=west,font=\small\bfseries] at (-0.8,2.35) {(a) O: fixed-observation accumulation};
\node[gain] (x) at (0,1.05) {$X$};
\node[gain] (og) at (3.15,1.82) {$-h_kv_{M,k}^{\mathrm O}$};
\node[mul] (om) at (3.15,1.05) {$\odot$};
\node[state] (oy) at (0,0) {$\mathsf Y_k^{\mathrm O}$};
\node[op] (os) at (7.15,0) {$+$};
\node[add] (od) at (7.15,-0.95) {$D_k^{\mathrm O}$};
\node[state] (on) at (10.7,0) {$\mathsf Y_{k+1}^{\mathrm O}$};
\draw[m] (x.east)--(om.west); \draw[m] (og.south)--(om.north);
\draw[m] (om.east)--node[above,font=\footnotesize] {$A_k=-h_kX\odot v_{M,k}^{\mathrm O}$} (7.15,1.05)--(os.north);
\draw[a] (oy.east)--(os.west); \draw[r] (od.north)--(os.south); \draw[a] (os.east)--(on.west);
\draw[black!25] (-0.8,-1.55)--(11.5,-1.55);
\node[anchor=west,font=\small\bfseries] at (-0.8,-1.98) {(b) R: recursive state modulation};
\node[state] (ry) at (0,-3.2) {$\mathsf Y_k$};
\node[gain] (rg) at (3.15,-2.48) {$G_k=1-h_kv_{M,k}$};
\node[mul] (rm) at (3.15,-3.2) {$\odot$};
\node[op] (rs) at (7.15,-3.2) {$+$};
\node[add] (rd) at (7.15,-4.15) {$D_k$};
\node[state] (rn) at (10.7,-3.2) {$\mathsf Y_{k+1}$};
\draw[a] (ry.east)--(rm.west); \draw[m] (rg.south)--(rm.north);
\draw[m] (rm.east)--node[above,font=\footnotesize] {$G_k\odot\mathsf Y_k$} (rs.west);
\draw[r] (rd.north)--(rs.south); \draw[a] (rs.east)--(rn.west);
\end{tikzpicture}
\end{adjustbox}
\par\vspace{2mm}
{\small\textbf{Algebraic contribution of the first additive increment}}\par\smallskip
\begin{adjustbox}{max width=\linewidth}
\small\begin{tabular}{lccc}
\toprule
 & After update 1 & After update 2 & After update 3\\
\midrule
O & $D_0^{\mathrm O}$ & $D_0^{\mathrm O}$ & $D_0^{\mathrm O}$\\
R & $D_0$ & $G_1\odot D_0$ & $G_2\odot G_1\odot D_0$\\
\bottomrule
\end{tabular}
\end{adjustbox}
\caption{Algebraic operations of one reverse Euler step for O and R. In O, $X$ is fixed and the increment $A_k$ is added to the state along with $D_k^{\mathrm O}=-h_kv_{R,k}^{\mathrm O}$. In R, $G_k$ acts directly on the evolving state before adding $D_k=-h_kv_{R,k}$. Both heads are evaluated jointly from the state, observation and time. The table describes the contribution of the first additive increment in the expansion along the realized trajectory, not an independently propagated signal or an additional network call.}
\label{fig:recursive_schematic}
\end{figure}


\paragraph{Why might early state modulation emerge without
branch targets?}
For the particular path in Eq.~\eqref{eq:flowse_path},
$\sigma_{\min}=0$ implies $Y_t=S+tU_t$.
Thus, for $t>0$ and a reverse Euler step of size
$0<h\leq t$,
\begin{align}
U_t
&=\frac{Y_t-S}{t}
=\frac{1}{t}Y_t-\frac{1}{t}S,
\label{eq:conditional_shrinkage}\\
Y_t-hU_t
&=\left(1-\frac{h}{t}\right)Y_t
+\frac{h}{t}S.
\label{eq:conditional_shrinkage_euler}
\end{align}
This identity exhibits one valid state-mediated decomposition
of the paired velocity: attenuation of the current-state
contribution and complementary correction toward the paired
target. The step-size restriction ensures that the oracle
Euler update is a convex combination of the current state
and paired endpoint in compressed coordinates.

At the unrestricted population optimum, the same identity
holds with $S$ replaced by $\E[S\mid W]$ in the velocity
and its Euler update. The target is unavailable at inference,
and the conditional identity does not prescribe the learned
head outputs. In particular, the aggregate loss admits
infinitely many allocations; neither this identity nor
the recursion itself guarantees multiplicative dominance
or contraction of the complete learned Euler map.

It does, however, motivate why a state-mediated
parameterization can express a shrinkage-and-reconstruction
field without explicit branch labels. The fixed-observation
parameterization has no corresponding direct multiplication
of the entire current state. We examine the resulting energy
profiles empirically in Sections~\ref{subsec:diagnostics}
and~\ref{subsec:recursive_results}.

\paragraph{Why factorize the velocity?} The multiplicative term expresses relative correction; the additive term avoids the constraint that all corrections must be proportional to the reference. Both admit arbitrary unrestricted spectral fields through $v_M=0$. The distinction is therefore an inductive bias in finite networks rather than a difference in unrestricted transport capacity. Aggregate supervision alone does not specify a physical or unique branch decomposition. Both are feedback systems through $Y_t$; recursive multiplication does not guarantee monotone improvement or imply training on model-generated errors.

\begin{figure}[t]
\centering
\begin{adjustbox}{max width=\linewidth}
\begin{tikzpicture}[>=Latex,font=\small,box/.style={draw=black!65,rounded corners=2pt,align=center,fill=black!3,inner sep=4pt,minimum height=8mm},mask/.style={box,draw=blue!65!black,fill=blue!5},res/.style={box,draw=teal!65!black,fill=teal!6},a/.style={->,semithick}]
\node[box,minimum width=113mm] (in) at (0,0) {$Y_t,X\in\C^{B\times1\times F\times L}$,\quad $t\in\mathbb R^B$};
\node[box,minimum width=113mm,fill=black!7] (net) at (0,-1.25) {\textbf{Shared NCSN++ encoder--decoder}\\[-1pt] $H_t\in\mathbb R^{B\times C_d\times F\times L}$};
\draw[a] (in)--(net);
\node[mask,minimum width=49mm] (mh) at (-2.85,-2.8) {\textbf{Multiplicative gain head}\\ $1\times1$ convolution};
\node[res,minimum width=49mm] (rh) at (2.85,-2.8) {\textbf{Additive head}\\ Original FlowSE route};
\draw[semithick] (net.south)--(0,-2.0);
\draw[a] (0,-2.0)-|(mh.north);\draw[a] (0,-2.0)-|(rh.north);
\node[mask,minimum width=49mm] (m) at (-2.85,-4.15) {$\Phi_t\odot v_M$\\ $\C^{B\times1\times F\times L}$};
\node[res,minimum width=49mm] (r) at (2.85,-4.15) {$v_R$\\ $\C^{B\times1\times F\times L}$};
\draw[a] (mh)--node[right,font=\footnotesize]{$v_M\in\C^{B\times1\times F\times L}$}(m);
\draw[a] (rh)--(r);
\draw[a,blue!65!black] (in.west)--(-6.7,0)--node[left,font=\footnotesize,align=center]{$\Phi_t=X$ (O)\\$\Phi_t=Y_t$ (R)}(-6.7,-4.15)--(m.west);
\node[box,minimum width=113mm] (sum) at (0,-5.55) {$v_\theta=\Phi_t\odot v_M+v_R\in\C^{B\times1\times F\times L}$};
\draw[a] (m.south)--([xshift=-28.5mm]sum.north);\draw[a] (r.south)--([xshift=28.5mm]sum.north);
\node[box,minimum width=113mm] (ode) at (0,-6.75) {\textbf{Euler update}\quad $Y_{t-h}=Y_t-hv_\theta\in\C^{B\times1\times F\times L}$};
\draw[a] (sum)--(ode);
\end{tikzpicture}
\end{adjustbox}
\caption{An overview of MaRFlow. A shared NCSN++ encoder--decoder conditions both heads on the transport state, noisy spectrum and time. The gain rate is applied to either the fixed observation (O) or current state (R). The combined velocity drives one Euler update of the single spectral ODE. $B,F,L,C_d$ denote batch size, frequency bins, frames and decoder width.}
\label{fig:arch}
\end{figure}
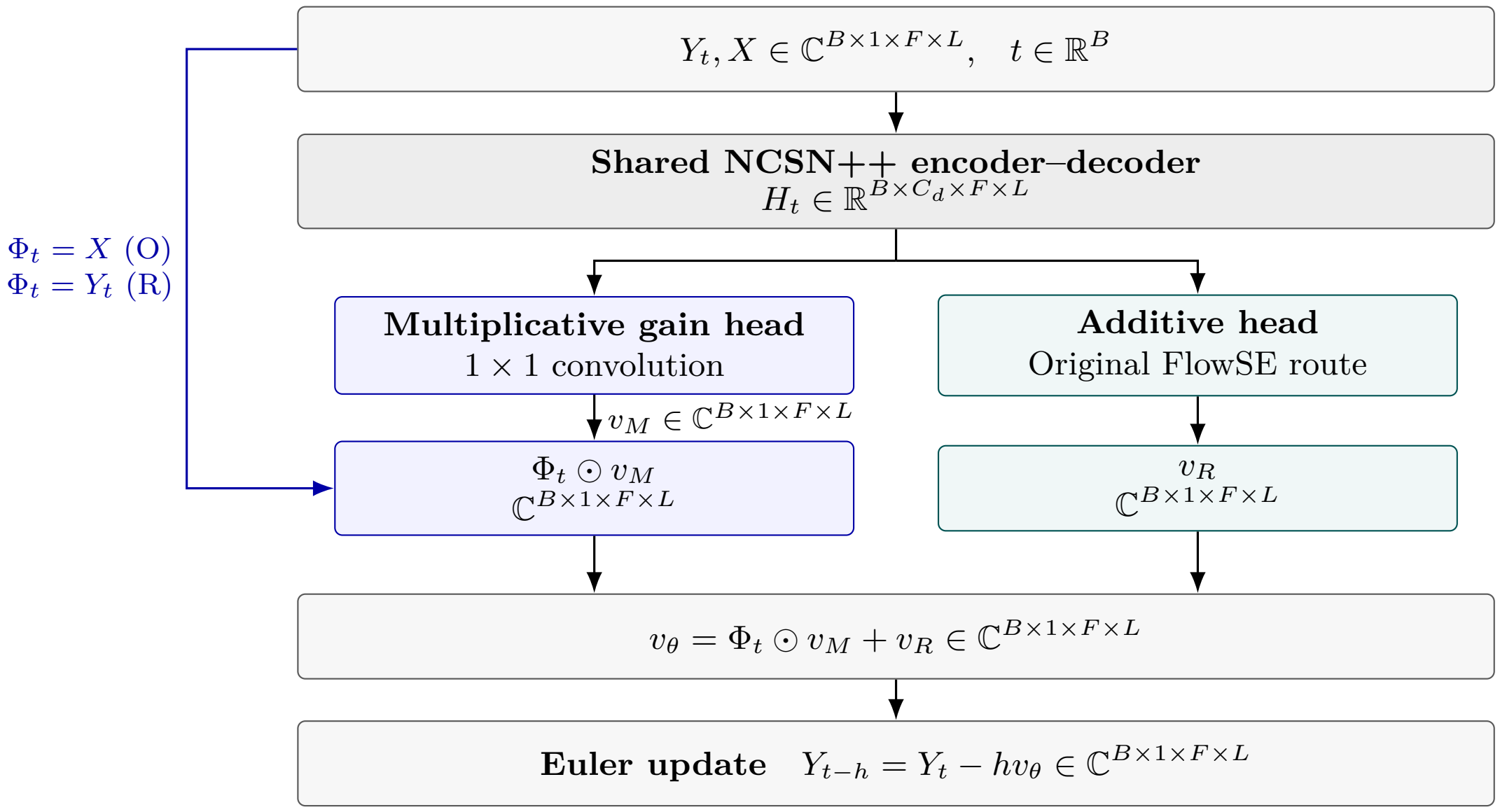

\subsection{Model architecture}\label{subsec:architecture}
\paragraph{Shared NCSN++ encoder--decoder.}
Both formulations retain the FlowSE backbone based on
NCSN++ \citep{song} and introduce a lightweight complex
gain head alongside its original spectral output route
(Fig.~\ref{fig:arch}).

Let $B$ denote the batch size and $C_d$ the number
of channels in the final full-resolution decoder
representation.
The batched spectral inputs have shape
$B\times1\times F\times L$, with a single complex
spectral channel, and the transport times are
represented by $t\in\mathbb R^B$.

The two heads share the activated full-resolution
decoder features
\begin{equation}
H_t\in\mathbb R^{B\times C_d\times F\times L},
\end{equation}
immediately before the original final $3\times3$
decoder projection.
The additive head retains the original FlowSE output
processing and time-dependent output scaling.

The gain head uses a $1\times1$ convolution to
produce two real-valued output channels, interpreted
as the real and imaginary components of the complex
gain rate. Both output routes follow the negative
output-sign convention inherited from FlowSE.
The resulting predictions have identical shapes:
\begin{equation}
v_M,v_R\in\C^{B\times1\times F\times L}.
\end{equation}
Unlike the additive route, the gain head does not
pass through the output pyramid or its final
time-dependent scaling; its dependence on transport
time enters through the shared decoder features.

\paragraph{Why a shared backbone?} Both rates depend on the same conditioning variables and must be evaluated at the same solver state. Sharing the backbone avoids a second decoder or sequential restoration network. A complex gain projection adds $2(C_d+1)$ parameters, i.e., 258 for $C_d=128$. The gain head is zero-initialized, so for identical original weights the initial aggregate field equals FlowSE's, while the new projection can receive gradients. All parameters are included in the optimizer and exponential moving average (EMA). Both coordinates are predicted in the same backbone call.

\subsection{MaRFlow-O: Terminal-displacement allocation}\label{subsec:oracle}
The decomposition in Eq.~\eqref{eq:factor} is nonunique.
To specify a training allocation, we split the paired
endpoint displacement $S-X$ using a terminal complex-mask
oracle, where an oracle denotes a deterministic supervision
target constructed from the paired noisy and clean
representations.

For an individual time--frequency bin, let
$m\in\C$ denote a candidate terminal complex mask.
The ideal ratio $S/X$ is ill-conditioned near
observation nulls. We therefore introduce an
identity-centered ridge penalty with coefficient
$\lambda>0$ and a maximum permitted mask magnitude
$m_{\max}\geq1$.
The regularized terminal mask is obtained by solving
\begin{align}
M^\star
&=\arg\min_{m\in\C,\;|m|\leq m_{\max}}
\{|mX-S|^2+\lambda|m-1|^2\}\\
&=\Pi_{m_{\max}}
\left(\frac{SX^*+\lambda}{|X|^2+\lambda}\right),
\label{eq:oracle}
\end{align}
where $M^\star$ denotes the optimal complex mask
and $\Pi_a:\C\rightarrow\C$ denotes Euclidean
projection onto the closed complex disk of radius
$a>0$ (Appendix~\ref{app:oracle}).
The neutral mask value is the multiplicative identity
$m=1$, favoring retention of the noisy observation
when relative correction is poorly conditioned.

Applying the per-bin solution across the spectral
array yields $M^\star\in\C^{F\times L}$.
Define the terminal gain correction relative to the
identity and its complementary additive displacement as
\begin{align}
D^\star&=M^\star-1,\\
R^\star&=S-M^\star\odot X,
\end{align}
respectively, with $D^\star,R^\star\in\C^{F\times L}$.

The corresponding multiplicative-reference-mediated
endpoint displacement and additive endpoint displacement
are
\begin{align}
A_M&=X\odot D^\star,\\
A_R&=R^\star.
\end{align}
Both contributions belong to $\C^{F\times L}$ and
satisfy the exact consistency identity
\begin{equation}
\boxed{A_M+A_R=S-X.}
\label{eq:decomp}
\end{equation}
This identity holds after clipping because the additive
complement is defined after the mask operation.

Equivalently, for a spectral bin, let $D,R\in\C$
denote candidate identity-relative gain corrections
and additive endpoint displacements, respectively.
The same oracle solves
\begin{equation}
\begin{aligned}
\min_{D,R}\quad
&|R|^2+\lambda|D|^2\\
\text{subject to}\quad
&XD+R=S-X,\\
&|1+D|\leq m_{\max}.
\end{aligned}
\end{equation}
The minimizing values are $D^\star$ and $R^\star$.

The terminal allocation is particularly natural for
observation-mediated restoration because the multiplicative
reference remains fixed throughout the transport trajectory.
Consequently, a decomposition of the paired endpoint displacement
can be converted into instantaneous branch targets without
requiring a state-dependent reallocation. This contrasts with
recursive state-mediated restoration, where the evolving reference
changes the spectral contribution associated with a given gain rate.


\paragraph{How does the ridge allocate restoration?}
Let $P=|X|^2$ denote the observed squared magnitude
in compressed spectral coordinates.
Ignoring the mask magnitude constraint, let
$D_{\mathrm{unc}}$ denote the unconstrained gain
correction, and let $A_{M,\mathrm{unc}}$ and
$A_{R,\mathrm{unc}}$ denote its associated
multiplicative-reference-mediated and additive
endpoint displacements, respectively.
The unconstrained solution gives
\begin{align}
D_{\mathrm{unc}}&=\frac{(S-X)X^*}{P+\lambda},
\label{eq:unclipped_gain}\\
A_{M,\mathrm{unc}}&=\frac{P}{P+\lambda}(S-X),\qquad
A_{R,\mathrm{unc}}=\frac{\lambda}{P+\lambda}(S-X).
\label{eq:allocation}
\end{align}
For $P\gg\lambda$, the paired displacement is allocated
predominantly to masking; for $P\ll\lambda$, to the additive
complement. At $P=\lambda$, both receive half the displacement
in the unclipped case. At $X=0$, $M^\star=1$, $A_M=0$, and
$A_R=S$. The unconstrained correction obeys
$|D_{\mathrm{unc}}|\le |S-X|/(2\sqrt\lambda)$; clipping yields
$|D^\star|\le m_{\max}+1$. The allocation is driven by observed
power, not local speech-to-noise ratio. It provides a
relative-correction coordinate system, not a detector of speech
reliability. We use $\lambda=0.01$ and $m_{\max}=3$ for the
primary configuration, in compressed spectral units.

\subsection{MaRFlow-O: Structured conditional flow matching}\label{subsec:training}
\paragraph{From terminal displacements to instantaneous targets.}
Let $q=1-t$ denote the restoration progress along
the reverse-time conditional path.
Using the endpoint consistency identity in
Eq.~\eqref{eq:decomp}, the conditional state can
be expressed as
\begin{align}
Y_t
&=X+qA_M+qA_R+\sigma(t)Z\\
&=(1-t)S+tX+\sigma(t)Z.
\label{eq:matchedpath}
\end{align}

Let $U_M,U_R\in\C^{F\times L}$ denote the
oracle gain-rate target and additive spectral
velocity target, respectively, and let
$U_Y\in\C^{F\times L}$ denote their assembled
aggregate velocity target.
Differentiation with respect to $t$, under the
reverse-time convention, gives
\begin{align}
U_M&=-D^\star,\qquad
U_R=-R^\star+\sigma'(t)Z,
\\
U_Y&=X\odot U_M+U_R\\
&=X-S+\sigma'(t)Z=U_t.\label{eq:branch_targets}
\end{align}
The artificial Gaussian velocity is assigned to the additive branch because it is additive in the modeled coordinates and not necessarily proportional to the observation. This is a modeling allocation, not a uniquely identifiable physical decomposition. The source, target, intermediate conditional states, and aggregate velocity are unchanged from FlowSE.

\paragraph{Compatibility with observation-mediated restoration.}
For a reverse Euler step of size $0<h\leq t$,
let $\Delta Y_M^{\mathrm O}$ and
$\Delta Y_R^{\mathrm O}$ denote the multiplicative
and additive spectral increments, respectively,
obtained by evaluating the observation-mediated
Euler update using the oracle velocity targets.
Both quantities belong to $\C^{F\times L}$ and
are given by
\begin{align}
\Delta Y_M^{\mathrm O}
&=-hX\odot U_M=hA_M,\\
\Delta Y_R^{\mathrm O}
&=-hU_R=hA_R-h\sigma'(t)Z.
\label{eq:observation_oracle_displacement}
\end{align}
Their sum satisfies
\begin{align}
\Delta Y_M^{\mathrm O}+\Delta Y_R^{\mathrm O}
&=h(S-X)-h\sigma'(t)Z\nonumber\\
&=-hU_t.
\label{eq:observation_oracle_euler}
\end{align}
Thus, the terminal-displacement allocation specifies how the
paired restoration difference is accumulated relative to the
fixed observation, while assigning the artificial perturbation
velocity to the additive coordinate. Because the multiplicative
reference remains unchanged, the prescribed endpoint allocation
does not require recomputation along the conditional path.

In the unclipped case, define the observation-dependent
allocation coefficient
\begin{equation}
\alpha_X=\frac{|X|^2}{|X|^2+\lambda}
\in[0,1]^{F\times L}.
\end{equation}
This coefficient specifies the fraction of the
paired endpoint displacement allocated to the
multiplicative spectral contribution in each bin.
The complementary fraction is $1-\alpha_X$.
The resulting oracle Euler increments are
\begin{align}
\Delta Y_M^{\mathrm O}
&=h\alpha_X\odot(S-X),\\
\Delta Y_R^{\mathrm O}
&=h(1-\alpha_X)\odot(S-X)-h\sigma'(t)Z.
\label{eq:observation_displacement_allocation}
\end{align}
The allocation of the paired endpoint displacement is therefore
determined by the fixed observation power, in contrast to the
state-dependent allocation developed for recursive restoration
in Section~\ref{subsec:recursive_training}.

\paragraph{Training objective.}
Let $\mathcal L(\theta)$ denote the training loss,
and let $w_Y,\beta_M,\beta_R\geq0$ denote the
weights associated with the aggregate velocity,
gain-rate and additive velocity regression terms,
respectively.
We jointly regress the assembled velocity and
the two reference-specific oracle targets:
\begin{align}
\mathcal L(\theta)=\E\big[&w_Y\norm{X\odot v_M+v_R-U_Y}_F^2
+\beta_M\norm{v_M-U_M}_F^2\nonumber\\
&+\beta_R\norm{v_R-U_R}_F^2\big].\label{eq:loss}
\end{align}
All three weights equal one in the primary supervised model. We retain the inherited sum-over-bins, batch-mean squared-error reduction up to common normalization. Gain and spectral errors use different coordinates, so equal weights need not imply equal gradient scales. Positive times and complex Gaussian draws are sampled according to the inherited CFM procedure. There is no waveform loss or solver differentiation. The aggregate-only fixed-reference variant, denoted \Oagg, uses $w_Y=1$ and $\beta_M=\beta_R=0$.

The terminal-displacement oracle therefore complements the
fixed-reference inductive bias of observation-mediated restoration
by prescribing a consistent allocation of the endpoint displacement.
For recursive restoration, however, the multiplicative reference
changes along the conditional path. This motivates an alternative
allocation defined locally in terms of the instantaneous state
and target velocity, as developed next.

\subsection{MaRFlow-R: Local instantaneous-velocity allocation}
\label{subsec:recursive_training}

\paragraph{Why a local oracle?}
The recursive parameterization introduced in
Section~\ref{subsec:recursive} provides an inductive bias toward
progressive relative correction of the evolving restoration state.
However, aggregate velocity supervision alone does not determine
how this correction should be allocated between the multiplicative
and additive coordinates.

In recursive restoration, the multiplicative reference is the
current conditional state $Y_t$, rather than the fixed observation
$X$. Reusing the terminal target $-D^\star$ from O would therefore
not retain its intended spectral contribution. Moreover, directly
dividing the desired spectral velocity by $Y_t$ is ill-conditioned
near state nulls. We instead seek a reference-dependent allocation
of the instantaneous velocity that balances additive correction
against the magnitude of the complex gain rate.


Specifically, for each complex time--frequency bin, let
$\gamma\in\C$ denote a candidate complex gain rate and
$r\in\C$ denote its complementary additive spectral velocity.
With $Y_t$ and $U_t$ denoting the corresponding scalar entries
of the conditional state and target velocity, respectively,
we pose the constrained allocation problem
\begin{equation}
\begin{aligned}
\min_{\gamma,r}\quad
& |r|^2+\lambda_{\mathrm{loc}}|\gamma|^2\\
\text{subject to}\quad
& Y_t\gamma+r=U_t,
\end{aligned}
\label{eq:local_constrained}
\end{equation}
where $\lambda_{\mathrm{loc}}>0$ controls the trade-off
between the magnitude of the additive spectral velocity
and that of the complex gain rate. The constraint ensures
that the two contributions exactly recover the prescribed
instantaneous velocity in compressed spectral coordinates.

This formulation favors expressing the instantaneous
correction through state modulation when the reference
supports it, while regularizing the gain rate near
low-magnitude states. The minimizing values of $\gamma$
and $r$ define the oracle targets $U_M^{\mathrm{loc}}$
and $U_R^{\mathrm{loc}}$, respectively.

Eliminating the additive coordinate yields the equivalent
regularized least-squares problem
\begin{align}
U_M^{\mathrm{loc}}
&=\arg\min_{\gamma}
\{|Y_t\gamma-U_t|^2
+\lambda_{\mathrm{loc}}|\gamma|^2\}
\nonumber\\
&=\frac{U_tY_t^*}{|Y_t|^2+\lambda_{\mathrm{loc}}},
\label{eq:local_gain}\\
U_R^{\mathrm{loc}}
&=U_t-Y_t\odot U_M^{\mathrm{loc}}.
\label{eq:local_add}
\end{align}
The strictly positive regularization coefficient ensures a unique
solution for every reference value, including state nulls.
The resulting reference-dependent and additive spectral targets
satisfy
\begin{equation}
\boxed{
Y_t\odot U_M^{\mathrm{loc}}+U_R^{\mathrm{loc}}=U_t.
}
\label{eq:local_consistency}
\end{equation}
Unlike the terminal oracle, this rule allocates the
\emph{total instantaneous velocity}, including the artificial
Gaussian component. It requires no terminal-mask clipping and
uses $\lambda_{\mathrm{loc}}=0.01$.

\paragraph{State-dependent allocation of restoration.}
Let $P_t=|Y_t|^2\in\mathbb R_{\geq0}^{F\times L}$
denote the instantaneous squared magnitude of the
conditional state in compressed spectral coordinates.
Substituting the local oracle solution into its
multiplicative and additive spectral contributions
gives
\begin{align}
Y_t\odot U_M^{\mathrm{loc}}
&=\frac{P_t}{P_t+\lambda_{\mathrm{loc}}}U_t,
\nonumber\\
U_R^{\mathrm{loc}}
&=\frac{\lambda_{\mathrm{loc}}}
{P_t+\lambda_{\mathrm{loc}}}U_t.
\label{eq:local_split}
\end{align}
Define the instantaneous state-dependent allocation
coefficient
\begin{equation}
\alpha_t
=\frac{P_t}{P_t+\lambda_{\mathrm{loc}}}
\in[0,1)^{F\times L}.
\label{eq:local_allocation_coefficient}
\end{equation}
Here, $\alpha_t$ specifies the fraction of the
instantaneous target velocity allocated to the
multiplicative spectral contribution in each bin,
and $1-\alpha_t$ specifies the complementary fraction
allocated to the additive coordinate.
The multiplicative and additive spectral targets are therefore
$\alpha_t\odot U_t$ and $(1-\alpha_t)\odot U_t$, respectively.
Both contributions are collinear with $U_t$, and their
coefficients sum to one.


For $P_t\gg\lambda_{\mathrm{loc}}$, the allocation favors
multiplicative correction, whereas for
$P_t\ll\lambda_{\mathrm{loc}}$, it favors additive correction.
At $P_t=\lambda_{\mathrm{loc}}$, the two contributions receive
equal fractions of the instantaneous velocity. At a state null,
the additive coordinate receives the entire target.

The gain target additionally satisfies
\begin{equation}
|U_M^{\mathrm{loc}}|
\leq
\frac{|U_t|}{2\sqrt{\lambda_{\mathrm{loc}}}},
\label{eq:local_gain_bound}
\end{equation}
preventing the singular behavior associated with unregularized
complex division. Importantly, the allocation adapts to the
instantaneous state power even with a constant ridge, without
requiring an explicitly prescribed temporal allocation schedule.
These coefficients allocate spectral velocity rather than
physical speech and noise components.

\paragraph{Compatibility with recursive restoration.}
Consider a reverse Euler step of size $0<h\leq t$.
Let $Y_{t-h}^{\mathrm{oracle}}\in\C^{F\times L}$
denote the state obtained by evaluating the recursive
Euler update at the sampled conditional state $Y_t$
using the local oracle targets
$U_M^{\mathrm{loc}}$ and $U_R^{\mathrm{loc}}$,
rather than the learned predictions.
The resulting update is
\begin{align}
Y_{t-h}^{\mathrm{oracle}}
&=(1-hU_M^{\mathrm{loc}})\odot Y_t
-hU_R^{\mathrm{loc}}
\nonumber\\
&=Y_t-h\alpha_t\odot U_t
-h(1-\alpha_t)\odot U_t
\nonumber\\
&=Y_t-hU_t.
\label{eq:local_oracle_euler}
\end{align}
The local oracle therefore distributes the prescribed
restoration displacement between recursive modulation and
additive correction without modifying their aggregate.

For the conditional path considered here, substituting
Eq.~\eqref{eq:conditional_shrinkage} into the local gain gives,
for $t>0$,
\begin{equation}
U_M^{\mathrm{loc}}
=\frac{1}{t}
\frac{|Y_t|^2-Y_t^*S}
{|Y_t|^2+\lambda_{\mathrm{loc}}}.
\label{eq:local_shrinkage}
\end{equation}
The corresponding additive target is
\begin{equation}
U_R^{\mathrm{loc}}
=\frac{1}{t}
\frac{\lambda_{\mathrm{loc}}}
{|Y_t|^2+\lambda_{\mathrm{loc}}}
\odot(Y_t-S).
\label{eq:local_additive_shrinkage}
\end{equation}
Let $\Delta Y_M^{\mathrm{loc}}$ and
$\Delta Y_R^{\mathrm{loc}}$ denote the spectral
increments contributed by the multiplicative and
additive local oracle velocities, respectively.
Both belong to $\C^{F\times L}$ and can be
expressed as
\begin{align}
\Delta Y_M^{\mathrm{loc}}
&=-hY_t\odot U_M^{\mathrm{loc}}
\nonumber\\
&=\frac{h}{t}\alpha_t\odot(S-Y_t),\\
\Delta Y_R^{\mathrm{loc}}
&=-hU_R^{\mathrm{loc}}
\nonumber\\
&=\frac{h}{t}(1-\alpha_t)\odot(S-Y_t).
\label{eq:local_restoration_displacement}
\end{align}
This establishes that both coordinates contribute to restoration
toward the paired clean endpoint, with their relative
responsibilities determined by the current state power.

The prescribed gain is thus a state- and target-dependent
complex correction rather than a uniform scalar contraction.
The local allocation complements the architectural inductive
bias of R by explicitly supervising relative correction
where the evolving reference supports it and additive
reconstruction where such modulation is insufficient or
poorly conditioned.


Crucially, this construction specifies an allocation of
the paired conditional velocity without modifying the
conditional probability path or its aggregate velocity target.
The construction neither requires an ODE rollout during
training nor assumes that recursive modulation alone is
sufficient for restoration. The clean target is used only
to construct supervision during training. The learned inference
dynamics need not reproduce the oracle allocation exactly.

\paragraph{Training.}
The locally supervised model, denoted
\textbf{MaRFlow-R (local)}, is trained by minimizing
$\mathcal L_{R,\mathrm{loc}}(\theta)$, the recursive
local-oracle training objective.
Using the same loss-weight notation as in
Eq.~\eqref{eq:loss}, we define
\begin{align}
\mathcal L_{R,\mathrm{loc}}(\theta)
=\E\big[
&w_Y\norm{Y_t\odot v_M+v_R-U_t}_F^2
\nonumber\\
&+\beta_M\norm{v_M-U_M^{\mathrm{loc}}}_F^2
\nonumber\\
&+\beta_R\norm{v_R-U_R^{\mathrm{loc}}}_F^2
\big].
\label{eq:recursive_local_loss}
\end{align}
The aggregate loss supervises the complete restoration field,
while the two auxiliary losses specify its reference-dependent
allocation. Since the oracle contributions sum exactly to
the original conditional velocity, the additional branch
losses preserve the unrestricted population-optimal
aggregate field, as established in
Section~\ref{subsec:population}.

All three weights are one, as in supervised O.
The architecture and solver are unchanged relative to
aggregate-only MaRFlow-R (R-agg), which uses $w_Y=1$,
$\beta_M=\beta_R=0$. The corresponding aggregate-only O
control is O-agg. Thus, within each reference family the
additional losses isolate the effect of structured supervision.
In comparisons of supervised O and R, the reference
\emph{and} the oracle differ by design.

\subsection{Population compatibility of reference-specific supervision}
\label{subsec:population}

We now use $U_M,U_R\in\C^{F\times L}$ generically
to denote the reference-specific oracle gain-rate
and additive velocity targets, respectively.
For O, these are the terminal-displacement-derived
targets in Eq.~\eqref{eq:branch_targets}; for R,
they are $U_M^{\mathrm{loc}}$ and
$U_R^{\mathrm{loc}}$ in
Eqs.~\eqref{eq:local_gain}--\eqref{eq:local_add}.
Both constructions satisfy
\begin{equation}
\Phi_t\odot U_M+U_R=U_t
\end{equation}
on the original conditional probability path.

Recall that $W=(Y_t,X,t)$ denotes the conditioning
variables available to the network.
Assuming finite second moments of the target
velocities, define the conditional mean branch targets
\begin{align}
\mu_M(W)&=\E[U_M\mid W],\\
\mu_R(W)&=\E[U_R\mid W],
\end{align}
where $\mu_M,\mu_R$ take values in
$\C^{F\times L}$.
Since the multiplicative reference $\Phi_t$ is
either $X$ or $Y_t$, it is measurable given $W$.
Taking conditional expectations therefore gives
\begin{equation}
\E[U_t\mid W]
=\Phi_t\odot\mu_M(W)+\mu_R(W).
\label{eq:expect}
\end{equation}

Let $e_M,e_R\in\C^{F\times L}$ denote the
deviations of the predicted branch outputs from
their respective conditional mean targets:
\begin{align}
e_M(W)&=v_M(W)-\mu_M(W),\\
e_R(W)&=v_R(W)-\mu_R(W).
\end{align}
Up to a prediction-independent term, the conditional
risk $\mathcal R_W$ is
\begin{equation}
\mathcal R_W=
w_Y\norm{\Phi_t\odot e_M+e_R}_F^2
+\beta_M\norm{e_M}_F^2
+\beta_R\norm{e_R}_F^2.
\end{equation}
For positive branch weights
$\beta_M,\beta_R>0$ and nonnegative aggregate
weight $w_Y\geq0$, its unique unrestricted minimizer
is $(v_M,v_R)=(\mu_M,\mu_R)$.
The corresponding assembled velocity equals the
original CFM population-optimal field.

Under aggregate-only supervision, with $w_Y>0$
and $\beta_M=\beta_R=0$, let
$\Delta(W)\in\C^{F\times L}$ denote an arbitrary
perturbation of the predicted gain rate.
The simultaneous transformation
\begin{equation}
(v_M,v_R)
\longmapsto
(v_M+\Delta,v_R-\Phi_t\odot\Delta)
\end{equation}
leaves the assembled velocity unchanged.
Consequently, the aggregate-only risk admits the
null direction $(\Delta,-\Phi_t\odot\Delta)$.
The branch losses remove this allocation ambiguity
relative to the chosen oracle.

For an individual time--frequency bin, define
the complex output-error vector
$e=(e_M,e_R)^\top\in\C^2$.
The corresponding conditional quadratic risk is
$e^{\mathrm H}H(\Phi_t)e$, where
$(\cdot)^{\mathrm H}$ denotes conjugate transpose
and $H(\Phi_t)\in\C^{2\times2}$ is the Hermitian
output-space risk matrix:
\begin{equation}
H(\Phi_t)=
\begin{pmatrix}
w_Y|\Phi_t|^2+\beta_M & w_Y\Phi_t^*\\
w_Y\Phi_t & w_Y+\beta_R
\end{pmatrix}
\succeq\min(\beta_M,\beta_R)I_2.
\label{eq:hessian}
\end{equation}
Here, $I_2$ denotes the $2\times2$ identity matrix,
and $\succeq$ denotes the positive-semidefinite
matrix ordering.
The lower bound establishes positive definiteness
when both branch weights are positive.
This result concerns the quadratic risk in the
output coordinates, not the Hessian with respect
to the neural network parameters $\theta$.

\paragraph{Population-level allocation in recursive restoration.}
The local oracle admits a more explicit characterization
of the optimal branch predictions.
Recall the original CFM population-optimal velocity
$v^\star(W)=\E[U_t\mid W]$.
Since the local reference $Y_t$ and its power are measurable
given $W$, taking conditional expectations in
Eqs.~\eqref{eq:local_gain} and~\eqref{eq:local_add} yields
\begin{align}
\mu_M^{\mathrm{loc}}(W)
&=\frac{Y_t^*}{|Y_t|^2+\lambda_{\mathrm{loc}}}
\odot v^\star(W),
\label{eq:population_local_gain}\\
\mu_R^{\mathrm{loc}}(W)
&=\frac{\lambda_{\mathrm{loc}}}
{|Y_t|^2+\lambda_{\mathrm{loc}}}
\odot v^\star(W).
\label{eq:population_local_add}
\end{align}
Consequently, the population-optimal spectral contributions
satisfy
\begin{align}
Y_t\odot\mu_M^{\mathrm{loc}}(W)
&=\alpha_t\odot v^\star(W),\nonumber\\
\mu_R^{\mathrm{loc}}(W)
&=(1-\alpha_t)\odot v^\star(W),
\label{eq:population_local_allocation}
\end{align}
where $\alpha_t=|Y_t|^2/(|Y_t|^2+\lambda_{\mathrm{loc}})$.

Thus, the state-dependent allocation prescribed by the local
oracle is preserved at the unrestricted population optimum,
rather than holding only for individual paired conditional
velocities. The oracle specifies a regularized decomposition
of the same optimal aggregate field into relative state
modulation and complementary additive correction.

This characterization does not prescribe a particular temporal
ordering of the branches, since the allocation depends on
the evolving state power. Nor does it imply that the learned
branches exactly realize the prescribed allocation along
inference trajectories.

\paragraph{Implications for finite-capacity learning.}
The proposed framework introduces inductive biases at two
distinct levels. The choice of multiplicative reference
determines how restoration contributions compose along the
inference trajectory, whereas reference-specific supervision
specifies how the instantaneous velocity is allocated between
the two output coordinates. The former is present even under
aggregate-only training; the latter resolves the allocation
ambiguity relative to the prescribed oracle.

Although neither changes the unrestricted aggregate velocity
capacity or its population-optimal target, the finite shared
network may not represent both conditional means simultaneously.
The auxiliary objectives modify the optimization landscape
and the allocation of shared model capacity, while the
aggregate loss directly supervises the field seen by the solver.
Their practical significance therefore lies in how they
influence finite-capacity approximation, optimization and
restoration dynamics. We investigate these effects empirically
in the following sections (see Appendix~\ref{app:risk}
for further theoretical details).

\subsection{Inference: comparing observation-mediated and recursive updates}
\label{subsec:inference}

All evaluated models start from the FlowSE Gaussian
source $\mathsf Y_0=X+\sigma_{\max}Z$, where
$Z\sim\mathcal{CN}(0,I)$ is the inference-time complex
Gaussian perturbation.

Let $K$ denote the number of network function evaluations
(NFE), which equals the number of Euler updates in our
implementation. We consider the reverse-time integration grid
$1=t_0>t_1>\cdots>t_K=0$, with positive step sizes
$h_k=t_k-t_{k+1}$ for $k=0,\ldots,K-1$.

For MaRFlow, at each solver time $t_k$, the shared
backbone jointly predicts the complex gain rate $v_{M,k}$
and additive spectral velocity $v_{R,k}$ conditioned on
$(\mathsf Y_k,X,t_k)$, where $\mathsf Y_k$ denotes the
current numerical approximation to the learned transport
state. The two Euler update rules are
\begin{align}
\text{O:}\qquad
\mathsf Y_{k+1}
&=\mathsf Y_k-h_kX\odot v_{M,k}-h_kv_{R,k},
\label{eq:euler}\\
\text{R:}\qquad
\mathsf Y_{k+1}
&=(1-h_kv_{M,k})\odot\mathsf Y_k-h_kv_{R,k}.
\label{eq:euler_R}
\end{align}

These updates implement the distinct restoration mechanisms
derived in Section~\ref{subsec:factorization}: fixed-reference
accumulation in O and recursive affine state modulation in R.
Despite their different composition rules, both formulations
integrate a single spectral state and require only one shared
network evaluation per Euler step. Neither requires a
separately integrated gain, an additional restoration network,
or sequential evaluation of the two branches.

For comparison, FlowSE directly predicts the aggregate
spectral velocity and applies the standard Euler update
\begin{equation}
\mathsf Y_{k+1}
=\mathsf Y_k-h_kv_\theta(\mathsf Y_k,X,t_k).
\end{equation}
All models use the same integration grid and matched NFE
for each evaluated inference budget.

\paragraph{Integration grid.}
For $K>1$, the solver evaluates the velocity field at
positive times from $t_0=1$ down to
$t_{K-1}=t_\epsilon=0.03$, where $t_\epsilon$ denotes
the final positive evaluation time. The final Euler
update transports the numerical state to $t_K=0$.
For $K=1$, a single Euler update spans the entire
interval from $t_0=1$ to $t_1=0$.

\paragraph{Waveform reconstruction.}
Let $\widehat S\in\C^{F\times L}$ denote the final
compressed spectral estimate obtained from $\mathsf Y_K$
after removing any inference-time frame padding, and
let $\widehat s\in\mathbb R^N$ denote the corresponding
enhanced waveform.

The waveform is reconstructed through amplitude
decompression, inverse short-time Fourier transform
(ISTFT), and restoration of the noisy normalization
scale:
\begin{equation}
\widehat s=
g\,\mathrm{ISTFT}\left[
\left(\frac{|\widehat S|}{c}\right)^{1/p}
e^{j\angle\widehat S}
\right].
\label{eq:waveform_reconstruction}
\end{equation}
The ISTFT uses the original analysis parameters and
reconstructs the waveform at its original length $N$.
Here, $g$, $c$ and $p$ are the noisy normalization
factor, spectral scaling coefficient and amplitude
compression exponent, respectively, as defined in
Section~\ref{subsec:background}.

Importantly, the synthesis operation is applied to
the assembled endpoint rather than independently to
the multiplicative and additive contributions.
The latter are algebraic corrections in compressed
spectral coordinates, not independently reconstructable
physical waveform components. Neither the clean
representation $S$ nor the oracle supervision targets
are required during inference.

\section{Experiments}\label{sec:experiments}
\paragraph{Datasets and experimental regimes.}
We distinguish two training regimes. The primary five-model study uses the original VoiceBank+DEMAND \citep{valentini} additive-noise mixtures and its 824-utterance test set. None of these five checkpoints receives explicit room-impulse-response (RIR) augmentation. Without adaptation, we evaluate them on the two DNS 2020 test subsets \citep{reddy}, each containing 150 utterances, with and without reverberation. The reverberant subset is therefore a zero-shot stress test for these checkpoints.

In a separate three-model study, FlowSE, O and R-local are trained from scratch on newly synthesized noisy reverberant mixtures using the same underlying VoiceBank clean-speech training/validation splits. Rather than reverberating the original VoiceBank+DEMAND mixtures, we construct new mixtures from clean VoiceBank speech, room impulse responses and separately sampled additive noise from the supplied point-source and isotropic-noise collections. We convolve the clean speech with an RIR and add noise at a prescribed SNR relative to the reverberant speech. The target remains reverberant speech: the task is additive-noise suppression under reverberant conditions, not dereverberation. Training uses dynamic augmentation; validation mixtures are fixed. Testing uses 824 newly synthesized mixtures of the held-out VoiceBank utterances with held-out RIR and noise recordings, denoted \emph{synthetic reverberant VoiceBank}, and the unchanged 150-utterance reverberant DNS subset. The latter permits a direct descriptive comparison with the zero-shot regime. We therefore use the term synthetic reverberant VoiceBank rather than VoiceBank+DEMAND with reverberation; the speech corpus is shared, but the noise and mixture construction differ from the original benchmark. All evaluation uses 16\,kHz audio and, in both reverberant sets, reverberant clean references. Appendix~\ref{app:implementation} specifies the mixture construction.

\paragraph{Evaluation.} We report utterance-averaged PESQ-WB, STOI, ESTOI, SI-SDR (dB), DNSMOS P.808 and P.835 overall (OVRL) \citep{pesq,stoi,estoi,sisdr,dnsmos,dnsmos835}.

\paragraph{Linguistic recoverability.}
To assess content preservation separately from acoustic restoration,
we use frozen Whisper Large-v3 \citep{whisper} and corpus-level
word error rate (WER), including clean and noisy anchors.
VoiceBank uses its original transcripts; DNS uses publicly
available, manually corrected Whisper-v2 transcripts%
\footnote{\url{https://github.com/Emrys365/DNS_text} \label{dnstext}}.
We normalize references and hypotheses with Whisper's English
text normalizer and compute $100(S+D+I)/N$, where $S,D,I$
are substitution, deletion and insertion counts and $N$ is
the number of reference words.

\paragraph{Inference and statistical protocol.} The primary evaluation uses five network function evaluations (NFE) and the first paired source draw; we additionally sweep $K\in\{1,2,3,5,10\}$ and use two additional source draws at five NFE on VoiceBank and non-reverberant DNS. Paired pointwise 95\% percentile bootstrap intervals use 5,000 utterance resamples, conditional on the trained checkpoints. For WER each resample recomputes the ratio of total errors to total reference words. All acoustic metrics are higher-is-better; WER is lower-is-better. We additionally report two-sided paired sign-flip permutation tests (20,000 permutations, plus-one correction) and Holm-adjusted $p$-values. For the additive-noise-trained checkpoints, the primary acoustic, zero-shot reverberant acoustic, and WER comparisons constitute separate families of 28, 18, and 18 tests, respectively. For the reverberation-trained study, the existing correction across the 36 acoustic comparisons and separately across the six WER comparisons is retained (Appendix~\ref{app:implementation}).

\paragraph{Training.} All eight reported training configurations (five additive-noise and three reverberation-aware) use a matched budget of \textbf{30 epochs each}. The five primary configurations are trained using seed 13, aligned 2.04-s crops, batch size six, Adam with learning rate $10^{-4}$ and EMA decay 0.999. Checkpoints are selected by validation SI-SDR at five NFE; all five primary selected checkpoints occur at epoch 30. FlowSE is retrained under the same protocol. The separate reverberant study retains the 30-epoch budget, seed, backbone, conditional path, losses and five-NFE EMA selection; the best epochs for FlowSE/O/R-local are 29/30/29 (Appendix~\ref{app:reverb_trained}). Only the three complete formulations are retrained in this study; no additional aggregate-only controls are used. The matched comparisons preserve the backbone, training budget and probability path. The two aggregate-only controls isolate the reference choice, while the explicitly supervised variants isolate the addition of reference-specific branch losses within each reference family. Further hyperparameters, preprocessing and evaluation details are provided in Appendix~\ref{app:implementation}.

\begin{table}[t]\centering\small
\caption{Primary model configurations. All use the same linear FlowSE conditional path and Euler solver. O and R-local are the two reference-specifically supervised formulations.}\label{tab:primary_design}
\fitTable{\begin{tabular}{lll}\toprule Model & Solver-facing velocity & Supervision\\\midrule
FlowSE & $v_\theta$ & Aggregate\\
O-agg & $X\odot v_M+v_R$ & Aggregate\\
\textbf{O} & $X\odot v_M+v_R$ & Aggregate + terminal allocation\\
R-agg & $Y_t\odot v_M+v_R$ & Aggregate\\
\textbf{R-local} & $Y_t\odot v_M+v_R$ & Aggregate + local allocation\\\bottomrule\end{tabular}}
\end{table}

\subsection{Results and Discussion}\label{subsec:results}
Table~\ref{tab:primary} presents absolute enhancement results on VoiceBank+DEMAND and non-reverberant DNS; the paired statistical results are reported in Appendix~\ref{app:extended} (Table~\ref{tab:paired}). All detailed statistical testing tables are provided in the appendices, while their interpretation remains in the main text. We then study input-SNR dependence, the reference and supervision choices, few-step inference, branch behavior, zero-shot reverberant transfer, matched reverberation-aware training, and linguistic recoverability across both training regimes. For compactness, O and R-local denote the supervised variants, and O-agg and R-agg denote their aggregate-only controls. Unless specified otherwise, results preceding Section~\ref{subsec:reverb_trained} refer to the additive-noise-trained checkpoints.

\subsubsection{Enhancement performance: in-domain and cross-domain evaluation}
\begin{table}[t]\centering\small
\caption{Enhancement on VoiceBank+DEMAND and non-reverberant DNS 2020 at five NFE, draw zero. Values are utterance means; SI-SDR in dB. Best unrounded means are bold.}\label{tab:primary}\fitTable{\MainResults}
\end{table}
\noindent\textbf{Do both supervised factorizations improve over direct velocity prediction (cf. Table~\ref{tab:primary})?} Yes. On VoiceBank, O and R-local attain PESQ of 2.62 and 2.59 versus FlowSE's 2.39, with SI-SDR of 17.85/17.84 versus 17.41\,dB. O leads PESQ and SI-SDR, while R-local leads STOI, DNSMOS P.808 and OVRL (0.932, 3.55 and 3.17). ESTOI is 0.840 for both at the displayed precision. On non-reverberant DNS the gains are larger: O reaches 1.97/14.00 PESQ/SI-SDR, while R-local reaches 2.00/14.11, compared with FlowSE's 1.70/12.64. R-local leads all six mean metrics on DNS. Notably, FlowSE's STOI is below that of noisy input (0.905 vs. 0.915), whereas both supervised formulations improve it to 0.932/0.934. These results support both reference-specific formulations rather than a preference for a fixed reference in all settings.

\noindent\textbf{How consistent are the gains (cf. Appendix Table~\ref{tab:paired})?} O improves PESQ on 92.8\%/98.7\% of VoiceBank/DNS utterances; R-local improves it on 88.2\%/96.0\%. The paired SI-SDR gains over FlowSE are 0.44 [0.40, 0.49] and 0.43 [0.39, 0.47] dB on VoiceBank, and 1.36 [1.16, 1.57] and 1.48 [1.25, 1.72] dB on DNS, for O and R-local respectively. The two supervised models have similar SI-SDR: R-local minus O is $-0.02$ [$-0.04$, 0.01] and $+0.12$ [$-0.05$, 0.29] dB on the two sets. PESQ favors O on VoiceBank ($-0.03$ for R-local minus O) and R-local on DNS ($+0.03$), with respective intervals [$-0.04$, $-0.02$] and [0.02, 0.05]. All eight PESQ and SI-SDR improvements of O and R-local over FlowSE remain significant after Holm correction ($p_{\rm Holm}\leq0.0014$); neither SI-SDR contrast between the supervised models is significant ($p_{\rm Holm}=0.3627$). The results indicate complementary strengths, not a universal ordering between the supervised formulations. The intervals quantify paired test-example variability, not variation across independent training runs.

\subsubsection{Enhancement across input SNR ranges}\label{subsec:snr}
We stratify the same five-NFE test outputs by the input SNR, using four half-open ranges with boundaries at 0, 5 and 10\,dB. Each range contains the same utterances across models; the noisy input is included as a reference. These are descriptive utterance means rather than separately trained SNR-specific models or independently generated test sets.

\begin{table}[!htbp]
\centering\small\setlength{\tabcolsep}{4pt}
\caption{Input-SNR-stratified enhancement on VoiceBank+DEMAND (five NFE, first inference draw). SI-SDR is in dB; PESQ is wideband. P.808 and OVRL are DNSMOS predictions. Sample counts refer to utterances within each range; bold indicates the highest unrounded mean in that range.}\label{tab:snr_vb}
\fitTable{\SNRVoiceBank}
\end{table}

\noindent\textbf{Does the benefit persist across input SNR (cf. Tables~\ref{tab:snr_vb} and~\ref{tab:snr_dns})?} Both supervised factorizations improve mean PESQ and SI-SDR over FlowSE in every input-SNR range on both datasets. On VoiceBank, O has higher mean PESQ than R-local in each range, while their SI-SDR is close; R-local gives higher OVRL throughout. On DNS, R-local improves or closely matches O's mean PESQ and SI-SDR across the four ranges and consistently improves OVRL. For instance, at 5--10 dB on DNS, R-local reaches 1.77/12.24 PESQ/SI-SDR versus O's 1.73/11.99 and FlowSE's 1.48/10.73. The benefit is not confined to a particular input-SNR interval.

\begin{table}[!htbp]
\centering\small\setlength{\tabcolsep}{4pt}
\caption{Input-SNR-stratified enhancement on the non-reverberant DNS 2020 test set (five NFE, first inference draw). The format follows Table~\ref{tab:snr_vb}. The negative-SNR range contains only five utterances.}\label{tab:snr_dns}
\fitTable{\SNRDNS}
\end{table}

\noindent\textbf{How should the low-SNR results be interpreted?} The negative-SNR strata contain only 16 VoiceBank and five DNS utterances, so their means should be read descriptively. The unequal strata complement the pooled paired comparisons; input SNR is distinct from the compressed spectral power used by the allocation oracles (Fig.~\ref{fig:snr}).

\subsubsection{Architectural choice: reference and structured supervision}
\noindent\textbf{Does the factorized interface alone improve enhancement?} O-agg improves PESQ by 0.02/0.05 and SI-SDR by 0.20/0.30\,dB over FlowSE on VoiceBank/DNS. The additional interface provides modest gains without an explicit allocation. These contrasts test the complete structured output interface, including the new gain head.

\noindent\textbf{What changes when the reference becomes recursive?} With aggregate-only supervision, R-agg trails O-agg in both PESQ and SI-SDR (SI-SDR differences $-0.25$/$-0.82$\,dB). The recursive interface alone therefore does not confer an advantage. Its state-dependent multiplicative contribution makes the allocation problem different from that of O, motivating local supervision rather than reuse of the terminal oracle.

\noindent\textbf{What is gained by reference-specific supervision?} O improves over O-agg by 0.21/0.23 PESQ and 0.24/1.06\,dB SI-SDR. The effect is even more pronounced for recursive restoration: R-local improves over R-agg by 0.24/0.38 PESQ and 0.48/2.00\,dB SI-SDR. The DNS SI-SDR interval for the latter gain is [1.74, 2.29] and the win rate is 97.3\%. The recursive model thus changes from a weaker aggregate-only configuration into a competitive supervised formulation without altering its architecture or Euler rule. Together the within-reference comparisons substantiate structured supervision as a useful inductive bias. The oracle families differ, so the comparison between the two supervised variants is a comparison of complete formulations, not a single-factor reference ablation.

These findings distinguish the availability of a restoration
mechanism from the ability to learn to exploit it.
Although the recursive parameterization provides an explicit
mechanism for progressive relative correction independently
of branch supervision, its aggregate-only realization does
not yield superior enhancement. Reference-specific local
supervision makes this parameterization empirically effective
under the studied training budget, without modifying its
recursive update rule or unrestricted population-optimal
aggregate field.

\subsubsection{Few-step inference and computational cost}
\label{subsec:nfe}
\begin{figure}[t]
\centering
\begin{minipage}[t]{.485\linewidth}\centering\small VoiceBank+DEMAND: PESQ\\\includegraphics[width=\linewidth]{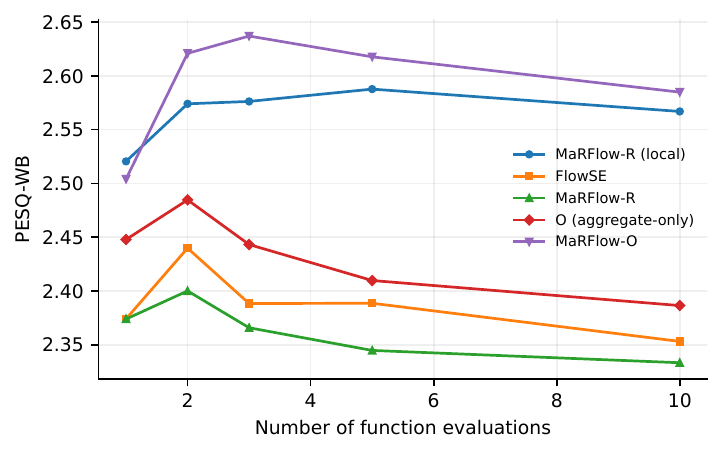}\end{minipage}\hfill
\begin{minipage}[t]{.485\linewidth}\centering\small DNS 2020: PESQ\\\includegraphics[width=\linewidth]{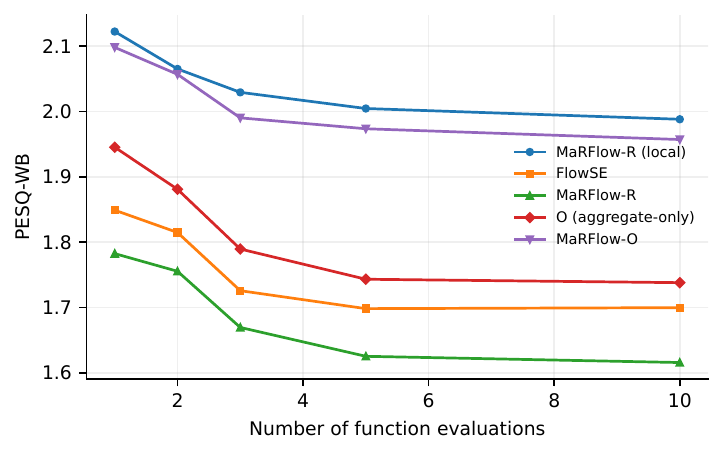}\end{minipage}\\[3pt]
\begin{minipage}[t]{.485\linewidth}\centering\small VoiceBank+DEMAND: SI-SDR\\\includegraphics[width=\linewidth]{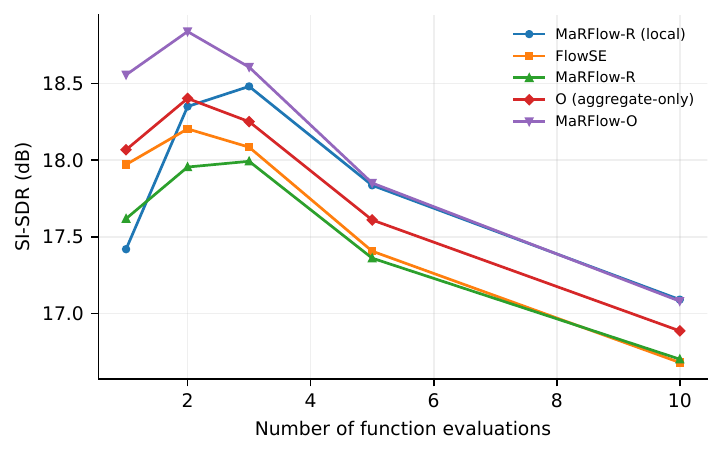}\end{minipage}\hfill
\begin{minipage}[t]{.485\linewidth}\centering\small DNS 2020: SI-SDR\\\includegraphics[width=\linewidth]{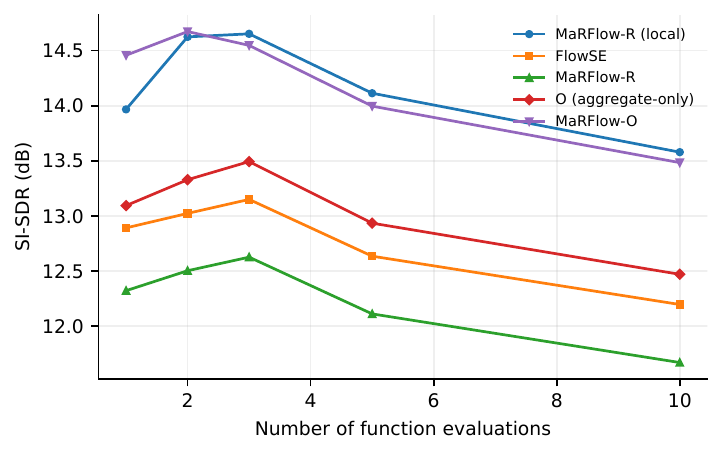}\end{minipage}
\caption{Enhancement quality against solver NFE, first inference draw. Each point is an independently evaluated endpoint at its indicated NFE, rather than an intermediate state of a fixed ten-step rollout. The five curves correspond to the five primary configurations.}\label{fig:nfe}
\end{figure}

\noindent\textbf{Does velocity factorization remain effective at
low NFE (cf. Fig.~\ref{fig:nfe})?}
O attains one-step PESQ/SI-SDR of 2.50/18.56 and 2.10/14.46
on VoiceBank/DNS, exceeding FlowSE's best scores across the
evaluated grid (2.44/18.20 and 1.85/13.15).
R-local achieves the strongest one-step PESQ (2.52/2.12),
although its one-step SI-SDR (17.42/13.97) is lower than O's.
With three evaluations, R-local reaches 18.48/14.65\,dB,
while O's peak SI-SDR is at two NFE (18.84/14.67).
R-local also achieves the highest mean DNSMOS OVRL from
two NFE onward on both datasets.

The recursive formulation exhibits a larger improvement
in SI-SDR when increasing the inference budget from one
to three evaluations: 1.06/0.68\,dB on VoiceBank/DNS,
compared with 0.05/0.09\,dB for O.
Consequently, the initial fidelity disadvantage of R-local
relative to O diminishes with additional evaluations,
and the two formulations attain comparable SI-SDR at
larger solver budgets. This behavior is consistent with
the inductive bias toward progressive relative correction
developed in Section~\ref{subsec:recursive}.
At one NFE, state-mediated modulation acts directly on
the perturbed source, without subsequent remodulation
of additive increments. Additional evaluations enable
the explicit multiplicative refinement of previously
introduced corrections. The endpoint comparisons therefore
establish few-step effectiveness and are consistent with
the proposed recursive restoration mechanism, but do not
isolate the effect of recursive composition from differences
in the learned fields, reference-specific supervision or
integration grids, nor establish a reduction in learned
field curvature.

\noindent\textbf{Does increasing NFE consistently improve enhancement?}
No. PESQ and SI-SDR frequently decrease beyond two or three evaluations, whereas DNSMOS OVRL tends to increase from one to three evaluations and then changes more modestly (Appendix~\ref{app:extended}). More evaluations may integrate the learned ODE more accurately without necessarily producing a better waveform according to a given metric. In particular, comparisons of separate endpoint budgets should not be confused with quality changes within a fixed rollout; the latter is shown separately in the appendix. The available measurements cannot distinguish field approximation, discretization, and objective mismatch as explanations for the endpoint trend.

\begin{table}[t]
\centering\small
\caption{Five-NFE inference cost on NVIDIA L4, uninstrumented waveform-to-waveform evaluation. Per-utterance medians: synchronized latency (ms), real-time factor (RTF), and peak allocated CUDA memory (MiB). Scoring excluded; one warm-up item per condition. FlowSE has 65,590,822 parameters; each MaRFlow model has 65,591,080.}\label{tab:complexity}
\fitTable{\ComplexityResults}
\end{table}
\noindent\textbf{What is the cost of the factorized interface (cf. Table~\ref{tab:complexity})?} The additional complex projection adds 258 parameters to approximately 65.59 million. At five NFE the median VoiceBank latency is 410.99/412.65/411.65 ms for FlowSE/O/R-local; DNS RTF is 0.174/0.175/0.175. Peak allocated memory is comparable (approximately 2.4 GiB). Hence the supervised recursive formulation provides its quality gains without an extra backbone evaluation or appreciable measured cost at matched NFE.

\subsubsection{Diagnostic analysis of the learned velocity decomposition}\label{subsec:diagnostics}

Throughout the following diagnostics, branch energy denotes
the squared Frobenius norm of a velocity contribution in
compressed complex spectral coordinates. In particular,
for $u=\Phi_t\odot v_M$ and $r=v_R$, we measure
$\norm{u}_F^2$ and $\norm{r}_F^2$. These quantities describe
the numerical magnitudes of the two coordinates of the
restoration field, rather than physical acoustic energies
or independently reconstructable waveform components.
All perceptual, fidelity and intelligibility scores are
instead evaluated on the assembled waveform estimates
obtained after inverse spectral compression.

For the local oracle, the velocity allocation coefficient
$\alpha_t$ in Eq.~\eqref{eq:local_allocation_coefficient}
should be distinguished from the corresponding per-bin
multiplicative energy share. For $U_t\neq0$, the latter is
\begin{equation}
\eta_M^{\mathrm{loc}}
=
\frac{\alpha_t^2}
{\alpha_t^2+(1-\alpha_t)^2}
=
\frac{P_t^2}{P_t^2+\lambda_{\mathrm{loc}}^2}.
\label{eq:local_oracle_energy_share}
\end{equation}
The pooled energy share used in our experiments is obtained
by summing the squared branch contributions before forming
their ratio, rather than averaging the per-bin allocation
coefficients. These diagnostics therefore measure agreement
with the prescribed velocity decomposition in squared
coordinate magnitude.

\noindent\textbf{Do the learned rates follow their prescribed allocation (cf. Appendix Figs.~\ref{fig:oracle} and~\ref{fig:localoracle})?} The terminal gain--oracle correlation at $t=0.5$ is 0.835/0.809 for O versus 0.218/0.269 for O-agg on VoiceBank/DNS. This difference persists over the sampled times. R-local attains correlations of 0.959/0.950 at $t=0.5$, with values ranging from 0.951--0.962 and 0.937--0.952 over the sampled times. The local and terminal target families differ and are not ranked by correlation. Rather, these results show that both explicit allocations induce their intended gain behavior on held-out conditional states, whereas aggregate-only O does not recover the terminal allocation. Beyond correlation, R-local closely reproduces the oracle's pooled multiplicative energy share: over all five sampled times and both datasets, the absolute difference is at most 0.78 percentage points (Appendix~\ref{app:extended}). For example, at $t=0.1$ the oracle/predicted shares are 28.46/29.23\% on VoiceBank and 29.51/29.89\% on DNS. At $t=0.9$ they are 99.43/99.49\% and 99.43/99.48\%. This establishes agreement with the prescribed allocation on held-out conditional states; it does not assert identical allocations on model-generated states.

\begin{figure}[t]
\centering
\begin{minipage}[t]{.485\linewidth}\centering\small VoiceBank+DEMAND\\\includegraphics[width=\linewidth]{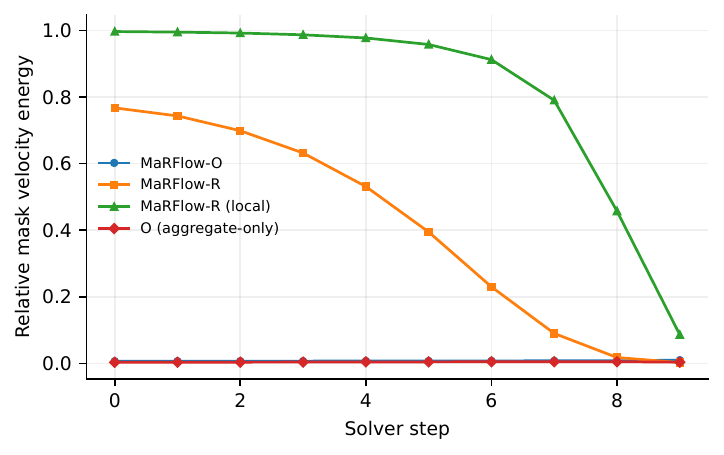}\end{minipage}\hfill
\begin{minipage}[t]{.485\linewidth}\centering\small DNS 2020\\\includegraphics[width=\linewidth]{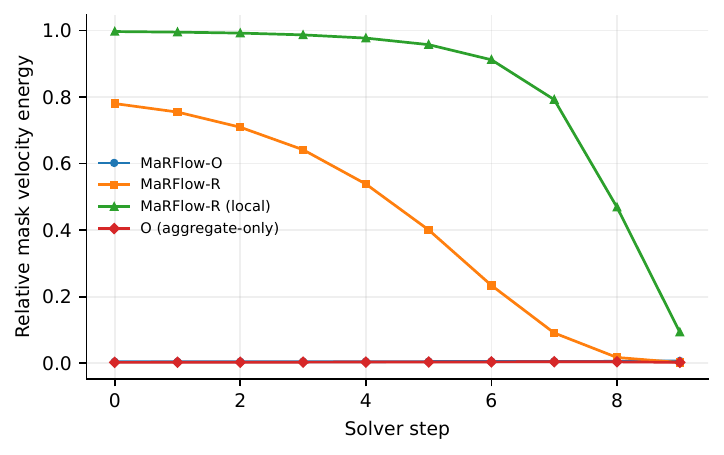}\end{minipage}\\[3pt]
\begin{minipage}[t]{.485\linewidth}\centering\small VoiceBank: power regions\\\includegraphics[width=\linewidth]{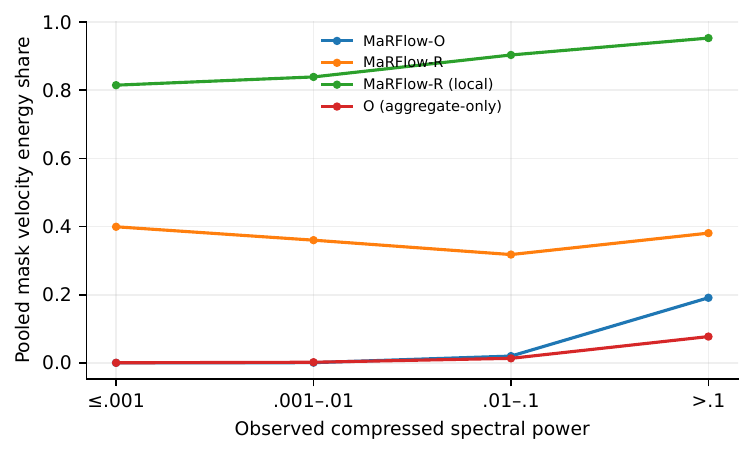}\end{minipage}\hfill
\begin{minipage}[t]{.485\linewidth}\centering\small DNS: power regions\\\includegraphics[width=\linewidth]{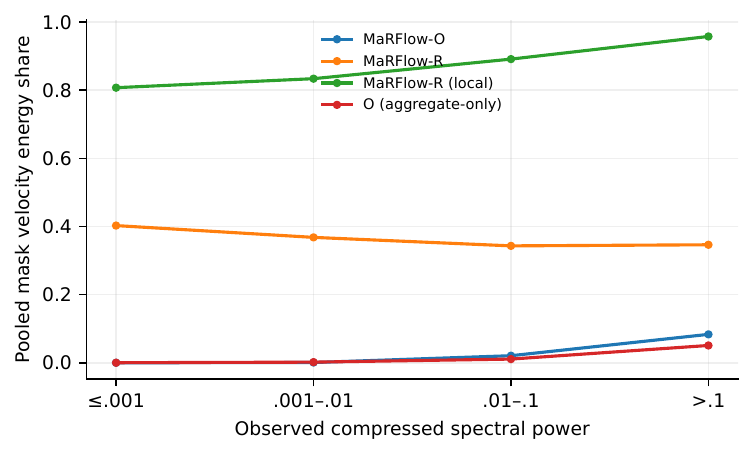}\end{minipage}
\caption{Pooled branch-energy diagnostics along fixed ten-NFE trajectories (top) and grouped by observed compressed spectral power (bottom). The multiplicative energy share is $E_M/(E_M+E_R)$, with $E_M=\sum|\Phi_t v_M|^2$ and $E_R=\sum|v_R|^2$. The bottom panels summarize power-region behavior across the trajectory. Shares exclude the branch cross-term and are not fractions of speech energy. Solver-step axes are zero-indexed.}\label{fig:branch_energy}
\end{figure}

\noindent\textbf{How does the multiplicative reference shape branch-energy allocation (cf. Fig.~\ref{fig:branch_energy})?} Both observation-mediated variants remain additive-dominant throughout inference: the pooled multiplicative energy share spans approximately 0.29--0.54\% for O-agg and 0.46--1.00\% for O across the two datasets. In contrast, both recursive variants exhibit early multiplicative dominance followed by increasing additive allocation. R-agg starts at 76.73\%/78.01\% and falls to 0.28\%/0.24\%, while R-local starts at 99.65\% on both datasets and falls to 8.63\%/9.37\% (VoiceBank/DNS). The qualitative distinction is therefore already present under aggregate-only supervision, suggesting a reference-dependent inductive bias rather than an effect that requires the oracles. Local supervision strengthens and sustains the multiplicative contribution in R. The O contribution is concentrated in higher observed-power bins, consistent with Eq.~\eqref{eq:allocation}. The R-local target instead depends on the power of $Y_t$, so observed-power stratification is a common diagnostic coordinate rather than its oracle's allocation variable. Branch energies omit the cross-term and should not be read as fractions of physical speech or noise.

\noindent\textbf{How do the two velocity contributions interact?} For $u_k=\Phi_{t_k}\odot v_{M,k}$ and $r_k=v_{R,k}$, let $E_M=\sum|u_k|^2$, $E_R=\sum|r_k|^2$ and $C=\operatorname{Re}\sum u_k^*r_k$, with sums over valid bins and utterances. The pooled normalized alignment $C/\sqrt{E_ME_R}$ and the combined energy $E_M+E_R+2C$ distinguish directional agreement from branch magnitude. Across the ten-step rollouts, alignment remains near zero in O and O-agg (absolute value below 0.035), but is positive in both recursive variants (Appendix Fig.~\ref{fig:alignment}). At the first update, R-agg has alignment 0.882/0.891, versus 0.633/0.632 for R-local (VoiceBank/DNS). The cross-term $2C/(E_M+E_R)$ accounts for 74.5/73.8\% of the sum of individual branch energies in R-agg, compared with 7.5/7.5\% in R-local and less than 0.5\% in magnitude for either O variant. The smaller initial cross-term in R-local reflects both alignment and its highly asymmetric energy allocation. Despite these differences, the pooled squared aggregate velocities of R-local and R-agg differ by at most 1.3\% across the first nine updates, relative to R-agg; at the final update R-local is 11.7/15.3\% higher. Thus the early change is principally one of branch allocation and interaction rather than aggregate energy scale. Similar aggregate energy does not imply identical vector fields, nor does near-zero pooled alignment establish binwise orthogonality or independence.

\begin{table}[t]
\centering\small
\caption{Inference-only branch suppression at five NFE, expressed as mean score drop (full minus intervened). Positive values indicate degradation. The checkpoint is unchanged; no model is retrained and both branches are evaluated with the modified state during the rollout. SI-SDR differences are in dB.}\label{tab:intervention}
\fitTable{\InterventionResults}
\end{table}
\noindent\textbf{Does energy dominance imply functional dominance (cf. Table~\ref{tab:intervention})?} In O-agg, suppressing the multiplicative and additive contributions decreases SI-SDR by 4.18/3.25 and 23.91/20.50\,dB, respectively (VoiceBank/DNS). In O, the corresponding drops are 7.39/2.79 and 27.21/24.63\,dB. Both observation-mediated variants thus rely primarily on additive correction. Despite its early multiplicative energy dominance, R-agg remains more dependent on additive correction: the drops are 10.81/5.38\,dB under multiplicative suppression and 22.81/19.62\,dB under additive suppression. R-local reverses this dependence, with drops of 26.06/23.41 and 2.77/2.95\,dB, respectively. Early energy dominance alone is therefore insufficient to establish functional dominance at the endpoint. These inference-only interventions change subsequent solver states and are not equivalent to training a single-branch architecture.

\noindent\textbf{What is induced by the reference, and what changes with oracle supervision?} Taken together, Fig.~\ref{fig:branch_energy} and Table~\ref{tab:intervention} distinguish two effects. First, the reference-dependent energy profiles emerge without explicit branch targets: fixed-observation modulation is energetically secondary in both O variants, whereas recursive state modulation dominates the early updates in both R variants. Second, explicit local supervision strengthens and sustains the recursive multiplicative contribution and is associated with a reversal of functional branch dependence relative to R-agg. The resulting R-local field relies primarily on state modulation, with complementary additive correction becoming increasingly prominent toward the endpoint; O remains functionally additive-dominant. The local oracle in Eq.~\eqref{eq:local_split} is consistent with the sustained multiplicative allocation because it assigns most of the instantaneous target velocity to modulation in high-state-power bins. It does not, however, uniquely explain the emergence of the qualitative energy profile, which also occurs without local supervision. These matched comparisons separate the observed reference effect from the additional effect of branch supervision within the studied configurations, without establishing universal allocation behavior or a physical partition into speech and noise.

\begin{table}[t]
\centering\small
\caption{Complementary branch contributions at five NFE, measured by the mean score difference between full inference and suppression of the indicated branch. Entries are VoiceBank/DNS. DNSMOS SIG, BAK and OVRL are predicted speech-signal, background-noise and overall quality, respectively. These are intervention effects along recomputed trajectories, not independently additive metric gains.}\label{tab:secondary_branch}
\begin{adjustbox}{max width=\linewidth}
\begin{tabular}{lrrrrr}
\toprule
Restored branch & PESQ-WB & ESTOI & SIG & BAK & OVRL\\
\midrule
O: multiplicative & 0.19/0.13 & 0.020/0.021 & 0.008/0.007 & 0.48/0.78 & 0.23/0.39\\
R-local: additive & 1.02/0.62 & 0.067/0.055 & $-0.016$/0.022 & 1.07/1.21 & 0.51/0.64\\
\bottomrule
\end{tabular}
\end{adjustbox}
\end{table}
\noindent\textbf{What do the complementary branches contribute
beyond SI-SDR (cf. Table~\ref{tab:secondary_branch})?}
Although O relies primarily on additive correction, restoring
its smaller multiplicative branch improves predicted
background-noise quality (BAK) by 0.48/0.78 on VoiceBank/DNS.
In R-local, restoring the additive branch improves PESQ
by 1.02/0.62 and ESTOI by 0.067/0.055, with BAK gains
of 1.07/1.21. In both cases, BAK gains substantially exceed
the changes in SIG. The corresponding pointwise 95\% paired
bootstrap intervals for BAK are [0.45, 0.52]/[0.68, 0.88]
for O and [1.05, 1.08]/[1.15, 1.27] for R-local.

The multiplicative-only intervention in R-local provides
a particularly informative contrast. Although its SI-SDR
remains substantially above that of the noisy input on
both datasets, its PESQ falls below the noisy baseline.
Restoring the complementary additive coordinate improves
both fidelity and perceptual quality scores, with particularly
large gains in PESQ and predicted background-noise quality.
Thus, functional dominance under a fidelity metric does
not establish the redundancy of the complementary coordinate.

These findings indicate that waveform fidelity, as measured
by SI-SDR, is more sensitive to suppression of the
multiplicative coordinate in R-local, whereas suppression
of its additive coordinate substantially affects perceptual
quality and intelligibility scores. The two coordinates
therefore exhibit complementary functional effects under
inference-time interventions, without necessarily
corresponding to distinct physical signal components.
In particular, the predicted BAK/SIG contrast does not
establish that one coordinate uniquely reconstructs speech
or removes physical noise.
\subsubsection{Recursive restoration: what happens to earlier updates?}\label{subsec:recursive_results}
\begin{figure}[t]
\centering
\begin{minipage}[t]{.485\linewidth}\centering\small VoiceBank: cumulative gain\\\includegraphics[width=\linewidth]{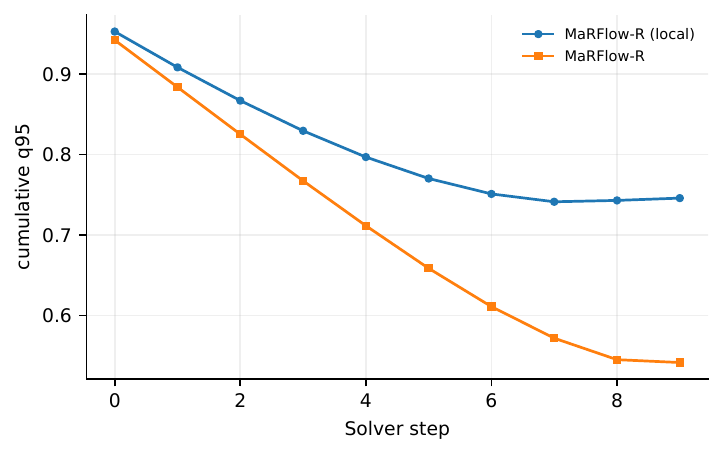}\end{minipage}\hfill
\begin{minipage}[t]{.485\linewidth}\centering\small DNS: cumulative gain\\\includegraphics[width=\linewidth]{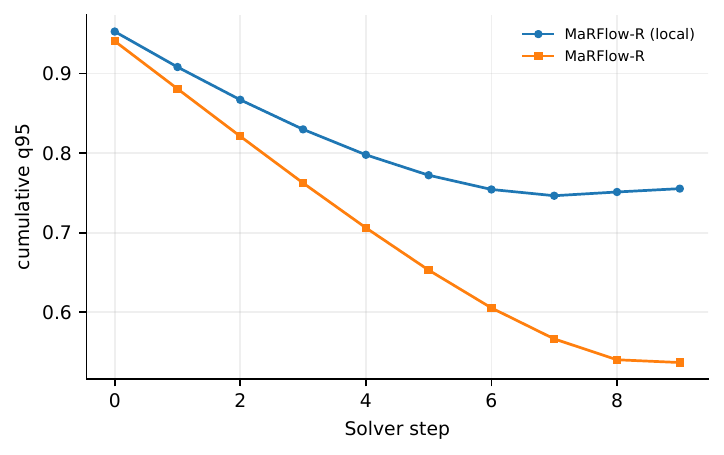}\end{minipage}\\[3pt]
\begin{minipage}[t]{.485\linewidth}\centering\small VoiceBank: additive-update propagation\\\includegraphics[width=\linewidth]{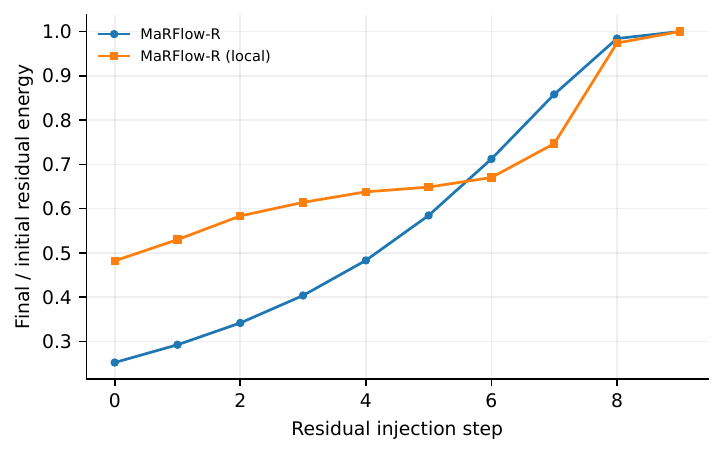}\end{minipage}\hfill
\begin{minipage}[t]{.485\linewidth}\centering\small DNS: additive-update propagation\\\includegraphics[width=\linewidth]{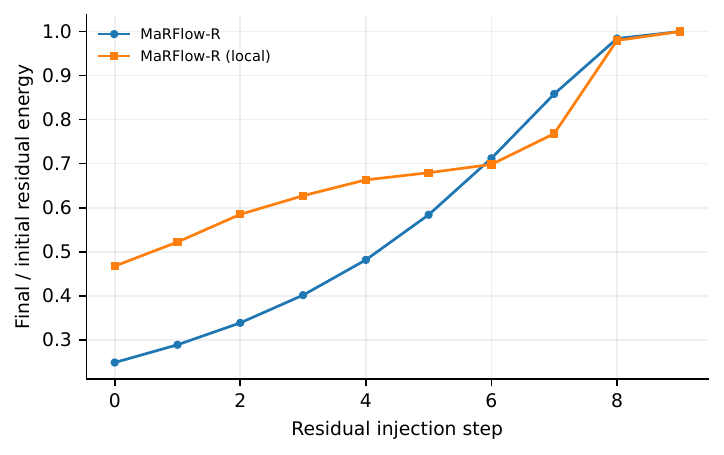}\end{minipage}
\caption{Recursive diagnostics for R-agg and R-local at ten NFE. Top: mean per-utterance 95th percentile of the magnitude of the cumulative complex gain product. Bottom: pooled final-to-initial energy ratio for each injected additive update after multiplication by subsequent gains along the realized trajectory (Eq.~\eqref{eq:recursive_expansion}). Steps are zero-indexed. The final injected update has no subsequent multiplicative modulation.}\label{fig:recursive}
\end{figure}
\begin{table}[t]
\centering\small
\caption{Final cumulative gain magnitude $|\prod_{k=0}^{9}G_k|$ along ten-NFE rollouts, reported as the mean of the per-utterance quantiles. A value below unity indicates attenuation of the corresponding initial-state contribution.}\label{tab:gain_quantiles}
\begin{tabular}{llrrrr}
\toprule
Dataset & Model & 5th & Median & 95th & 99th\\
\midrule
VoiceBank & R-agg & 0.367 & 0.427 & 0.541 & 0.587\\
 & R-local & 0.097 & 0.232 & 0.746 & 1.098\\
DNS & R-agg & 0.368 & 0.428 & 0.537 & 0.579\\
 & R-local & 0.100 & 0.238 & 0.755 & 1.108\\
\bottomrule
\end{tabular}
\end{table}
\noindent\textbf{Does local supervision merely reduce
attenuation, or induce more heterogeneous modulation?} Both recursive models multiply the evolving state, but the full distribution of $|\prod_k G_k|$ reveals different behavior (Table~\ref{tab:gain_quantiles}). Compared with R-agg, R-local has lower 5th-percentile and median cumulative gains, but larger 95th- and 99th-percentile gains; its mean per-utterance 99th percentile exceeds unity. Indeed, each of the 974 paired utterances exhibits both a lower median and a higher 95th percentile under R-local. The effect is therefore not uniformly weaker attenuation: local supervision is associated with stronger attenuation in the lower part of the gain distribution and greater preservation or amplification in its upper tail. This more heterogeneous spectral modulation is consistent across both datasets. These are complex-gain magnitude statistics in compressed spectral coordinates; they do not identify whether particular attenuated or preserved bins contain speech or noise.

\noindent\textbf{What happens to the earlier additive updates (cf. Fig.~\ref{fig:recursive})?} The first additive increment retains 48.2\%/46.7\% of its pooled energy under subsequent gains in R-local, compared with 25.2\%/24.8\% in R-agg. Crucially, R-local's first injected additive energy is only 2.45\%/2.60\% of R-agg's; its final propagated energy is 4.67\%/4.89\% of R-agg's. Thus R-local introduces much smaller early additive increments while retaining a larger fraction of their energy, rather than carrying more absolute additive energy to the endpoint. The final increment has no subsequent multiplication, giving a retention ratio of one by construction. These algebraic diagnostics describe contributions along each realized trajectory and do not establish that gain propagation alone causes the quality improvement.

Here, retention refers to the ratio of squared magnitudes
of an algebraically propagated compressed-coordinate
increment before and after subsequent multiplicative
modulation. It does not measure the fraction of physical
waveform energy retained, since inverse amplitude compression
is nonlinear and individual additive increments need not
correspond to independently meaningful waveform components.

\noindent\textbf{When does the benefit of local allocation become visible?} Despite the substantial difference in branch allocation from the first step, intermediate waveform SI-SDR remains relatively close early in the fixed ten-NFE rollouts (Appendix Fig.~\ref{fig:trajectory}). After six completed updates, the mean R-local advantage over R-agg is only 0.03/0.20\,dB on VoiceBank/DNS; after nine updates it grows to 0.38/1.77\,dB, and at the endpoint to 0.39/1.91\,dB. This is an analysis of the same ten-step trajectory, not a comparison of independent shorter solver budgets. The substantial late separation in endpoint fidelity, especially on DNS, coexists with different early branch allocations; it does not localize the causal effect of supervision to the late updates.

\noindent\textbf{What is the main takeaway?}
The multiplicative reference determines how restoration
contributions compose and is associated with distinct
branch-energy profiles even without branch targets.
Local supervision strengthens functional reliance on
recursive state modulation and is associated with more
heterogeneous cumulative gains and greater fractional
retention of substantially smaller early additive increments.
These observations characterize how the learned recursive
model exploits its structural capacity to remodulate earlier
corrections.

The complementary additive coordinate remains important
for perceptual quality and intelligibility, as established
through waveform-level evaluation under branch suppression.
Observation-mediated restoration exhibits a different,
additive-dominant organization with a functionally important
multiplicative complement. Together, these observations
support matching structured supervision to the algebra
of the restoration field. They do not establish that
individual coordinates recover distinct physical signal
components or that recursive propagation alone causes
the observed performance improvements.


\subsubsection{Zero-shot cross-domain stress test: reverberant DNS}\label{subsec:reverb}
\begin{table}[t]\centering\small
\caption{Zero-shot DNS 2020 with reverberation (150 utterances), additive-noise-trained checkpoints, five NFE, first draw. Intrusive metrics use the released reverberant reference. Best overall means in bold.}\label{tab:reverb}\fitTable{\ReverbResults}
\end{table}
\noindent\textbf{Does the advantage transfer to reverberant conditions (cf. Table~\ref{tab:reverb})?} Unlike the non-reverberant DNS evaluation, this setting introduces room-induced reverberation in addition to the cross-dataset shift in speakers and noise conditions. 
The evaluation target is the released reverberant clean
speech rather than its anechoic counterpart. Thus, the
task remains additive-noise suppression in a reverberant
environment, not joint denoising and dereverberation.
The desired restoration is the removal of additive
interference while retaining the reverberant target speech.
\textbf{Crucially, reverberation is an unseen corruption for these five additive-noise-trained checkpoints:} none of these checkpoints is trained with RIR-based reverberation augmentation or adapted to the reverberant DNS condition. This experiment therefore probes extrapolation beyond the additive-noise enhancement task on which all five models are trained. The transfer picture is mixed. R-local obtains the highest DNSMOS P.808 and OVRL (3.31 and 2.56), with O reaching 3.18/2.42 and FlowSE 3.22/2.32. However, all enhancement models substantially degrade intrusive metrics relative to the noisy input (PESQ 1.82, SI-SDR 9.19\,dB). O and R-local attain PESQ near 1.09 and SI-SDR of 2.46/2.01\,dB. R-agg gives the highest SI-SDR among enhanced outputs (3.73\,dB), but remains well below the noisy signal. The degradation across all five models points to a shared difficulty in transferring to unseen reverberant conditions, rather than a failure specific to multiplicative--additive velocity factorization. This is a stress-test regime in which predicted perceptual quality and fidelity diverge, rather than a successful transfer result in the same sense as non-reverberant DNS.

\noindent\textbf{What does paired analysis reveal
(cf. Appendix Table~\ref{tab:reverbpaired})?}
O and R-local improve OVRL over the noisy input by
1.03 [0.96, 1.09] and 1.17 [1.10, 1.24], respectively.
R-local additionally improves OVRL over O by
0.14 [0.11, 0.17] and over FlowSE by
0.24 [0.21, 0.28].
However, R-local's SI-SDR is 7.19\,dB below the noisy
input and 0.45\,dB below O, with the corresponding
intervals excluding zero.
These opposing trends highlight why a non-intrusive
quality predictor cannot substitute for reference fidelity.
As shown in Section~\ref{subsec:wer}, recognition accuracy also deteriorates
substantially in this regime. We therefore do not interpret
the DNSMOS improvements as evidence of successful
restoration of the target reverberant speech.

\subsubsection{Speech Enhancement Under Reverberant Conditions: Matched Training}
\label{subsec:reverb_trained}
The preceding stress test establishes a failure of zero-shot
transfer, but cannot determine whether the factorized restoration
fields remain effective when reverberation is represented in the
training distribution. We therefore train FlowSE, O and R-local
independently from scratch on the common reverberant-mixture
protocol described in Section~\ref{sec:experiments}. We retain
both original oracles, the 30-epoch budget, and five-NFE inference.
The held-out synthetic reverberant VoiceBank mixtures assess
within-domain generalization to new distortion recordings; the
unchanged reverberant DNS test set enables a cross-dataset test
and a direct descriptive contrast with the zero-shot results.
This experiment compares complete supervised formulations,
not the isolated reference and supervision effects established
by the five-model additive-noise study.

\begin{table}[t]\centering\small
\caption{Enhancement after reverberation-aware training, five NFE and first paired source draw. The VoiceBank mixtures use RIR-convolved clean speech and separately sampled additive noise, rather than the original VoiceBank+DEMAND mixtures. Both sets use reverberant speech references. Best unrounded means are bold.}\label{tab:reverb_trained}
\fitTable{\ReverbTrainedResults}
\end{table}
\noindent\textbf{Does exposure to reverberation change the fidelity--quality trade-off (cf. Tables~\ref{tab:reverb} and~\ref{tab:reverb_trained})?}
On the same reverberant DNS cohort, both supervised models now
improve all six mean acoustic metrics relative to noisy speech.
O attains 2.09/12.73 PESQ/SI-SDR, and R-local attains 2.16/12.44,
versus 1.66/9.67 for FlowSE and 1.82/9.19 for noisy input.
Relative to their additive-noise-trained counterparts, the
SI-SDR of O and R-local increases by approximately 10.27 and
10.43\,dB, respectively; FlowSE improves by approximately
6.52\,dB. Nevertheless, the reverberation-trained FlowSE still
reduces PESQ, STOI and ESTOI relative to noisy speech.
The corresponding DNSMOS OVRL scores decrease relative to
zero-shot enhancement for all three models (2.32 to 1.73,
2.42 to 1.90 and 2.56 to 1.97 for FlowSE/O/R-local), while
remaining above the noisy score of 1.39; P.808 similarly decreases
relative to zero-shot enhancement. Thus better fidelity and
recognition (Section~\ref{subsec:wer}) do not require higher predicted
perceptual scores than the failed zero-shot outputs. These
contrasts show the importance of training-distribution coverage
and the benefit of factorization within the new matched protocol;
they do not isolate the effect of RIR exposure from other changes
in the synthesized training mixtures.

\noindent\textbf{Which formulation benefits most from the matched reverberant setting?}
On synthetic reverberant VoiceBank, R-local leads all six acoustic
means, with 2.39 PESQ and 16.29\,dB SI-SDR, versus 2.16/15.41
for O and 1.84/13.63 for FlowSE. O and R-local improve SI-SDR over
noisy input by 7.86 and 8.74\,dB, and PESQ by 0.28 and 0.52.
FlowSE increases SI-SDR by 6.07\,dB but decreases PESQ, STOI and
ESTOI, as on reverberant DNS. On DNS, R-local leads PESQ, STOI,
ESTOI, P.808 and OVRL, whereas O has the highest mean SI-SDR.
R-local therefore shows a particularly strong advantage under reverberant training, while the higher DNS SI-SDR for O motivates retaining both reference choices.

\noindent\textbf{Are the improvements consistent across utterances (cf. Appendix Tables~\ref{tab:reverb_trained_pair_vb} and~\ref{tab:reverb_trained_pair_dns})?}
On synthetic VoiceBank, the paired SI-SDR gains over FlowSE are
1.79 [1.64, 1.94] and 2.66 [2.49, 2.84]\,dB for O and R-local;
on DNS they are 3.06 [2.69, 3.46] and 2.77 [2.34, 3.20]\,dB.
The corresponding PESQ gains are 0.32 [0.30, 0.34] and
0.55 [0.52, 0.58] on VoiceBank, and 0.43 [0.37, 0.48] and
0.50 [0.44, 0.56] on DNS. R-local exceeds O in synthetic VoiceBank SI-SDR
by 0.88 [0.78, 0.98]\,dB (86.5\% wins), but its DNS difference
is $-0.29$ [$-0.63$, 0.02]\,dB (56.0\% wins). The latter
interval includes zero despite the higher mean for O; the
majority win rate and negative mean illustrate that the sign
of a typical paired difference need not determine its mean.
R-local also improves PESQ over O by 0.23 [0.21, 0.25] and
0.075 [0.041, 0.109] on the two sets.
All 24 paired comparisons of either supervised model against
FlowSE have positive pointwise intervals; for R-local versus O,
11 of 12 intervals are positive, with DNS SI-SDR the exception
(Appendix Tables~\ref{tab:reverb_trained_pair_vb}--\ref{tab:reverb_trained_pair_dns}).
The formal paired tests support the same pattern: all 24 MaRFlow-versus-FlowSE differences remain significant after Holm correction ($p_{\rm Holm}\leq0.0018$), as do 11 of the 12 R-local-versus-O differences in favor of R-local. The sole exception is DNS SI-SDR ($p_{\rm Holm}=0.0744$), for which we do not establish superiority or equivalence. These results provide particularly strong evidence that the acoustic gains under reverberation-aware training are not confined to a small subset of utterances.

\noindent\textbf{Do the gains persist across acoustic conditions (cf. Appendix~\ref{app:reverb_trained})?}
Both MaRFlow variants improve PESQ and SI-SDR over FlowSE in
every SNR, RIR, noise and speaker stratum evaluated (where
applicable); R-local leads O in every synthetic VoiceBank
stratum for these two metrics. On DNS, R-local leads PESQ in all
four SNR ranges, but O leads SI-SDR in the three nonnegative-SNR
ranges. Detailed descriptive condition results and their
exceptions are in Appendix~\ref{app:reverb_trained}.
These intervals and subgroup means describe the evaluated
checkpoints and cohorts, not independent training seeds or
population-level speaker generalization.

\begin{table}[t]\centering\small
\caption{Inference-only branch suppression for reverberation-trained models at five NFE. Format as in Table~\ref{tab:intervention}; VB denotes synthetic reverberant VoiceBank. Positive values denote degradation.}\label{tab:reverb_trained_intervention}
\fitTable{\ReverbTrainedIntervention}
\end{table}
\noindent\textbf{Does the learned restoration organization survive the change in training distribution (cf. Appendix Fig.~\ref{fig:reverb_trained_branches} and Table~\ref{tab:reverb_trained_intervention})?}
The observation-mediated model remains additive-dominant over
all ten updates, with multiplicative shares of 0.77--1.17\%
and 0.54--0.85\% on synthetic VoiceBank/DNS. R-local starts at
99.64/99.63\%, remains above 90\% through the seventh evaluation,
and falls to 9.80/13.59\% at the final evaluation, with additive
energy becoming dominant at the penultimate evaluation on both
sets. This reproduces the qualitative distinction observed in
the additive-noise-trained models (Fig.~\ref{fig:branch_energy}),
including the late transition toward additive correction.
Functional dependence also persists: suppressing the multiplicative
versus additive contribution decreases SI-SDR by 5.78 versus
25.10\,dB on synthetic VoiceBank and 2.31 versus 22.70\,dB on
DNS for O. The corresponding R-local drops are 24.63 versus
3.12\,dB and 20.77 versus 2.18\,dB. These patterns establish
that the characteristic organization of the two supervised
fields is not confined to the original additive-noise training
regime. Since aggregate-only controls are not retrained here,
we do not separately attribute the persistence to the reference
or the oracle, nor infer gain propagation or oracle agreement
on reverberant trajectories from these measurements alone.

\begin{table}[t]\centering\small
\caption{Complementary branch effects after reverberation-aware training, in the format of Table~\ref{tab:secondary_branch}. Entries are synthetic VoiceBank/DNS.}\label{tab:reverb_trained_secondary}
\fitTable{\ReverbTrainedSecondary}
\end{table}
\noindent\textbf{Do the smaller branches retain their complementary function?}
The additive-only O outputs already improve SI-SDR over noisy
input (9.64/10.42 versus 7.55/9.19\,dB), consistent with the
large functional dependence on additive correction.
The multiplicative-only R-local outputs improve SI-SDR over
noisy input (13.17/10.26 versus 7.55/9.19\,dB) but reduce PESQ
(1.60/1.63 versus 1.88/1.82). Restoring the additive coordinate
improves PESQ by 0.80/0.53, ESTOI by 0.062/0.074 and BAK by
1.71/1.08. The smaller multiplicative branch of O improves
PESQ by 0.13/0.11 and BAK by 1.40/0.79.
Unlike the nearly unchanged SIG in the original non-reverberant
complementarity analysis, both complementary branches now also
improve predicted signal quality: SIG gains are 0.57/0.53 for O
and 0.63/0.74 for R-local. BAK gains still exceed SIG gains.
On DNS, O's small ESTOI improvement has an interval containing
zero, so the benefit of its smaller branch is not equally
resolved by every metric. This variation in the magnitude
of complementary effects coexists with stable functional
branch ordering under SI-SDR suppression. Full absolute
intervention scores and paired intervals are provided in
Appendix~\ref{app:reverb_trained}; these effects do not identify
physical signal components or establish that either branch
could be replaced by an independently trained estimator.

\noindent\textbf{Main takeaway.}
Taken together, the matched reverberant experiments show that the acoustic advantage of reference-specific velocity factorization is not restricted to additive-noise training. Both supervised formulations improve all six mean acoustic metrics over direct velocity prediction on both reverberant datasets, with particularly strong gains for R-local on synthetic reverberant VoiceBank. The diagnostic analyses further show that their distinct branch organization and functional complementarity persist under the second training distribution. This supports the design premise of matching velocity supervision to the multiplicative reference, while retaining the distinction between the two complete formulations.

\subsubsection{Linguistic recoverability under a frozen recognizer}\label{subsec:wer}
\begin{table}[t]\centering\small
\caption{Corpus WER (\%) at five NFE for additive-noise-trained models. Clean and noisy signals are reference anchors; bold indicates the lowest WER among enhancement models. DNS transcripts are manually corrected ASR references.}\label{tab:wer}\fitTable{\WERResults}
\end{table}
\noindent\textbf{Does factorization improve linguistic recoverability (cf. Table~\ref{tab:wer})?} Relative to FlowSE, both supervised variants reduce WER on VoiceBank (3.18/3.26\% versus 3.67\%) and non-reverberant DNS (6.69/7.16\% versus 8.78\%). The gains are particularly meaningful on DNS, where R-agg reaches 10.00\% and local supervision reduces this to 7.16\%. Nonetheless, the noisy inputs achieve 2.42\% and 4.33\%, respectively, and are better than any enhanced output. Thus the factorized fields improve ASR recoverability relative to direct velocity prediction, but do not improve recognition relative to unprocessed speech on these test sets. The clean anchors (1.39\% and 2.36\%) contextualize the recognizer's own errors.

\noindent\textbf{How does recognition behave in the reverberant stress test?} WER rises from 10.00\% on noisy input to 95.45\% for FlowSE, 94.18\% for O, and 105.00\% for R-local. R-agg has the lowest WER among enhanced outputs (83.54\%), but remains much worse than the noisy signal. WER can exceed 100\% because insertions add errors beyond the reference word count. The increased substitutions, deletions and insertions (Appendix~\ref{app:extra}) and poor intrusive scores indicate a severe failure of content recoverability under this transfer setting despite higher predicted perceptual quality. These results should not be conflated with perceptual improvement on the non-reverberant datasets.

\noindent\textbf{Are the recognition differences supported by paired analysis (cf. Appendix Table~\ref{tab:werpaired})?} The VoiceBank WER differences relative to FlowSE are small, with intervals containing zero. On non-reverberant DNS, O and R-local reduce WER by 2.09 and 1.62 percentage points relative to FlowSE, with intervals excluding zero. R-local also reduces WER by 2.83 points relative to R-agg. The contrast between O and R-local is smaller (0.47 points favoring O; interval includes zero). On reverberant DNS, both supervised models are substantially worse than noisy input, and R-local has 10.82 points higher mean WER than O, although the latter difference is not significant after Holm correction ($p_{\rm Holm}=0.3192$). These differences concern a fixed recognizer and do not identify the physical or linguistic cause of individual recognition errors. The results underscore that objective enhancement quality and ASR accuracy need not improve together.
\FloatBarrier

\begin{table}[t]\centering\small
\caption{Corpus WER (\%) after reverberation-aware training. Bold indicates the lowest WER among enhancement models.}\label{tab:reverb_trained_wer}
\fitTable{\ReverbTrainedWER}
\end{table}

\noindent\textbf{Is linguistic recoverability restored as well (cf. Table~\ref{tab:reverb_trained_wer} and Appendix Table~\ref{tab:reverb_trained_werpaired})?}
On reverberant DNS, WER falls from 95.45/94.18/105.00\% for
additive-noise-trained FlowSE/O/R-local to 26.28/18.27/18.80\%
after reverberation-aware training. Relative to the retrained
FlowSE, the signed WER differences are $-8.01$ [$-10.59$, $-5.50$]
and $-7.48$ [$-9.93$, $-5.15$] percentage points for O and
R-local, respectively. The R-local minus O
contrast is 0.52 [$-1.22$, 2.25] points, so the two supervised
models are not clearly distinguished by WER. Error counts show
large reductions in substitutions, deletions and insertions
relative to the zero-shot regime. On synthetic reverberant
VoiceBank, R-local has the lowest enhanced WER (20.22\%),
compared with 21.56\% for FlowSE and 22.74\% for O; all three
pairwise intervals include zero and none of the Holm-adjusted tests reject at 0.05. On DNS, both comparisons against FlowSE remain significant after correction ($p_{\rm Holm}=3.00\times10^{-4}$), whereas the comparison between O and R-local does not ($p_{\rm Holm}=1$). O's higher mean WER than
FlowSE coexists with higher acoustic scores, with insertions
contributing to the difference (Appendix~\ref{app:reverb_trained}).
Importantly, neither supervised model surpasses the noisy WER
on either set (12.97\% and 10.00\%). Reverberation-aware training
therefore substantially mitigates the zero-shot content-recovery
failure, but the residual recognition gap demonstrates that
improved acoustic metrics are insufficient to guarantee
linguistic recoverability.

\noindent\textbf{Could a larger training budget mitigate the remaining WER gap?}
All models are trained for only 30 epochs to isolate the relative effectiveness of the proposed parameterizations under a common optimization budget. The validation curves (Figs.~\ref{fig:validation} and~\ref{fig:validation_reverb}) show that the selected checkpoints occur at or close to the end of this budget. This motivates investigating whether longer training, especially on diverse reverberant mixtures, can improve linguistic recoverability. However, the validation criterion is SI-SDR rather than WER, and we do not measure recognition as a function of training duration. Thus, the remaining WER gap may reflect limited optimization or a mismatch between the CFM restoration objective and recognition, among other factors; a benefit from longer training remains an empirical question. The present results establish comparative effectiveness under matched 30-epoch training rather than the best achievable WER of either formulation.

\FloatBarrier

\section{Conclusion}\label{sec:conclusion}
We presented \textbf{MaRFlow}, a multiplicative--additive
factorization of the instantaneous conditional speech restoration
field, with two complementary formulations: observation-mediated
modulation of the noisy spectrum and recursive modulation of the
evolving state. The distinction manifests in their Euler updates:
fixed-reference increments accumulate additively, whereas recursive
gains directly reweight the initial state and previously introduced
additive increments. To supervise these distinct restoration
mechanisms, we derived an identity-regularized terminal-displacement
oracle for O and a local instantaneous-velocity oracle for R,
both preserving the original conditional path and its unrestricted
population-optimal aggregate velocity.

Under matched training, both supervised formulations improve
over FlowSE on VoiceBank+DEMAND and non-reverberant DNS.
Local supervision is particularly consequential for recursive
restoration, improving the aggregate-only recursive model by
2.00\,dB SI-SDR on DNS and yielding the strongest mean results
there at five NFE. Both formulations exhibit complementary
few-step restoration characteristics at comparable inference
cost, with only 258 additional parameters over FlowSE.

Our mechanistic analyses distinguish the structural inductive
bias introduced by the multiplicative reference from the
supervision-induced bias introduced by the oracle. Additive
dominance in O and early multiplicative dominance in R emerge
even under aggregate-only supervision. Local supervision
sustains recursive modulation, increases functional reliance
on the multiplicative coordinate, and induces more heterogeneous
cumulative gains with greater fractional retention of early
additive increments. Branch interventions further demonstrate
the importance of complementary additive correction in R-local
and multiplicative correction in O, establishing their functional
complementarity in spectral restoration.

Under matched reverberation-aware training, both supervised
formulations outperform FlowSE across all six acoustic metrics
on synthetic reverberant VoiceBank and reverberant DNS, with
paired bootstrap intervals and Holm-adjusted permutation tests
supporting all 24 improvements. On reverberant DNS, O and
R-local improve SI-SDR by 3.06 and 2.77\,dB and reduce WER
by 8.01 and 7.48 percentage points over FlowSE, respectively.
R-local achieves the strongest mean performance across all
six acoustic metrics on synthetic reverberant VoiceBank and
five on DNS. Furthermore, the characteristic branch organization
and functional complementarity of the two formulations persist
across training regimes.

The linguistic recoverability analysis demonstrates improvements
over FlowSE on DNS, while the remaining WER gap relative to
noisy speech motivates further investigation into linguistic
content-preserving supervision and extended training.
Collectively, our theoretical, empirical and mechanistic
findings support reference-specific velocity factorization
and supervision as complementary inductive biases for
conditional flow matching, providing a principled means of
organizing continuous restoration dynamics without altering
the underlying conditional probability path.

\section{Limitations and Clarifications}
\label{sec:limitations}

\paragraph{Scope of the theoretical results.}
The proposed factorizations preserve the original conditional
probability path and its unrestricted population-optimal
aggregate velocity under reference-specific supervision.
These properties establish compatibility with the underlying
CFM transport formulation, while the practical benefits of
structured velocity supervision are assessed empirically.
Similarly, the positive definiteness of the conditional risk
concerns the output coordinates rather than the neural
parameter-space optimization problem.

\paragraph{Interpretation of the velocity decomposition.}
The proposed oracles define regularized algebraic allocations
in compressed complex spectral coordinates rather than a
physical decomposition of the acoustic mixture. Consequently,
the individual velocity contributions and their energy
statistics are interpreted within the chosen spectral
representation. The oracle agreement, branch intervention
and trajectory analyses provide complementary insights into
the learned restoration mechanisms, with the intervention
results characterizing the functional dependence of the
trained models on their respective velocity coordinates.

\paragraph{Experimental scope and generalizability.}
Our experiments isolate the effects of reference choice and
structured supervision within a matched FlowSE setting,
using a common backbone, conditional path, training budget
and numerical solver. All models are trained for 30 epochs
using a single training seed; the paired statistical analyses
therefore characterize test-example variability conditional
on the trained models. The synthetic reverberant VoiceBank
evaluation additionally involves shared speakers, RIRs and
noise recordings. The reverberant experiments assess noise
suppression in reverberant environments using reverberant
clean references. Extending the proposed factorizations to
other backbones, conditional paths, spectral representations
and larger training budgets remains a direction for future
investigation.

\paragraph{Acoustic restoration and linguistic recoverability.}
While structured velocity supervision improves acoustic
restoration under matched training conditions, improvements
in acoustic metrics do not necessarily translate into
improved linguistic recoverability. DNSMOS provides learned
perceptual quality estimates, whereas WER indirectly measures
linguistic content recoverability under a fixed ASR model.
Their complementary nature motivates evaluating restoration
across multiple criteria. Given the observed validation
convergence characteristics, investigating longer training
budgets and linguistic content-preserving objectives
constitutes a natural extension of the present work.

\section{Broader Implications and Future Work}
\label{sec:future}

\paragraph{Generalization to conditional audio recovery.}
The proposed framework is not intrinsically restricted to
speech enhancement. Its central principle of structuring
conditional restoration through multiplicative modulation
and complementary additive correction motivates its
investigation in broader audio recovery tasks, including
target-speaker extraction and universal sound separation.

\paragraph{Training scale and linguistic content preservation.}
The consistent benefits of structured velocity supervision
under matched training budgets motivate its investigation
at larger training scales and across more diverse acoustic
conditions. Further investigation into the relationship
between acoustic restoration and linguistic content
recoverability constitutes an important direction for
future work.

\paragraph{Velocity allocation and transport dynamics.}
The present study demonstrates the significance of reference
choice and structured supervision in organizing conditional
restoration dynamics. Further investigation into the
relationship between velocity allocation, representation
structure and transport dynamics may inform the development
of more effective structured conditional flow matching
frameworks.

\section{Broader Impact}

Computationally efficient speech restoration has potential
applications in communication systems, assistive listening,
speech interfaces and audio accessibility. The ability to
obtain improved enhancement quality with few network function
evaluations, without appreciable additional inference cost
over the studied backbone, may be useful in applications
with constrained computational budgets.

However, perceptual improvement does not necessarily imply
faithful recovery of the underlying speech content.
Our recognition and zero-shot reverberant transfer experiments
demonstrate that improvements in predicted perceptual
quality can coexist with substantial deterioration in
reference fidelity and linguistic recoverability. Even under
matched reverberant training, where both supervised models
improve all six mean acoustic scores, linguistic recoverability remains
worse than the noisy input.
This distinction is particularly consequential in clinical,
forensic, legal and other sensitive applications, where
alterations to speech content may affect downstream
interpretation and decision-making.

Accordingly, enhanced recordings should not be treated
as verified reconstructions of the original acoustic
event. Deployment in consequential settings should include
appropriate validation under representative acoustic
conditions, preservation of the original recordings and
transparent disclosure of algorithmic restoration.

More broadly, speech restoration and source-conditioned
audio extraction may enable recovery of conversations
or acoustic events without the knowledge or consent of
the individuals involved. Applications of these technologies
should therefore incorporate appropriate privacy safeguards,
consent mechanisms and responsible data-handling practices.

These considerations motivate evaluating generative audio
restoration systems not only for perceptual quality and
computational efficiency, but also for content preservation,
robustness under distribution shift and suitability for
their intended deployment conditions.

\section{Declaration of LLM Usage}
A large language model was used to assist with manuscript organization, language editing, and presentation of experimental results.

\section{Acknowledgment.} \label{ack}
I thank my doctoral supervisor, Prof. Vipul Arora
(Department of Electrical Engineering, KU Leuven, Belgium;
\texttt{vipul.arora@kuleuven.be}) for providing access to
the GPU computing infrastructure at MADHAV Lab, IIT Kanpur, which was used to
conduct the experiments presented in this work.

\bibliographystyle{tmlr}
\bibliography{main}

@article{ephraim,
  title={Speech enhancement using a minimum-mean square error short-time spectral amplitude estimator},
  author={Ephraim, Yariv and Malah, David},
  journal={IEEE Transactions on acoustics, speech, and signal processing},
  volume={32},
  number={6},
  pages={1109--1121},
  year={1984},
  publisher={IEEE}
}

@inproceedings{erdogan,
  title={Phase-sensitive and recognition-boosted speech separation using deep recurrent neural networks},
  author={Erdogan, Hakan and Hershey, John R and Watanabe, Shinji and Le Roux, Jonathan},
  booktitle={2015 IEEE International Conference on Acoustics, Speech and Signal Processing (ICASSP)},
  pages={708--712},
  year={2015},
  organization={IEEE}
}

@article{williamson,
  title={Complex ratio masking for monaural speech separation},
  author={Williamson, Donald S and Wang, Yuxuan and Wang, DeLiang},
  journal={IEEE/ACM transactions on audio, speech, and language processing},
  volume={24},
  number={3},
  pages={483--492},
  year={2015},
  publisher={IEEE}
}

@article{wangchen,
  title={Supervised speech separation based on deep learning: An overview},
  author={Wang, DeLiang and Chen, Jitong},
  journal={IEEE/ACM transactions on audio, speech, and language processing},
  volume={26},
  number={10},
  pages={1702--1726},
  year={2018},
  publisher={IEEE}
}

@inproceedings{lee,
  title={Flowse: Flow matching-based speech enhancement},
  author={Lee, Seonggyu and Cheong, Sein and Han, Sangwook and Shin, Jong Won},
  booktitle={ICASSP 2025-2025 IEEE International Conference on Acoustics, Speech and Signal Processing (ICASSP)},
  pages={1--5},
  year={2025},
  organization={IEEE}
}

@article{lipman,
  title={Flow matching for generative modeling},
  author={Lipman, Yaron and Chen, Ricky TQ and Ben-Hamu, Heli and Nickel, Maximilian and Le, Matt},
  journal={arXiv preprint arXiv:2210.02747},
  year={2022}
}

@article{liu,
  title={Flow straight and fast: Learning to generate and transfer data with rectified flow},
  author={Liu, Xingchao and Gong, Chengyue and Liu, Qiang},
  journal={arXiv preprint arXiv:2209.03003},
  year={2022}
}

@article{richter,
  title={Speech enhancement and dereverberation with diffusion-based generative models},
  author={Richter, Julius and Welker, Simon and Lemercier, Jean-Marie and Lay, Bunlong and Gerkmann, Timo},
  journal={IEEE/ACM Transactions on Audio, Speech, and Language Processing},
  volume={31},
  pages={2351--2364},
  year={2023},
  publisher={IEEE}
}

@article{jukic,
  title={Schr$\backslash$" odinger Bridge for Generative Speech Enhancement},
  author={Juki{\'c}, Ante and Korostik, Roman and Balam, Jagadeesh and Ginsburg, Boris},
  journal={arXiv preprint arXiv:2407.16074},
  year={2024}
}

@article{cmgan,
  title={CMGAN: Conformer-based metric GAN for speech enhancement},
  author={Cao, Ruizhe and Abdulatif, Sherif and Yang, Bin},
  journal={arXiv preprint arXiv:2203.15149},
  year={2022}
}

@inproceedings{d2former,
  title={D2former: A fully complex dual-path dual-decoder conformer network using joint complex masking and complex spectral mapping for monaural speech enhancement},
  author={Zhao, Shengkui and Ma, Bin},
  booktitle={ICASSP 2023-2023 IEEE International Conference on Acoustics, Speech and Signal Processing (ICASSP)},
  pages={1--5},
  year={2023},
  organization={IEEE}
}

@article{cross,
  title={Flowing straighter with conditional flow matching for accurate speech enhancement},
  author={Cross, Mattias and Ragni, Anton},
  journal={arXiv preprint arXiv:2508.20584},
  year={2025}
}

@article{cascade,
  title={Speech Enhancement based on cascaded two flows},
  author={Lee, Seonggyu and Cheong, Sein and Han, Sangwook and Kim, Kihyuk and Shin, Jong Won},
  journal={arXiv preprint arXiv:2508.06842},
  year={2025}
}

@inproceedings{rddm,
  title={Residual denoising diffusion models},
  author={Liu, Jiawei and Wang, Qiang and Fan, Huijie and Wang, Yinong and Tang, Yandong and Qu, Liangqiong},
  booktitle={2024 IEEE/CVF Conference on Computer Vision and Pattern Recognition (CVPR)},
  pages={2773--2783},
  year={2024},
  organization={IEEE}
}

@misc{valentini,
author="Valentini-Botinhao, Cassia",
title="Noisy reverberant speech database for training speech enhancement algorithms and TTS models",
publisher="University of Edinburgh",
year="2017",
month="01",
DOI="10.7488/ds/2139",
URL="https://cir.nii.ac.jp/crid/1882836540917659264"
}

@article{reddy,
  title={The interspeech 2020 deep noise suppression challenge: Datasets, subjective testing framework, and challenge results},
  author={Reddy, Chandan KA and Gopal, Vishak and Cutler, Ross and Beyrami, Ebrahim and Cheng, Roger and Dubey, Harishchandra and Matusevych, Sergiy and Aichner, Robert and Aazami, Ashkan and Braun, Sebastian and others},
  journal={arXiv preprint arXiv:2005.13981},
  year={2020}
}

@inproceedings{pesq,
  title={Perceptual evaluation of speech quality (PESQ)-a new method for speech quality assessment of telephone networks and codecs},
  author={Rix, Antony W and Beerends, John G and Hollier, Michael P and Hekstra, Andries P},
  booktitle={2001 IEEE international conference on acoustics, speech, and signal processing. Proceedings (Cat. No. 01CH37221)},
  volume={2},
  pages={749--752},
  year={2001},
  organization={IEEE}
}

@inproceedings{stoi,
  title={A short-time objective intelligibility measure for time-frequency weighted noisy speech},
  author={Taal, Cees H and Hendriks, Richard C and Heusdens, Richard and Jensen, Jesper},
  booktitle={2010 IEEE international conference on acoustics, speech and signal processing},
  pages={4214--4217},
  year={2010},
  organization={IEEE}
}

@article{estoi,
  title={An algorithm for predicting the intelligibility of speech masked by modulated noise maskers},
  author={Jensen, Jesper and Taal, Cees H},
  journal={IEEE/ACM Transactions on Audio, Speech, and Language Processing},
  volume={24},
  number={11},
  pages={2009--2022},
  year={2016},
  publisher={IEEE}
}

@inproceedings{sisdr,
  title={SDR--half-baked or well done?},
  author={Le Roux, Jonathan and Wisdom, Scott and Erdogan, Hakan and Hershey, John R},
  booktitle={ICASSP 2019-2019 IEEE International Conference on Acoustics, Speech and Signal Processing (ICASSP)},
  pages={626--630},
  year={2019},
  organization={IEEE}
}

@inproceedings{dnsmos,
  title={DNSMOS: A non-intrusive perceptual objective speech quality metric to evaluate noise suppressors},
  author={Reddy, Chandan KA and Gopal, Vishak and Cutler, Ross},
  booktitle={ICASSP 2021-2021 IEEE International Conference on Acoustics, Speech and Signal Processing (ICASSP)},
  pages={6493--6497},
  year={2021},
  organization={IEEE}
}

@inproceedings{dnsmos835,
  title={DNSMOS P. 835: A non-intrusive perceptual objective speech quality metric to evaluate noise suppressors},
  author={Reddy, Chandan KA and Gopal, Vishak and Cutler, Ross},
  booktitle={ICASSP 2022-2022 IEEE international conference on acoustics, speech and signal processing (ICASSP)},
  pages={886--890},
  year={2022},
  organization={IEEE}
}

@article{song,
  title={Score-based generative modeling through stochastic differential equations},
  author={Song, Yang and Sohl-Dickstein, Jascha and Kingma, Diederik P and Kumar, Abhishek and Ermon, Stefano and Poole, Ben},
  journal={arXiv preprint arXiv:2011.13456},
  year={2020}
}

@article{
codecsep,
title={CodecSep:  Prompt-Driven Universal Sound Separation on Neural Audio Codec Latents},
author={Adhiraj Banerjee and Vipul Arora},
journal={Transactions on Machine Learning Research},
issn={2835-8856},
year={2026},
url={https://openreview.net/forum?id=r63GX9hKhC},
note={}
}

@article{convtasnet,
  title={Conv-tasnet: Surpassing ideal time--frequency magnitude masking for speech separation},
  author={Luo, Yi and Mesgarani, Nima},
  journal={IEEE/ACM transactions on audio, speech, and language processing},
  volume={27},
  number={8},
  pages={1256--1266},
  year={2019},
  publisher={IEEE}
}

@article{audiosep,
  title={Separate anything you describe},
  author={Liu, Xubo and Kong, Qiuqiang and Zhao, Yan and Liu, Haohe and Yuan, Yi and Liu, Yuzhuo and Xia, Rui and Wang, Yuxuan and Plumbley, Mark D and Wang, Wenwu},
  journal={IEEE Transactions on Audio, Speech and Language Processing},
  volume={33},
  pages={458--471},
  year={2024},
  publisher={IEEE}
}

@inproceedings{flowsep,
  title={Flowsep: Language-queried sound separation with rectified flow matching},
  author={Yuan, Yi and Liu, Xubo and Liu, Haohe and Plumbley, Mark D and Wang, Wenwu},
  booktitle={ICASSP 2025-2025 IEEE International Conference on Acoustics, Speech and Signal Processing (ICASSP)},
  pages={1--5},
  year={2025},
  organization={IEEE}
}

@inproceedings{
ryoo2026,
title={Compositional Flow Matching with Factored Velocity Fields},
author={Avery Hee-Woon Ryoo and Dane Malenfant and Matthew G Perich and Guillaume Lajoie},
booktitle={ICML 2026 Workshop on Structured Probabilistic Inference {\&} Generative Modeling},
year={2026},
url={https://openreview.net/forum?id=P1b6pr3632}
}

@article{chi2026,
  title={Disentangled representation learning via flow matching},
  author={Chi, Jinjin and Liu, Taoping and Yin, Mengtao and Li, Ximing and Jing, Yongcheng and Shen, Jialie and Rutkowski, Leszek and Tao, Dacheng},
  journal={arXiv preprint arXiv:2602.05214},
  year={2026}
}

@inproceedings{whisper,
  title={Robust speech recognition via large-scale weak supervision},
  author={Radford, Alec and Kim, Jong Wook and Xu, Tao and Brockman, Greg and McLeavey, Christine and Sutskever, Ilya},
  booktitle={International conference on machine learning},
  pages={28492--28518},
  year={2023},
  organization={PMLR}
}

@misc{dnstext,author={{DNS\_text contributors}},title={DNS Challenge test-set transcriptions},year={2026},note={Community transcript repository},url={https://github.com/Emrys365/DNS_text}}

@article{frse2026,
  title={FRSE: Flow-matching with residual-decoupled learning for speech enhancement},
  author={Cai, Zhisheng and Sun, Jiayi and Chen, Feilong and Leng, Yan and Sun, Miao and Ding, Biyun and Sun, Chengli},
  journal={Speech Communication},
  pages={103449},
  year={2026},
  publisher={Elsevier}
}

@article{flowsep2,
  title={FlowSep 2: Self-Supervised Flow Matching for Language-Queried Audio Source Separation},
  author={Yuan, Yi and Liu, Xubo and Liu, Haohe and Kang, Xiyuan and Plumbley, Mark D and Wang, Wenwu},
  journal={arXiv preprint arXiv:2608.22111},
  year={2026}
}

@article{hajiali2025,
  title={Improving progressive generation with decomposable flow matching},
  author={Haji-Ali, Moayed and Menapace, Willi and Skorokhodov, Ivan and Sahni, Arpit and Tulyakov, Sergey and Ordonez, Vicente and Siarohin, Aliaksandr},
  journal={Advances in Neural Information Processing Systems},
  volume={38},
  pages={163844--163885},
  year={2026}
}

@article{blockwise2025,
  title={Blockwise flow matching: Improving flow matching models for efficient high-quality generation},
  author={Park, Dogyun and Lee, Taehoon and Joo, Minseok and Kim, Hyunwoo J},
  journal={Advances in Neural Information Processing Systems},
  volume={38},
  pages={84751--84784},
  year={2026}
}

@article{lin2026arex,
  title={AREX: Affine-Residual Exponential Integrator for Few-Step Sampling in Flow Matching},
  author={Lin, Shizheng and Lim, Soon Hoe and Erichson, N Benjamin},
  journal={arXiv preprint arXiv:2610.03483},
  year={2026}
}
\appendix
\raggedbottom

\section{Derivation of the regularized oracle and allocation bounds}
\label{app:oracle}

For a compressed spectral bin, write
$c_X=|X|^2+\lambda>0$ and
$m_0=(SX^*+\lambda)/c_X$.
Completing the square gives
\begin{equation}
|mX-S|^2+\lambda|m-1|^2
=c_X|m-m_0|^2+\mathrm{const.}
\end{equation}
The unique constrained minimizer is
$M^\star=\Pi_{m_{\max}}(m_0)$, where
$\Pi_c(z)=z\min(1,c/|z|)$ for $z\ne0$
and $\Pi_c(0)=0$.

Subtracting the identity from the unclipped minimizer yields
\begin{equation}
D_{\mathrm{unc}}
=\frac{(S-X)X^*}{|X|^2+\lambda}.
\end{equation}
Since $2\sqrt{\lambda}|X|\leq |X|^2+\lambda$,
the unclipped correction satisfies
\begin{equation}
|D_{\mathrm{unc}}|
\leq\frac{|S-X|}{2\sqrt{\lambda}}.
\end{equation}
Projection additionally ensures
$|D^\star|\leq m_{\max}+1$.
Defining the additive complement after projection preserves
$XD^\star+R^\star=S-X$ exactly.
The unclipped solution yields the allocation in
Eq.~\eqref{eq:allocation}; clipping can change the directions
and relative magnitudes of the two contributions.

\paragraph{Illustrative relative distortion.}
Consider an illustrative relationship $X=aS$ between two
nonzero compressed-coordinate spectral bins, with
$a\in\C\setminus\{0\}$.
In the high-power unclipped regime,
$|X|^2\gg\lambda$, we obtain
\begin{equation}
D_{\mathrm{unc}}
=
\frac{|X|^2}{|X|^2+\lambda}(a^{-1}-1)
\approx a^{-1}-1.
\end{equation}
Thus, when clipping is inactive, the corresponding
multiplicative velocity target satisfies
$U_M\approx1-a^{-1}$, expressing a relative correction
that is approximately independent of the observation
magnitude in this regime.
This illustration does not assume a physical multiplicative
corruption model or linear mixing in compressed coordinates.


\section{Population compatibility and output-space conditioning}
\label{app:risk}

For a random complex array $U$ with finite second moment,
squared-error regression admits the conditional decomposition
\begin{align}
\E[\norm{v-U}_F^2\mid W]
&=\norm{v-\E[U\mid W]}_F^2
\nonumber\\
&\quad+
\E[\norm{U-\E[U\mid W]}_F^2\mid W].
\end{align}
The second term is independent of the prediction $v$.

For either reference-specific oracle, the branch targets
satisfy
\begin{equation}
\Phi_t\odot U_M+U_R=U_t.
\end{equation}
Since $\Phi_t$ is measurable given $W$, linearity of
conditional expectation gives
\begin{equation}
\E[U_t\mid W]
=
\Phi_t\odot\E[U_M\mid W]
+\E[U_R\mid W].
\end{equation}
Defining $\mu_M=\E[U_M\mid W]$,
$\mu_R=\E[U_R\mid W]$,
$e_M=v_M-\mu_M$, and $e_R=v_R-\mu_R$,
the prediction-dependent part of the conditional risk is
\begin{align}
\mathcal R_W
&=w_Y\norm{\Phi_t\odot e_M+e_R}_F^2
\nonumber\\
&\quad+\beta_M\norm{e_M}_F^2
+\beta_R\norm{e_R}_F^2.
\end{align}

For an individual complex bin, define
\begin{equation}
J=(\Phi_t,1),
\qquad
e=(e_M,e_R)^\top.
\end{equation}
The conditional quadratic risk is $e^{\mathrm H}He$, where
\begin{equation}
H=w_YJ^{\mathrm H}J+\operatorname{diag}(\beta_M,\beta_R).
\end{equation}
For $w_Y\geq0$ and $\beta_M,\beta_R>0$,
\begin{equation}
H\succeq\min(\beta_M,\beta_R)I\succ0.
\end{equation}
Its determinant is
\begin{equation}
\det H=
w_Y\beta_R|\Phi_t|^2
+\beta_M(w_Y+\beta_R)>0.
\end{equation}
Consequently, the unique unrestricted minimizer is
$(v_M,v_R)=(\mu_M,\mu_R)$, whose assembled velocity
equals the original population-optimal aggregate field.

With aggregate-only supervision, $\beta_M=\beta_R=0$
and $w_Y>0$, the risk retains the null direction
\begin{equation}
(e_M,e_R)=(\Delta,-\Phi_t\Delta),
\end{equation}
for arbitrary complex $\Delta$.
The auxiliary objectives therefore resolve the allocation
ambiguity relative to the prescribed conditional mean targets.

This result applies to any conditional path for which the
branch targets have finite second moments, satisfy exact
aggregate consistency and use a reference measurable given
the conditioning variables. It does not require linearity
of the underlying signal representation.
The positive-definiteness result concerns the output-space
conditional risk and does not establish positive definiteness
or favorable conditioning of the neural parameter-space
Hessian.

\section{Derivation of the recursive local oracle}
\label{app:local}

For an individual compressed spectral bin, let
$Y_t,U_t\in\C$ denote the conditional state and its
instantaneous target velocity, respectively.
Let $\gamma\in\C$ be a candidate complex gain rate and
$r\in\C$ its complementary additive spectral velocity.
The reference-specific allocation problem introduced
in Eq.~\eqref{eq:local_constrained} is
\begin{equation}
\min_{\gamma,r}
|r|^2+\lambda_{\mathrm{loc}}|\gamma|^2
\quad\text{subject to}\quad
Y_t\gamma+r=U_t,
\end{equation}
where $\lambda_{\mathrm{loc}}>0$.
Eliminating $r=U_t-Y_t\gamma$ gives
\begin{equation}
\min_{\gamma}
|Y_t\gamma-U_t|^2
+\lambda_{\mathrm{loc}}|\gamma|^2.
\end{equation}
Define
\begin{equation}
c_t=|Y_t|^2+\lambda_{\mathrm{loc}}>0,
\qquad
\gamma_0=\frac{U_tY_t^*}{c_t}.
\end{equation}
Completing the square yields
\begin{align}
&|Y_t\gamma-U_t|^2
+\lambda_{\mathrm{loc}}|\gamma|^2
\nonumber\\
&\qquad=c_t|\gamma-\gamma_0|^2+\mathrm{const.}
\end{align}
Since $c_t>0$, the unique minimizer is
$U_M^{\mathrm{loc}}=\gamma_0$.
The corresponding additive complement is
$U_R^{\mathrm{loc}}=U_t-Y_t\gamma_0$,
which gives the exact target consistency and
state-dependent allocation in
Eqs.~\eqref{eq:local_gain}--\eqref{eq:local_split}.

Furthermore, since
\begin{equation}
2\sqrt{\lambda_{\mathrm{loc}}}|Y_t|
\leq |Y_t|^2+\lambda_{\mathrm{loc}},
\end{equation}
the gain magnitude satisfies
\begin{equation}
|U_M^{\mathrm{loc}}|
=
\frac{|U_t||Y_t|}
{|Y_t|^2+\lambda_{\mathrm{loc}}}
\leq
\frac{|U_t|}{2\sqrt{\lambda_{\mathrm{loc}}}}.
\end{equation}
This bound controls the oracle gain target for a given
instantaneous velocity, rather than imposing a bound on
the predicted neural gain or cumulative Euler gain.

At a state null, $Y_t=0$, the regularized solution remains
well-defined:
\begin{equation}
U_M^{\mathrm{loc}}=0,
\qquad
U_R^{\mathrm{loc}}=U_t.
\end{equation}
Thus, the additive coordinate retains the ability to express
the entire target velocity when multiplicative correction
is unavailable. All allocations are defined in the modeled
compressed spectral coordinates and do not require a
physical decomposition of the acoustic mixture.

\section{Extended experimental results}\label{app:extended}

\subsection{Paired statistical testing}
\begin{table}[H]\centering\small
\caption{Paired differences at five NFE, draw zero. Differences are first minus second model; Win is the percentage of strictly positive differences. Intervals are pointwise 95\% percentile bootstrap intervals. We report two-sided paired sign-flip $p$-values and Holm-adjusted $p$-values across all 28 tests in this table.}\label{tab:paired}\fitTable{\PairedResults}
\end{table}
\subsection{Additional oracle agreement diagnostics}
\begin{figure}[H]
\centering
\begin{minipage}[t]{.485\linewidth}\centering\small VoiceBank+DEMAND\\\includegraphics[width=\linewidth]{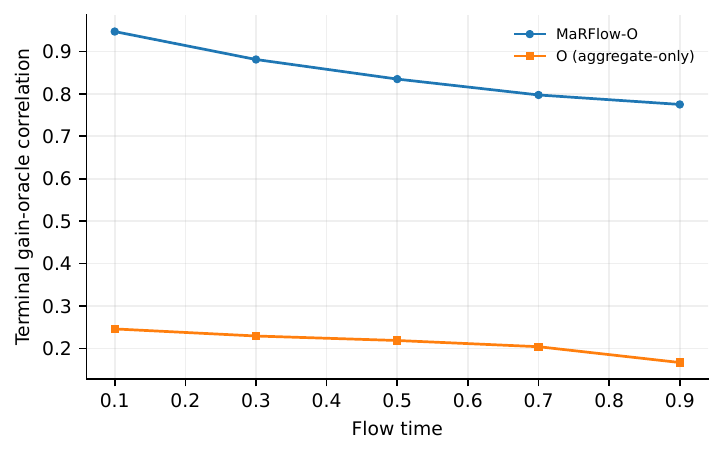}\end{minipage}\hfill
\begin{minipage}[t]{.485\linewidth}\centering\small DNS 2020\\\includegraphics[width=\linewidth]{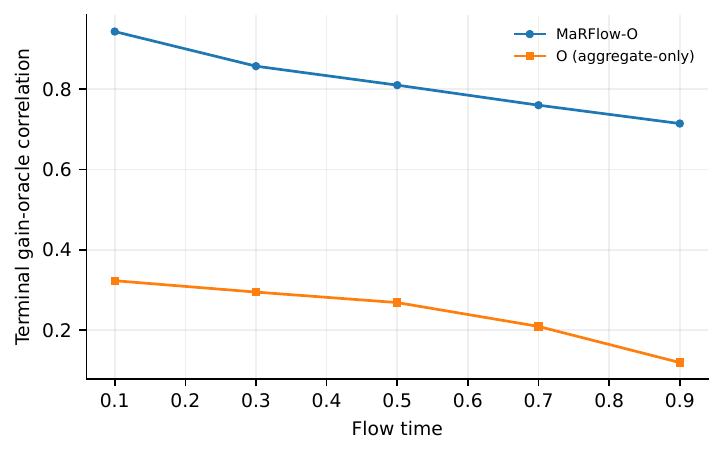}\end{minipage}
\caption{Mean per-utterance centered gain--oracle correlation on sampled held-out conditional paths. Flow time is in the reverse-time convention ($t=1$ source, $t=0$ target). Both O variants are compared against the same terminal-oracle target. These are not correlations along model-generated rollouts.}\label{fig:oracle}
\end{figure}
\begin{figure}[H]\centering
\begin{minipage}[t]{.485\linewidth}\centering\small VoiceBank+DEMAND\\\includegraphics[width=\linewidth]{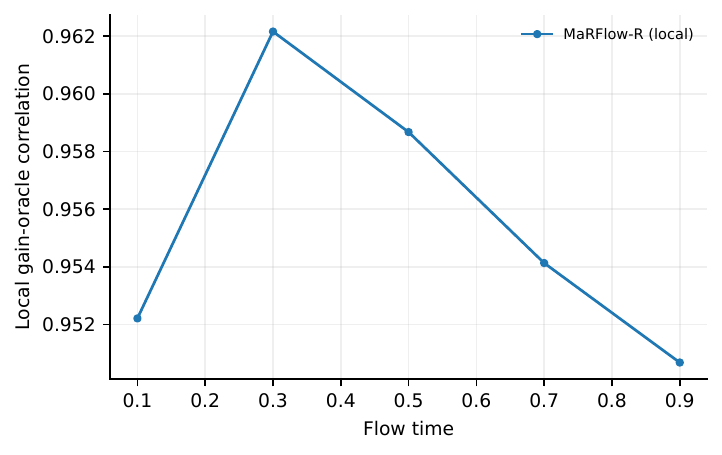}\end{minipage}\hfill
\begin{minipage}[t]{.485\linewidth}\centering\small DNS 2020\\\includegraphics[width=\linewidth]{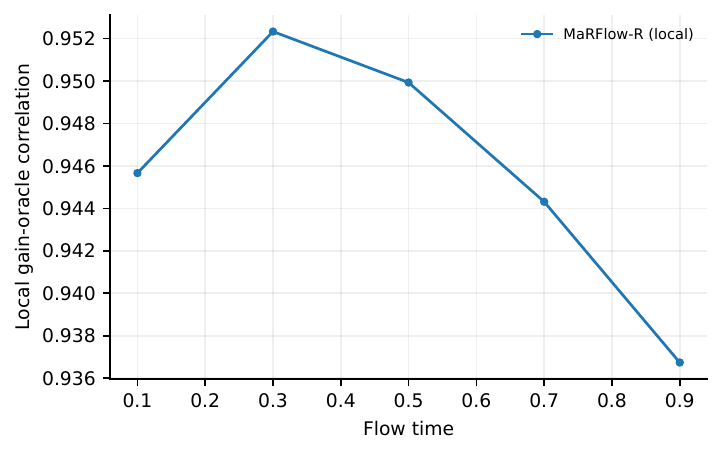}\end{minipage}
\caption{Centered gain--oracle correlation for R-local, evaluated on held-out conditional states at five flow times. The target is the local total-velocity allocation, distinct from the terminal allocation in Fig.~\ref{fig:oracle}.}\label{fig:localoracle}
\end{figure}
\subsection{Additional training and inference results}

\begin{figure}[H]
\centering\includegraphics[width=.63\linewidth,height=1.9in,keepaspectratio]{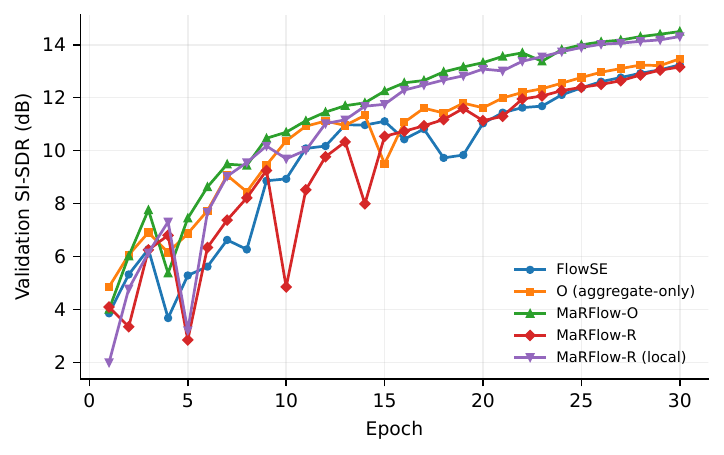}
\caption{Recorded validation SI-SDR by training epoch. All five best recorded checkpoints occur at epoch 30, the end of the matched training budget.}\label{fig:validation}
\end{figure}
\paragraph{Training curves.} The final validation SI-SDR values for FlowSE, O-agg, O, R-agg and R-local are 13.24, 13.47, 14.50, 13.16 and 14.31\,dB (Fig.~\ref{fig:validation}). Both supervised variants improve over their corresponding aggregate-only controls under the matched budget.

\begin{table}[H]\centering\small
\caption{Complete endpoint PESQ-WB sweep over NFE, first inference draw.}\label{tab:nfe}\fitTable{\NFEPESQ}\end{table}
\begin{table}[H]\centering\small
\caption{Complete endpoint SI-SDR (dB) sweep over NFE, first inference draw.}\label{tab:nfesdr}\fitTable{\NFESDR}\end{table}
\begin{table}[H]\centering\small
\caption{Complete endpoint STOI sweep over NFE, first inference draw.}\label{tab:nfestoi}\fitTable{\NFESTOI}\end{table}
\begin{table}[H]\centering\small
\caption{Complete endpoint DNSMOS OVRL sweep over NFE, first inference draw.}\label{tab:nfemos}\fitTable{\NFEMOS}\end{table}
\paragraph{Sensitivity to inference draws.} Across the three paired Gaussian source draws at five NFE, O improves SI-SDR over FlowSE by 0.445/0.454/0.431\,dB on VoiceBank and 1.360/1.393/1.424\,dB on non-reverberant DNS. The corresponding gains for R-local are 0.430/0.431/0.419 and 1.478/1.556/1.579\,dB. The R-local PESQ gains are 0.199/0.197/0.194 and 0.306/0.296/0.308, respectively. The benefit of both supervised fields persists across the sampled source perturbations.

\paragraph{Endpoint sweeps.} Tables~\ref{tab:nfe}--\ref{tab:nfemos} give the absolute means at the five evaluated inference budgets. These are not the successive intermediate states of a single integration. The test grid is descriptive; it should not be treated as a validation-based selection procedure.

\newcommand{\pairplots}[4]{\begin{minipage}[t]{.485\linewidth}\centering\small #1\\\includegraphics[width=\linewidth,height=1.45in,keepaspectratio]{#2.pdf}\end{minipage}\hfill\begin{minipage}[t]{.485\linewidth}\centering\small #3\\\includegraphics[width=\linewidth,height=1.45in,keepaspectratio]{#4.pdf}\end{minipage}}
\begin{figure}[H]\centering
\pairplots{VoiceBank: STOI}{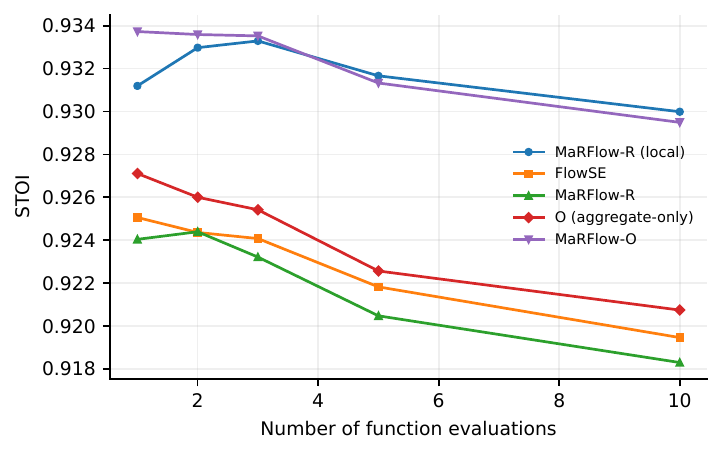}{DNS: STOI}{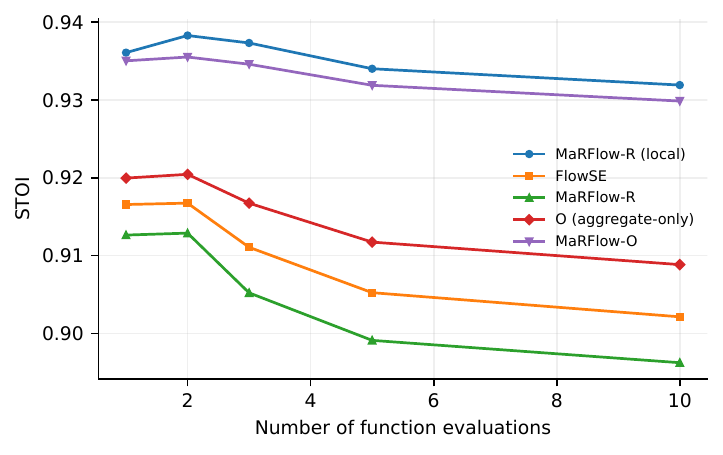}\\[3pt]
\pairplots{VoiceBank: DNSMOS OVRL}{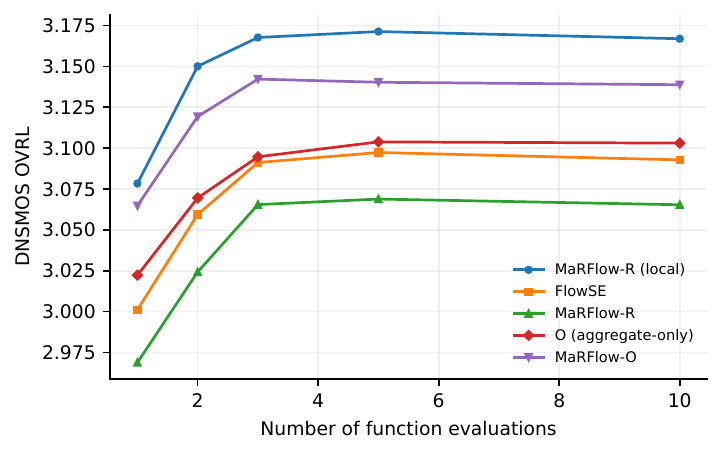}{DNS: DNSMOS OVRL}{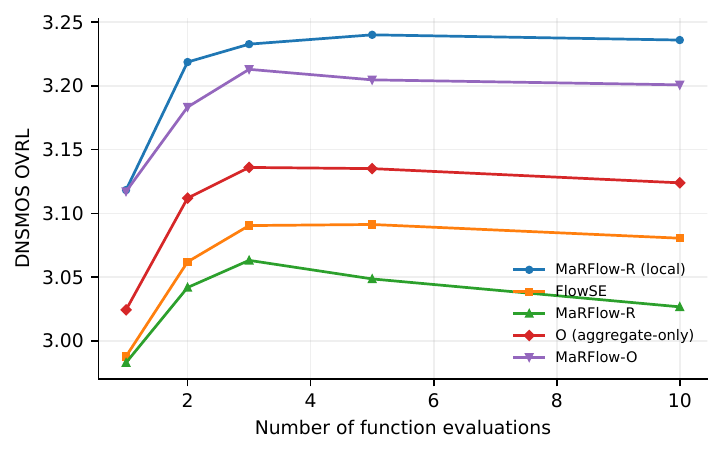}
\caption{Additional endpoint metrics as a function of NFE. DNSMOS OVRL may favor more evaluations than PESQ or SI-SDR.}\label{fig:extra_nfe}
\end{figure}
\begin{figure}[H]\centering
\pairplots{VoiceBank: PESQ}{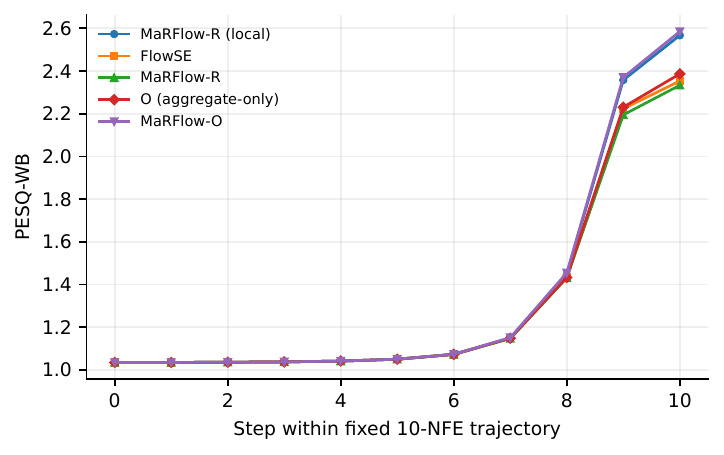}{DNS: PESQ}{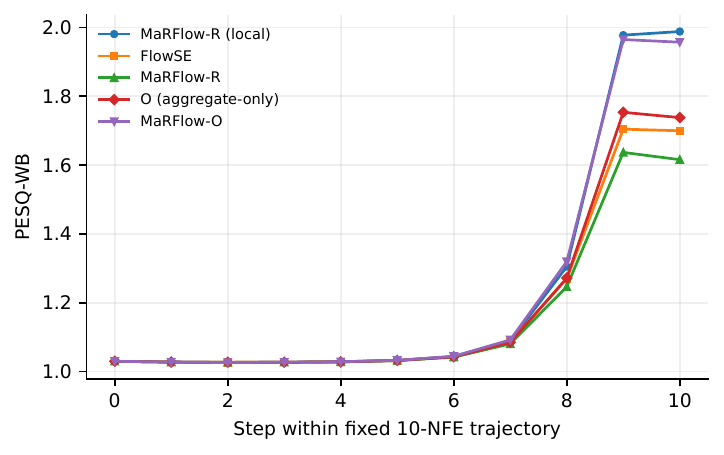}\\[3pt]
\pairplots{VoiceBank: SI-SDR}{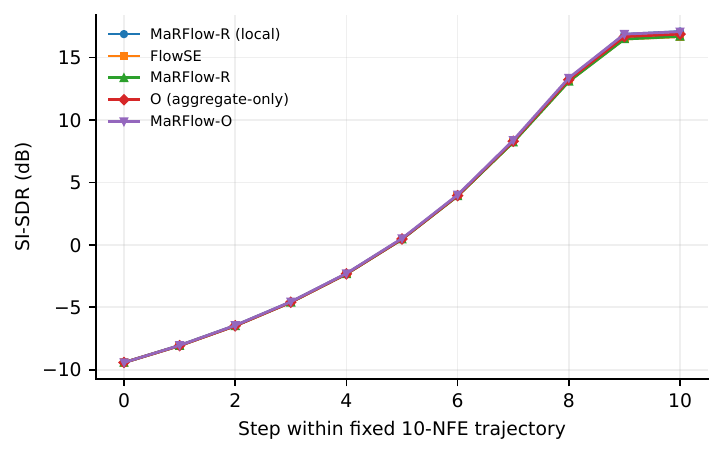}{DNS: SI-SDR}{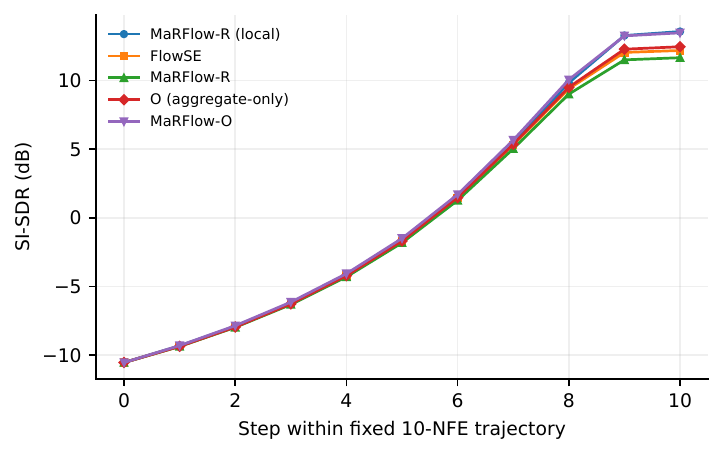}
\caption{Intermediate waveform scores within fixed ten-NFE trajectories. Step zero is the Gaussian-perturbed source, not the unperturbed noisy waveform. These trajectories are distinct from the endpoint NFE sweep.}\label{fig:trajectory}
\end{figure}
\begin{figure}[H]\centering
\pairplots{VoiceBank: PESQ}{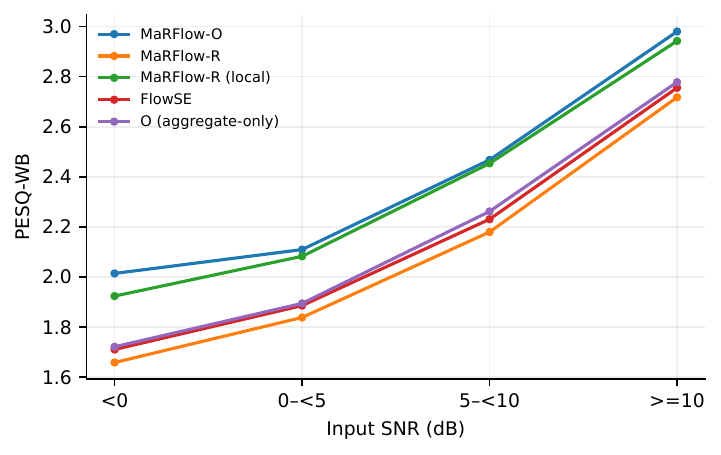}{DNS: PESQ}{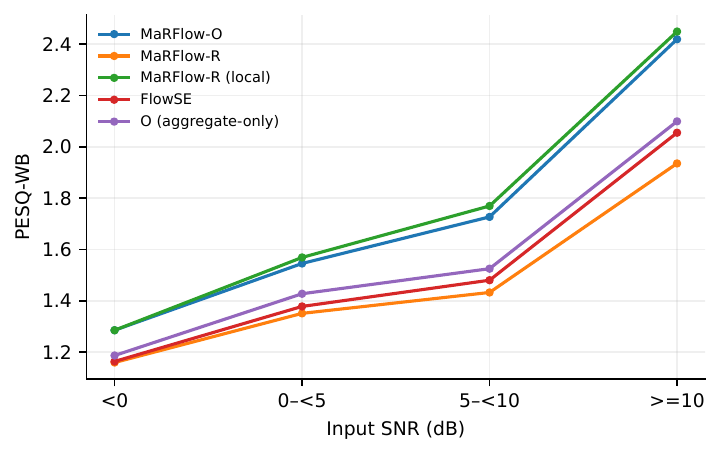}\\[3pt]
\pairplots{VoiceBank: SI-SDR}{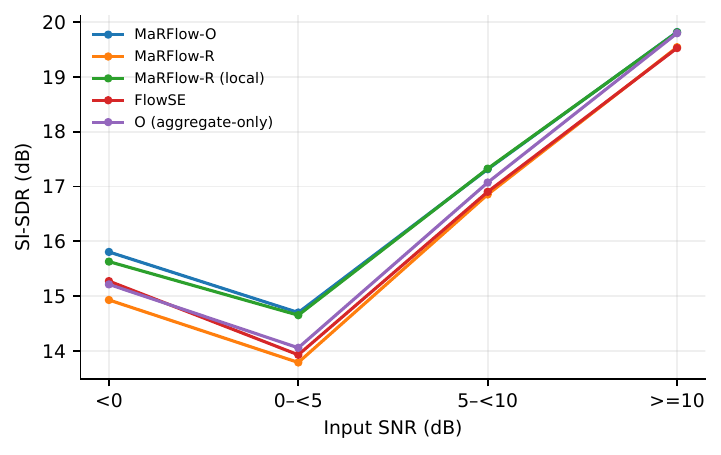}{DNS: SI-SDR}{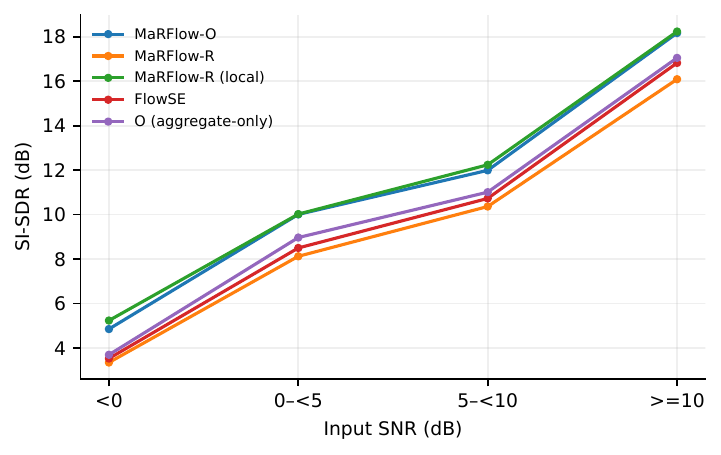}
\caption{Descriptive five-NFE mean scores grouped by input SNR. Groups have unequal sample counts; the plotted means do not establish statistical differences between SNR strata.}\label{fig:snr}
\end{figure}
\paragraph{Trajectory and SNR analyses.} Figure~\ref{fig:trajectory} reports fixed-budget intermediate quality, whereas Fig.~\ref{fig:snr} stratifies the primary endpoints by input SNR. The latter indicates how the observed performance varies with input difficulty; the DNS group below 0\,dB contains only five samples, so we avoid subgroup-level robustness claims.

\begin{figure}[H]\centering
\pairplots{VoiceBank: PESQ}{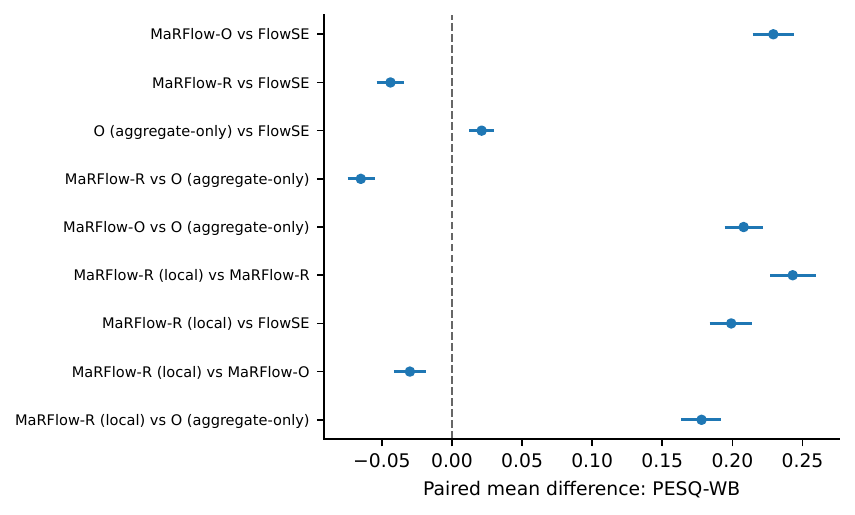}{DNS: PESQ}{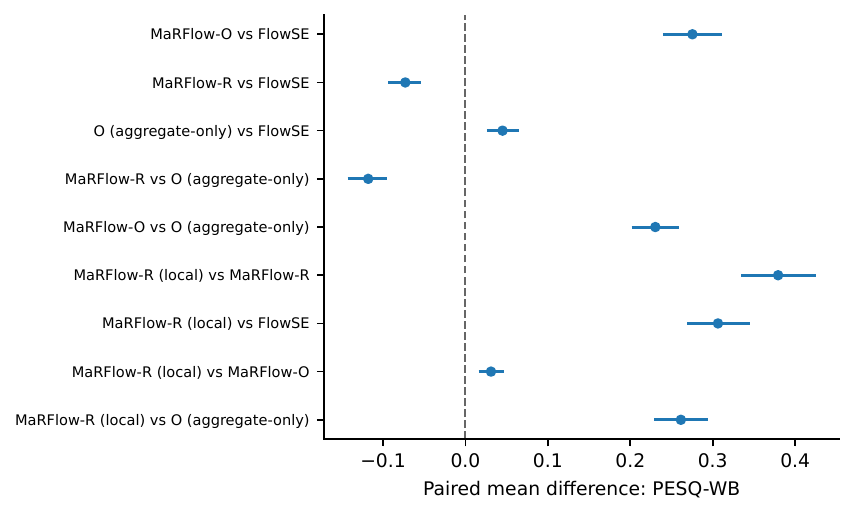}\\[3pt]
\pairplots{VoiceBank: SI-SDR}{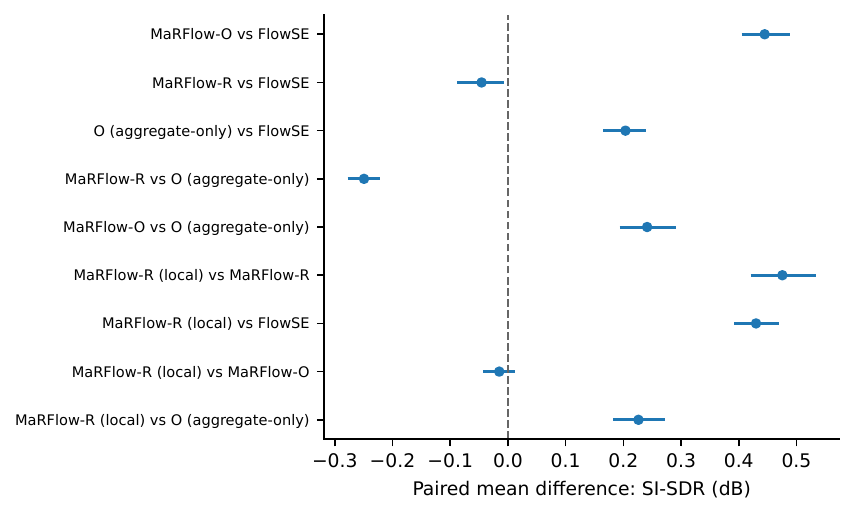}{DNS: SI-SDR}{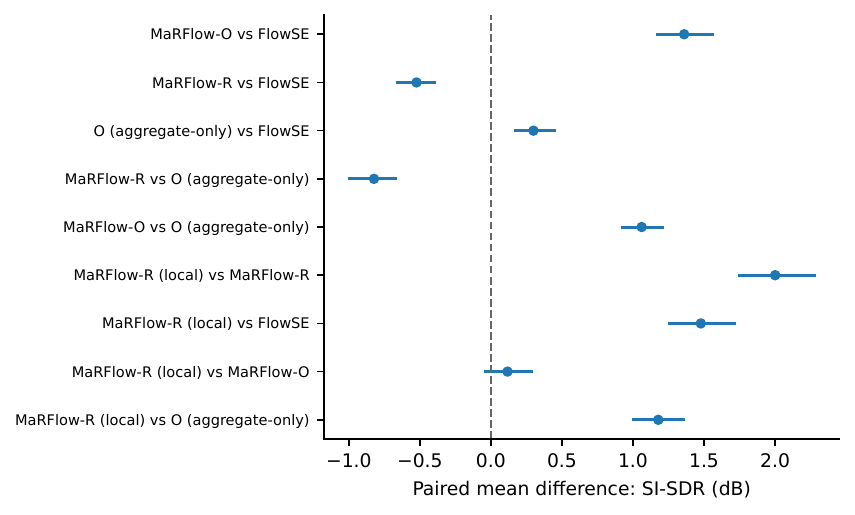}
\caption{Paired mean differences and pointwise bootstrap intervals for the primary comparisons (five NFE, first draw). The plots complement Table~\ref{tab:paired}; the intervals are not training-seed uncertainty.}\label{fig:paired}
\end{figure}
\begin{figure}[H]\centering
\pairplots{VoiceBank: oracle energy share}{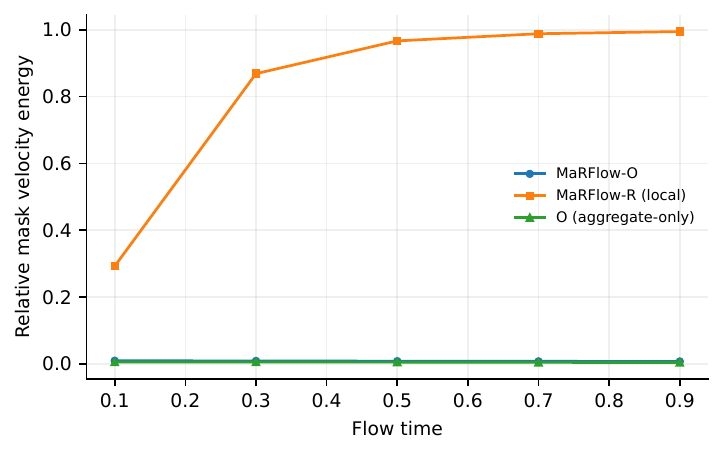}{DNS: oracle energy share}{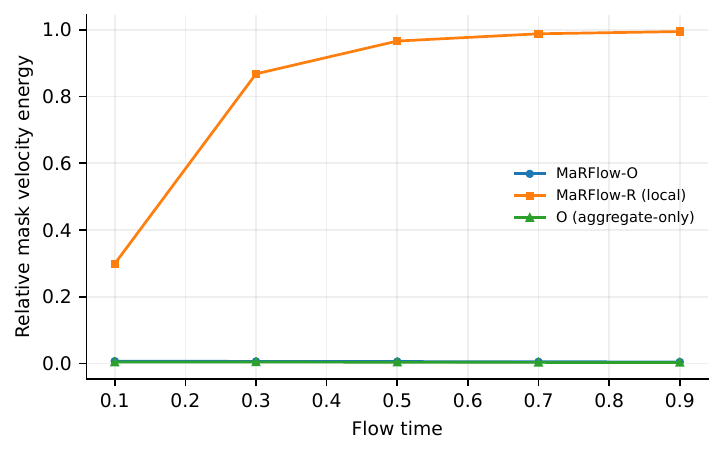}\\[3pt]
\pairplots{VoiceBank: oracle-path power regions}{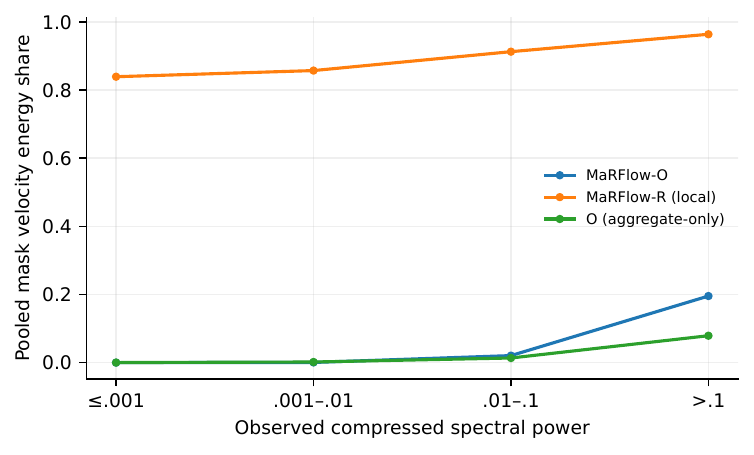}{DNS: oracle-path power regions}{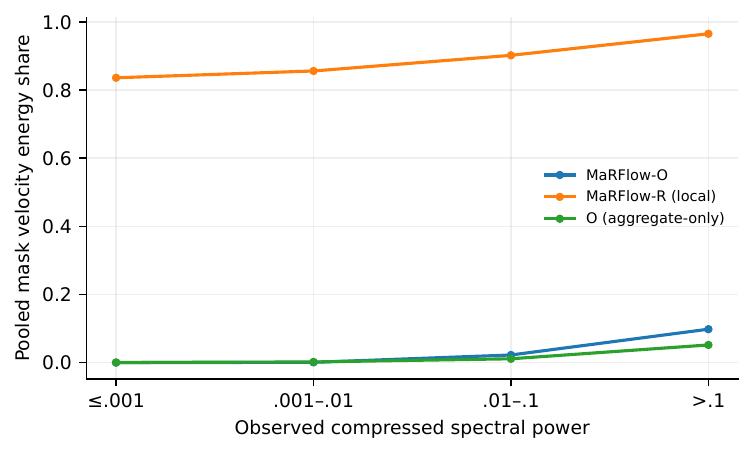}
\caption{Branch-energy allocation on sampled conditional paths, separate from the inference trajectories in Fig.~\ref{fig:branch_energy}. The power-dependent oracle allocation is evaluated in compressed spectral coordinates.}\label{fig:oracle_share}
\end{figure}
\begin{figure}[H]\centering
\pairplots{VoiceBank: branch alignment}{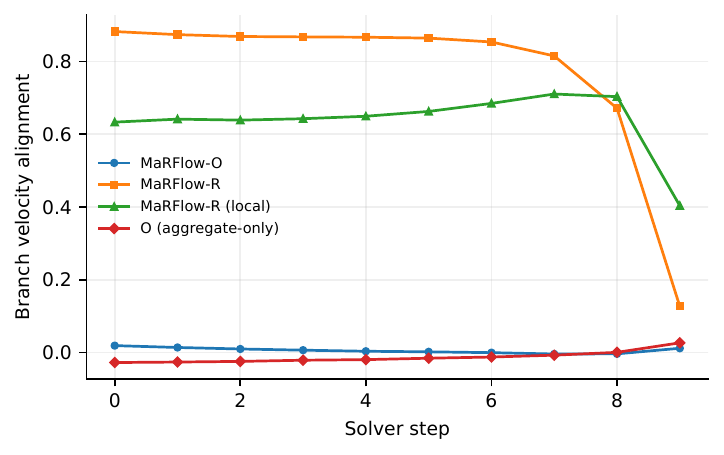}{DNS: branch alignment}{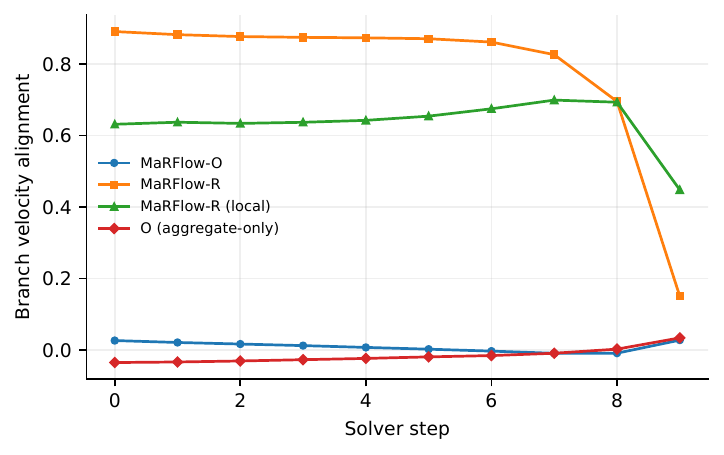}\\[3pt]
\pairplots{VoiceBank: conditional-path alignment}{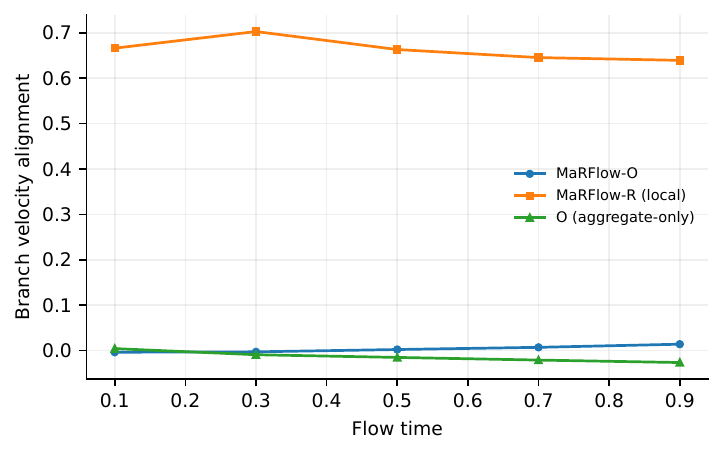}{DNS: conditional-path alignment}{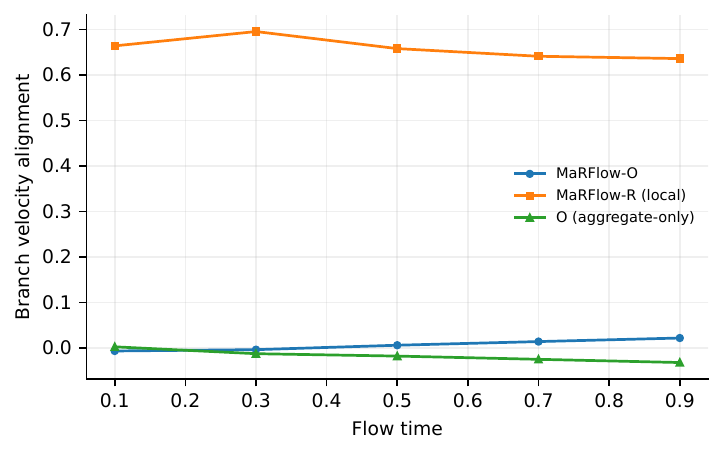}
\caption{Pooled normalized real inner product between the multiplicative and additive spectral velocities, on model rollouts (top) and conditional-path samples (bottom). Nonzero alignment indicates that branch energies alone do not add to the squared combined velocity.}\label{fig:alignment}
\end{figure}
\paragraph{Additional decomposition diagnostics.} Figures~\ref{fig:oracle_share} and~\ref{fig:alignment} distinguish sampled conditional-path measurements from actual ODE rollout measurements. For $u=\Phi_t\odot v_M$ and $r=v_R$, the combined energy is $\norm{u+r}_F^2=\norm{u}_F^2+\norm{r}_F^2+2\operatorname{Re}\langle u,r\rangle_F$. Consequently, the reported branch shares are descriptive measures of relative numerical energy, not an additive partition of recovered speech or a causal measure of the two routes' importance.
\FloatBarrier

\section{Additional zero-shot reverberant and linguistic results}\label{app:extra}

\subsection{Paired statistical testing}
\begin{table}[H]\centering\small
\caption{Paired acoustic differences on reverberant DNS. For each comparison, the first row gives the mean difference and pointwise 95\% bootstrap interval (first minus second; higher favors the first model); the second gives the two-sided paired sign-flip $p$-value and Holm-adjusted $p$-value in brackets (18-test family).}\label{tab:reverbpaired}\fitTable{\ReverbPaired}
\end{table}
\begin{table}[H]\centering\small
\caption{Paired differences in corpus WER (percentage points), with pointwise 95\% bootstrap intervals. Negative differences favor the first model. The second row of each comparison reports the two-sided paired sign-flip $p$-value and Holm-adjusted $p$-value in brackets (18-test family).}\label{tab:werpaired}\fitTable{\WERPaired}
\end{table}
\subsection{Additional acoustic and recognition results}

\begin{figure}[H]\centering
\pairplots{DNS reverb: SI-SDR}{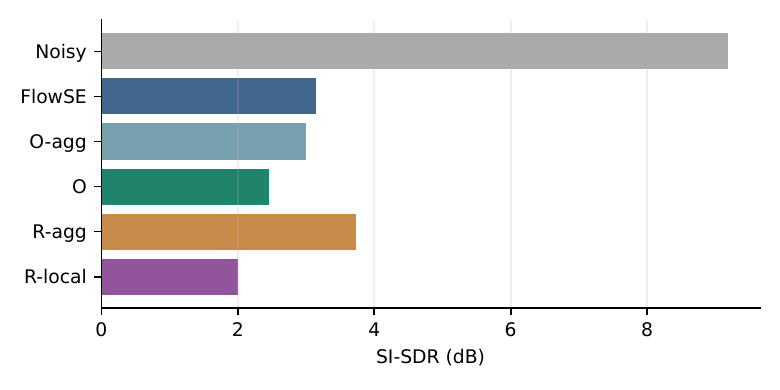}{DNS reverb: OVRL}{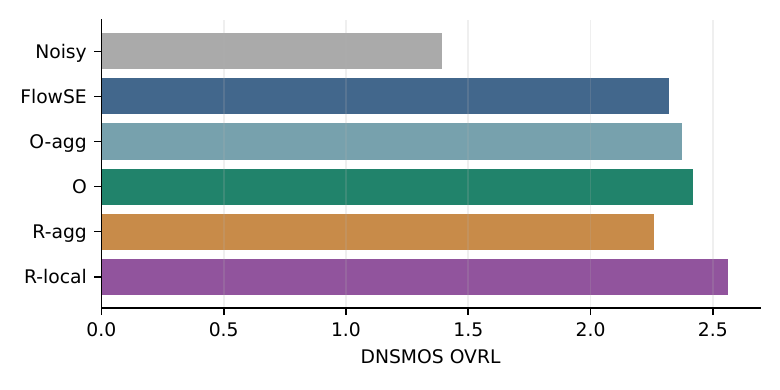}
\caption{Zero-shot reverberant DNS at five NFE for additive-noise-trained checkpoints. The two metrics reveal opposing fidelity and predicted perceptual quality trends.}\label{fig:reverbextra}
\end{figure}
\begin{figure}[H]\centering
\pairplots{VoiceBank: WER}{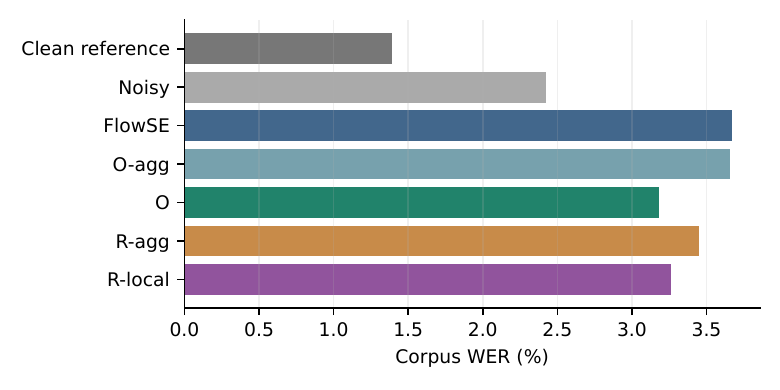}{DNS non-reverb: WER}{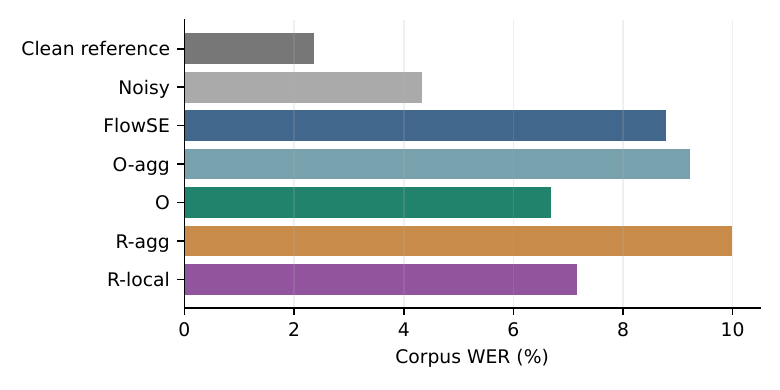}\\[3pt]
\includegraphics[width=.48\linewidth,height=1.6in,keepaspectratio]{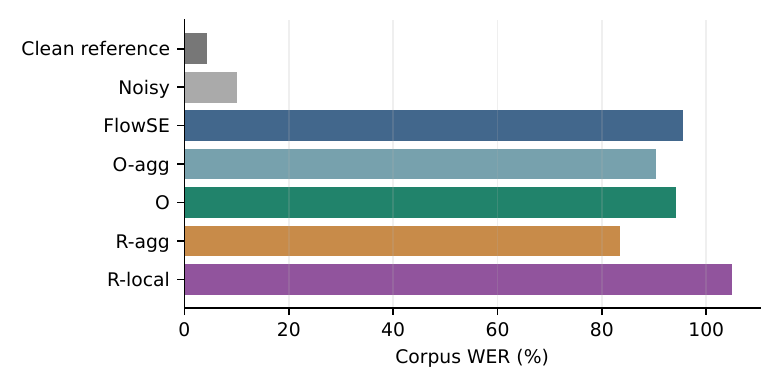}
\caption{Corpus WER (\%) for the clean and noisy anchors and all five enhancement models. The reverberant condition uses a different horizontal scale.}\label{fig:werextra}
\end{figure}
\begin{table}[H]\centering\small
\caption{Corpus-normalized substitution, deletion and insertion rates (\%) and total WER. All three error types use the same total reference-word denominator.}\label{tab:werbreak}\fitTable{\WERBreakdown}
\end{table}
\FloatBarrier

\section{Extended reverberation-trained results}
\label{app:reverb_trained}

\subsection{Paired statistical testing}
\begin{table}[H]\centering\small
\caption{Paired statistical analysis on \textbf{synthetic reverberant VoiceBank} ($n=824$), with all models trained for 30 epochs. Mean differences, pointwise 95\% bootstrap intervals, strict win rates, two-sided sign-flip permutation $p$-values and Holm-adjusted $p$-values are reported for the six primary acoustic metrics. Higher differences favor the first model.}\label{tab:reverb_trained_pair_vb}
\fitTable{\ReverbTrainedPairedVB}
\end{table}
\begin{table}[H]\centering\small
\caption{Paired statistical analysis on \textbf{reverberant DNS 2020} ($n=150$), under matched reverberation-aware training. Format as in Table~\ref{tab:reverb_trained_pair_vb}; Holm correction is shared across both acoustic tables.}\label{tab:reverb_trained_pair_dns}
\fitTable{\ReverbTrainedPairedDNS}
\end{table}
\begin{table}[H]\centering\small
\caption{Paired differences in corpus WER (percentage points) for the reverberation-trained models; negative favors the first model. Pointwise 95\% bootstrap intervals, two-sided paired sign-flip $p$-values and Holm-adjusted $p$-values (separate six-test WER family).}\label{tab:reverb_trained_werpaired}
\fitTable{\ReverbTrainedWERPaired}
\end{table}
\subsection{Additional restoration diagnostics}
\begin{figure}[H]\centering
\begin{minipage}[t]{.485\linewidth}\centering\small Synthetic reverberant VoiceBank\\\includegraphics[width=\linewidth]{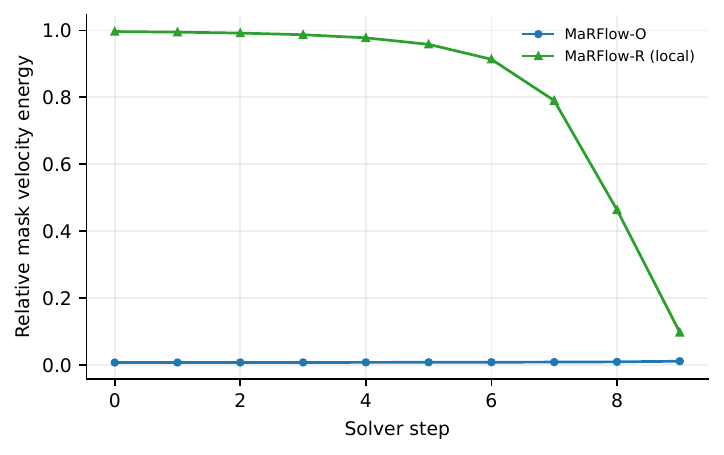}\end{minipage}\hfill
\begin{minipage}[t]{.485\linewidth}\centering\small Reverberant DNS\\\includegraphics[width=\linewidth]{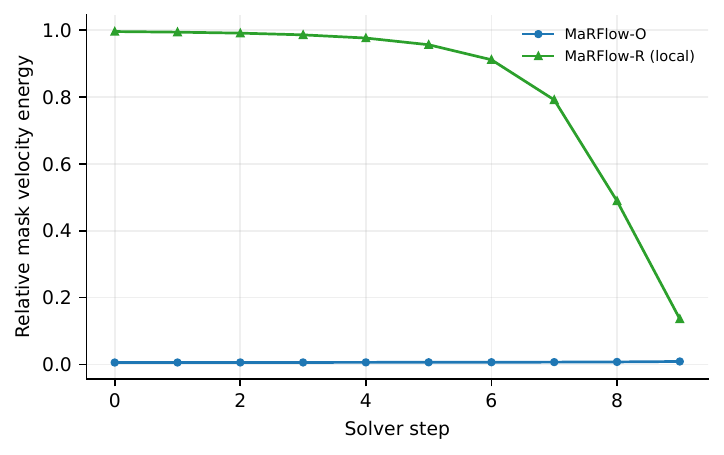}\end{minipage}
\caption{Pooled multiplicative velocity energy share for the reverberation-trained supervised models along ten-NFE trajectories. As in Fig.~\ref{fig:branch_energy}, shares omit the cross-term and describe compressed-coordinate squared magnitudes, not physical speech energy.}\label{fig:reverb_trained_branches}
\end{figure}
\subsection{Training with reverberation and condition-specific results}

\begin{figure}[H]\centering
\includegraphics[width=.63\linewidth,height=1.9in,keepaspectratio]{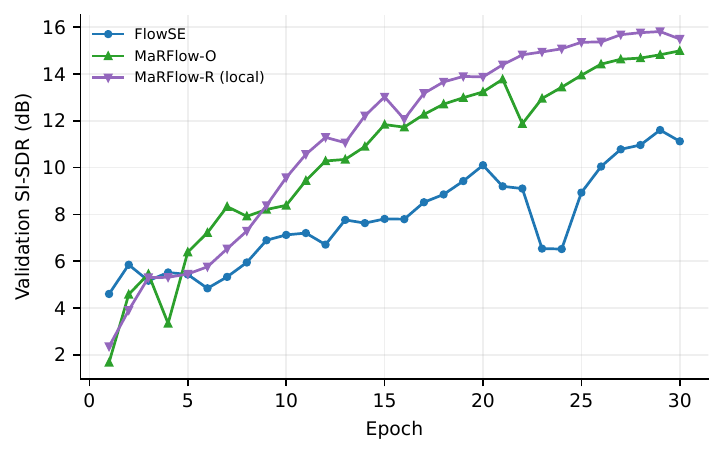}
\caption{Recorded validation SI-SDR for the three reverberation-trained models, evaluated on the fixed reverberant validation set at five NFE.}\label{fig:validation_reverb}
\end{figure}
\begin{table}[H]\centering\small
\caption{Validation checkpoint selection for the reverberation-trained models. All training runs complete 30 epochs; SI-SDR is in dB.}\label{tab:validation_reverb}
\fitTable{\ReverbTrainedValidation}
\end{table}
\paragraph{Validation convergence.} The best validation SI-SDR
values of 11.61, 14.99 and 15.82\,dB for FlowSE/O/R-local
occur at epochs 29, 30 and 29, respectively. The final values
are 11.13, 14.99 and 15.50\,dB. The supervised models improve
best validation SI-SDR by 3.39 and 4.21\,dB over FlowSE, and
R-local exceeds O by 0.82\,dB. Selection uses only the fixed
validation set; the better R-local validation score does not
imply that it must lead SI-SDR on the cross-domain DNS test.

\paragraph{Paired acoustic comparisons.} On synthetic VoiceBank,
O and R-local exceed FlowSE on PESQ for 91.1\% and 96.5\% of
utterances and on SI-SDR for 91.0\% and 96.2\%, respectively.
The corresponding DNS percentages are 98.0/97.3\% for PESQ
and 96.7/91.3\% for SI-SDR. R-local's OVRL advantages over O
are 0.097 [0.085, 0.109] and 0.078 [0.037, 0.119],
respectively (see the rounded values in the tables). Thus the
stronger mean results are not restricted to a few large gains,
although the paired intervals remain conditional on the
trained checkpoints and utterance cohorts.

\paragraph{Condition dependence.} Tables~\ref{tab:reverb_trained_strata_vb} and~\ref{tab:reverb_trained_strata_dns} summarize the condition means. Both supervised models improve
all six mean metrics over FlowSE in each synthetic VoiceBank
RIR, SNR and speaker stratum. Across noise strata, the same
holds except that R-local's mean P.808 for car noise is lower
than FlowSE's (3.16 versus 3.19); O attains 3.20. For car noise,
O also has slightly lower STOI and ESTOI than the noisy input.
R-local leads O in all six metrics in each RIR and speaker
stratum, and in at least five of six metrics in every noise and SNR
stratum; the P.808 exception is car noise and, by a much
smaller margin, the 15-dB SNR group. In synthetic VoiceBank,
R-local's SI-SDR advantage over O is 1.16\,dB at 0\,dB SNR
and 0.65\,dB at 15\,dB, with an OVRL advantage of 0.17 and
0.05, respectively. The advantage is present for every RIR
identifier (SI-SDR gain 0.61--1.08\,dB) and all seven noise
recordings (0.57--1.85\,dB). On reverberant DNS, both supervised
models improve all six means over both FlowSE and the noisy
input in every SNR group. R-local leads O on the five non-SI-SDR
metrics throughout; the SI-SDR advantage switches from R-local
below 0\,dB to O at nonnegative SNR, where the largest mean
difference is 0.61\,dB at $\geq10$\,dB.
Only seven DNS utterances have negative SNR, and the synthetic
VoiceBank results involve just two test speakers and shared
distortion recordings. We interpret these patterns descriptively,
not as tests of interactions with room, speaker or SNR.

\begin{table}[H]\centering\small\setlength{\tabcolsep}{3pt}
\caption{Synthetic reverberant VoiceBank condition means at five NFE. In each metric cell the order is noisy / FlowSE / O / R-local (N/F/O/R). RIR identifiers are the supplied filenames, not independently measured reverberation times. Counts are shared across models.}\label{tab:reverb_trained_strata_vb}
\fitTable{\ReverbTrainedStrataVB}
\end{table}
\begin{table}[H]\centering\small\setlength{\tabcolsep}{3pt}
\caption{Reverberant DNS condition means at five NFE for the reverberation-trained models. Format as in Table~\ref{tab:reverb_trained_strata_vb}.}\label{tab:reverb_trained_strata_dns}
\fitTable{\ReverbTrainedStrataDNS}
\end{table}
\begin{table}[H]\centering\small
\caption{Absolute five-NFE intervention scores for synthetic reverberant VoiceBank. Add. only and Mult. only retain only the indicated velocity contribution at each update.}\label{tab:reverb_trained_suppabs_vb}
\fitTable{\ReverbTrainedSuppAbsVB}
\end{table}
\begin{table}[H]\centering\small
\caption{Absolute five-NFE intervention scores for reverberant DNS. Same format as Table~\ref{tab:reverb_trained_suppabs_vb}.}\label{tab:reverb_trained_suppabs_dns}
\fitTable{\ReverbTrainedSuppAbsDNS}
\end{table}
\begin{table}[H]\centering\small
\caption{Mean full-minus-intervened score differences and pointwise 95\% paired bootstrap intervals on synthetic reverberant VoiceBank.}\label{tab:reverb_trained_supp_vb}
\fitTable{\ReverbTrainedSuppressionVB}
\end{table}
\begin{table}[H]\centering\small
\caption{Mean full-minus-intervened score differences and pointwise 95\% paired bootstrap intervals on reverberant DNS. Same format as Table~\ref{tab:reverb_trained_supp_vb}.}\label{tab:reverb_trained_supp_dns}
\fitTable{\ReverbTrainedSuppressionDNS}
\end{table}
\paragraph{Complementarity and uncertainty.} For the secondary
multiplicative contribution of O, the paired BAK intervals are
[1.35, 1.46] and [0.68, 0.91] on synthetic VoiceBank/DNS.
For the complementary additive coordinate in R-local, they
are [1.67, 1.74] and [0.96, 1.19]. Suppressing the primary
branch of either model yields negative mean SI-SDR on both
sets, while retaining the primary branch alone yields mean
SI-SDR above that of noisy input. The branch-suppression
results therefore distinguish functional dependence from the
numerical magnitude of a velocity coordinate, as in the
original study. All intervals in Tables~\ref{tab:reverb_trained_supp_vb}
--\ref{tab:reverb_trained_supp_dns} are pointwise; the interval
for O's multiplicative contribution to DNS ESTOI includes zero
($-0.0005$ to 0.0081 before rounding).

\begin{table}[H]\centering\small
\caption{Corpus-normalized substitution, deletion and insertion rates (\%) and WER after reverberation-aware training. The wet clean and noisy anchors are unchanged across models.}\label{tab:reverb_trained_werbreak}
\fitTable{\ReverbTrainedWERBreakdown}
\end{table}
\paragraph{Recognition error composition.} On reverberant DNS,
O and R-local reduce all three error components relative to
reverberation-trained FlowSE, but none reaches the noisy
anchor in any component. On synthetic VoiceBank, O has fewer
substitutions and deletions than FlowSE but more insertions
(5.73 versus 3.63\% of reference words); R-local reduces
substitutions and deletions relative to FlowSE while increasing
insertions more modestly (4.12\%). These are corpus-normalized
counts from the frozen recognizer, not diagnoses of the
underlying acoustic causes.
\FloatBarrier

\section{Implementation and statistical protocol}\label{app:implementation}
\paragraph{Spectral representation and data.} Audio is mono at 16\,kHz. Paired signals are jointly scaled to avoid peaks above 0.95, then normalized by the noisy peak for the spectral model. Training uses aligned random 32,640-sample crops (2.04 s), zero-padding shorter signals and shuffling examples each epoch. The centered STFT uses 510-point FFT, a periodic Hann window, hop 128, and 256 frequency bins and 256 frames per training segment. Amplitude compression uses exponent 0.5 and scale 0.15. Inference right-pads the frame dimension to a multiple of 64, then applies the inverse compression, ISTFT at the original waveform length, and restores the noisy scale.

\paragraph{Reverberant mixture construction.}
The separate reverberant training protocol uses the original 10,700 training and 872 validation clean-speech utterances. Let $s_{\mathrm{dry}}$ be the original speech, $h$ a selected unit-energy RIR channel, and $n$ a selected noise channel. We form the wet target $s_{\mathrm{wet}}=(h*s_{\mathrm{dry}})_{1:N}$, truncating the convolution to the original length, and the input $x=s_{\mathrm{wet}}+a n$, with $a$ set by the desired SNR relative to wet speech. Convolution precedes cropping; SNR is computed over the unpadded crop for training or the full utterance for evaluation. Both signals receive common amplitude scaling. Noise recordings are added directly without further convolution; reverberation modifies the speech reference and is not itself the additive noise term. The original DEMAND benchmark mixtures are not reused. The training RIR and additive-noise pools are the supplied real RIR/isotropic-noise and point-source-noise collections; recognized isotropic noise recordings, when present, form a second equally weighted noise pool. The sampling seed depends on seed 13, epoch and training row, so the mixture realizations are matched across models and vary across epochs. Training SNR is uniform on $[-5,20]$\,dB. Validation uses fixed mixtures from the training distortion pools, not test pools. The synthetic VoiceBank test uses six held-out RIR files, seven held-out noise recordings and SNRs $\{0,5,10,15\}$\,dB, with the 168 possible condition combinations assigned as evenly as possible in randomized order, one mixture per utterance. The preparation procedure checks for identical files across distortion pools by SHA-256; this does not establish disjoint physical rooms or acoustically distinct recordings. The DNS reverberant test uses the released wet pairs unchanged apart from common scaling. The training and inference protocol remains otherwise matched within the three-model study; no ridge sweep or additional trained variants are included.

\paragraph{Architecture and optimization.} NCSN++ has base width 128, channel multipliers $(1,1,2,2,2,2,2)$, two residual blocks per resolution, attention at frequency resolution 16, Fourier time embedding with scale 16, Swish activation, BigGAN residual blocks and zero dropout. The original output route retains its time scaling. The gain head is a zero-initialized $1\times1$ projection of the activated full-resolution decoder features (128 input channels, two output channels). The reference has 65,590,822 parameters; each factorized model has 65,591,080. We use Adam ($10^{-4}$), EMA decay 0.999, no gradient clipping for these models, batch size six, two data-loader workers, 30 epochs and seed 13. The complex squared-error objective is reduced as half the sum over non-batch dimensions followed by a batch mean. Time is uniform over $[0.03,1]$, and complex standard Gaussian draws have independent real and imaginary components of variance $1/2$. We use $\sigma_{\min}=0$, $\sigma_{\max}=0.487$; all supervised branch weights and aggregate weights are one. The O oracle uses $\lambda=0.01$ and $m_{\max}=3$; R-local uses $\lambda_{\mathrm{loc}}=0.01$ without clipping. Validation uses five-NFE SI-SDR and EMA parameters, with the best checkpoint used for testing. The FlowSE implementation is based on revision \texttt{f6b479d13fecc6cb6f12394f46dfc6799fb479b6}.

\paragraph{Evaluation reproducibility.} We evaluate full utterances, one at a time, using NVIDIA L4 GPUs. Software versions are PyTorch 2.10.0+cu128, torchaudio 2.10.0+cu128, pesq 0.0.4, pystoi 0.4.1, torchmetrics 1.9.0 and onnxruntime-gpu 1.22.0. The primary acoustic, zero-shot reverberant and ASR evaluations use the same five additive-noise-trained checkpoints. The matched reverberant evaluation uses the separately validation-selected three checkpoints. Both regimes use source seed 13, draw zero and five NFE for primary endpoint and ASR comparisons. The non-reverberant NFE sweeps use $1,2,3,5,10$ evaluations and additional draws at five NFE. The reverberant DNS pairs come from the 16-kHz DNS Challenge 2020 DevTest distribution (revision \texttt{a093fc1f5d3b3d5cefe0200f5c6f8813469f683e}); the clean reference is the released reverberant speech. The noisy/clean pairs use shared waveform scaling. VoiceBank uses the released text transcripts; the DNS transcript references are from DNS\_text$^{\ref{dnstext}}$, with IDs matched one-to-one. The recognizer is Whisper Large-v3, English transcription, beam size five, temperature zero, FP16 and no conditioning on previous text. We use openai-whisper 20250625 and JiWER 4.0.0 with Whisper's English text normalization.

\paragraph{Acoustic scoring.} We use wideband PESQ and the standard and extended PySTOI implementations at 16\,kHz, and compute SI-SDR after removing the waveform means. DNSMOS uses the non-personalized models, averaging complete 9.01-s windows with a one-second hop; shorter utterances are repeated until at least one window is available. Each window is zero-padded to 10\,s for the metric interface. Source draws are seeded per utterance and draw index by hashing the evaluation seed, utterance identifier and draw index with SHA-256 and mapping the first eight bytes to a nonnegative 63-bit seed.

\paragraph{Paired statistics.} For acoustic metrics we bootstrap paired utterance differences with replacement (5,000 replicates) and report the 2.5th and 97.5th percentiles and strict win rates. For WER the paired difference is $100\sum_i(E_{A,i}-E_{B,i})/\sum_i N_i$, with word-error counts $E_{A,i},E_{B,i}$ and reference lengths $N_i$. Each replicate resamples the aligned utterance and recomputes the ratio. Intervals are pointwise, conditional on the trained checkpoints and assume independent utterances. Inference draws are paired by utterance and draw index. The reverberation-trained analysis uses the same paired means, 5,000-resample pointwise intervals and strict win rates for acoustic comparisons, and the same corpus-ratio bootstrap for WER. We additionally use two-sided paired sign-flip tests with 20,000 Monte Carlo permutations and plus-one correction, retaining the existing tests for the reverberation-trained comparisons. For the additive-noise-trained checkpoints, the tests are computed from the aligned utterance-level results (random seed 13). For acoustic metrics the statistic is the mean paired difference; for WER it is the paired corpus error-count difference divided by the shared reference-word count. For the additive-noise-trained checkpoints, Holm correction is applied separately to all 28 tests in Table~\ref{tab:paired}, all 18 tests in Table~\ref{tab:reverbpaired}, and all 18 tests in Table~\ref{tab:werpaired}. For the reverberation-trained checkpoints, the 36 acoustic tests (three comparisons, six metrics and two datasets) form one Holm family; the six WER tests form a separate family. We report raw and adjusted $p$-values, with rejection at adjusted $p<0.05$. The minimum attainable unadjusted Monte Carlo $p$-value is $1/20001$ (approximately $5.00\times10^{-5}$). These tests are conditional on the checkpoints and require independent, sign-exchangeable paired units under the null; shared speakers and distortion recordings can weaken this assumption. Bootstrap intervals remain pointwise and are not simultaneous confidence intervals. Shared speakers, noise recordings and RIRs may induce dependencies not accounted for by utterance resampling; condition breakdowns and between-training-regime contrasts are descriptive.

\paragraph{Diagnostics and timing.}
Oracle diagnostics sample five times along the prescribed
conditional path: $t=0.1$, 0.3, 0.5, 0.7 and 0.9;
O and R-local use their respective target families.
Rollout branch statistics use ten NFE, exclude right padding
and use the actual multiplicative reference for each model.
Power bins use the compressed observation power with
boundaries $0.001,0.01,0.1$.

Branch energy share is defined as
\begin{equation}
\frac{\sum|\Phi_t\odot v_M|^2}
{\sum|\Phi_t\odot v_M|^2+\sum|v_R|^2},
\end{equation}
and branch alignment is measured by the normalized real
inner product between the two spectral velocity contributions.
These quantities describe squared magnitudes and directional
agreement in compressed spectral coordinates, rather than
physical acoustic energies.

Cumulative gain quantiles are computed per utterance and
then averaged across utterances. Additive-update propagation
is evaluated by multiplying each injected additive increment
by the subsequent gain product along the realized trajectory.
The corresponding retention ratio measures the pooled squared
magnitude of the propagated increment relative to its injected
squared magnitude in compressed coordinates.

Branch suppression uses five NFE and the same trained
checkpoint, with subsequent predictions evaluated on the
intervened solver states. In the separate reverberant study,
branch-energy profiles use ten NFE and branch suppression
uses five NFE for O and R-local only, with the same pooled
energy and paired bootstrap definitions. No additional
reverberant oracle agreement, gain-propagation or endpoint
NFE sweep is evaluated. Timing is synchronized, includes
waveform transforms and solver operations, excludes scoring,
and uses one warm-up utterance per uninstrumented endpoint
condition. Latency, RTF and peak allocated CUDA memory
are summarized by medians.

\end{document}